\RequirePackage{fix-cm}
\documentclass[11pt]{article}
\newcounter{ourcount}
\usepackage{amsmath,amsfonts,amssymb,graphicx,epsfig,pdflscape,multirow,theorem,gensymb}
\usepackage[T1]{fontenc}
\usepackage{lscape,arydshln,float}
\usepackage[matrix,arrow,curve]{xy} 
\numberwithin{equation}{section}
\usepackage[nosort]{cite}
\usepackage[usenames]{color}
\usepackage{pstricks}
\usepackage{pst-plot}
\usepackage[backref=false]{hyperref} 
\usepackage[capitalise,noabbrev]{cleveref}  
\usepackage{float} 
\usepackage[11pt]{moresize} 
\hypersetup{
colorlinks=true,
citecolor=red,
linkcolor=darkblue,
urlcolor=darkblue
} 
\definecolor{darkblue}{rgb}{0,0,.8}
\definecolor{red}{rgb}{1,0,0}

\theorembodyfont{\itshape} 
\theoremheaderfont{\scshape}
\theoremstyle{plain}

\numberwithin{equation}{section}

\newcommand{\nc}{\newcommand}

\def\arxiv#1#2{\href{http://arxiv.org/abs/#1/#2}{\textsf{arXiv:#1/#2}}}
\nc{\ir}{\mathrm{i}}
\nc{\dd}{\mathrm{d}}   
\nc{\eE}{\mathsf{e}}
\nc{\bib}{\bibitem}
\nc{\be}{\begin{equation}}
\nc{\ee}{\end{equation}}
\nc{\bea}{\begin{eqnarray}}
\nc{\eea}{\end{eqnarray}}
\nc{\chit}{\raisebox{0.25ex}{$\chi$}}
\nc{\dtl}{\mathsf{dTL}}
\nc{\pdtl}{\mathsf{pdTL}}

\nc{\Dbh}{\mbox{\boldmath $\widehat D$}}
\nc{\Dh}{\mbox{$\hat D$}}
\nc{\Dbb}{\mbox{\boldmath $\bar D$}}
\nc{\Dbm}{\mbox{\boldmath $\mathcal D$}}
\nc{\Dbt}{\mbox{\boldmath $\tilde{D}$}}
\nc{\Tbt}{\mbox{\boldmath $\tilde{T}$}}
\nc{\Tbh}{\mbox{\boldmath $\widehat{T}$}}

\nc{\setS}{\mathcal S}

\nc{\db}{\mbox{\boldmath $d$}}
\nc{\Ab}{\mbox{\boldmath $A$}}
\nc{\Bb}{\mbox{\boldmath $B$}}
\nc{\Cb}{\mbox{\boldmath $C$}}
\nc{\Db}{\mbox{\boldmath $D$}}
\nc{\eb}{\mbox{\boldmath $e$}}
\nc{\Fb}{\mbox{\boldmath $F$}}
\nc{\Fbt}{\mbox{\boldmath $\tilde{F}$}}
\nc{\fb}{\mbox{\boldmath $f$}}
\nc{\fbt}{\mbox{\boldmath $\tilde{f}$}}
\nc{\Gb}{\mbox{\boldmath $G$}}
\nc{\Hb}{\mbox{\boldmath $H$}}
\nc{\Ib}{\mbox{\boldmath $I$}}
\nc{\Jb}{\mbox{\boldmath $J$}}
\nc{\Kb}{\mbox{\boldmath $K$}}
\nc{\Lb}{\mbox{\boldmath $L$}}
\nc{\Mb}{\mbox{\boldmath $M$}}
\nc{\Pb}{\mbox{\boldmath $P$}}
\nc{\Qb}{\mbox{\boldmath $Q$}}
\nc{\Rb}{\mbox{\boldmath $R$}}
\nc{\Tbb}{\mbox{\boldmath $\bar T$}}
\nc{\Tbm}{\mbox{\boldmath $\mathcal T$}}
\nc{\tb}{\mbox{\boldmath $t$}}
\nc{\Ub}{\mbox{\boldmath $U$}}
\nc{\Vb}{\mbox{\boldmath $V$}}
\nc{\Wb}{\mbox{\boldmath $W$}}
\nc{\xb}{\mbox{\boldmath $x$}}
\nc{\yb}{\mbox{\boldmath $y$}}
\nc{\Zb}{\mbox{\boldmath $Z$}}
\nc{\Lambdab}{\boldsymbol{\Lambda}}
\nc{\T}{\mbox{\boldmath $T$}}
\nc{\D}{\mbox{\boldmath $D$}}
\nc{\B}{\mbox{\boldmath $B$}}
\nc{\diagD}{\mbox{\boldmath $\Lambda$}}
\nc{\Ta}{\mbox{\boldmath $T$}^a}
\nc{\Tb}{\mbox{\boldmath $T$}^b}
\nc{\Tk}{\mbox{\boldmath $T$}^\kappa}

\def\nn#1{\mbox{\boldmath $\mathbf n$}_{#1}}

\def\half {\mbox{$\textstyle {1 \over 2}$}}

\nc{\even}{\textrm{ even}}
\nc{\odd}{\textrm{ odd}}

\nc{\Atwotwo}{\mbox{$A_2^{\textrm{\fontsize{7pt}{7pt}\selectfont $(2)$}}$}}
\nc{\Aoneone}{\mbox{$A_1^{\textrm{\fontsize{7pt}{7pt}\selectfont $(1)$}}$}}
\nc{\amf}{\mbox{$\mathfrak a$}}
\nc{\bmf}{\mbox{$\mathfrak b$}}
\nc{\cmf}{\mbox{$\mathfrak c$}}
\nc{\dmf}{\mbox{$\mathfrak d$}}
\nc{\fmf}{\mbox{$\mathfrak f$}}
\nc{\gmf}{\mbox{$\mathfrak g$}}
\nc{\Amf}{\mbox{$\mathfrak A$}}
\nc{\Bmf}{\mbox{$\mathfrak B$}}
\nc{\Cmf}{\mbox{$\mathfrak C$}}
\nc{\Dmf}{\mbox{$\mathfrak D$}}
\nc{\Fmf}{\mbox{$\mathfrak F$}}
\nc{\asf}{\mbox{$\mathsf a$}}
\nc{\bsf}{\mbox{$\mathsf b$}}
\nc{\csf}{\mbox{$\mathsf c$}}
\nc{\dsf}{\mbox{$\mathsf d$}}
\nc{\fsf}{\mbox{$\mathsf f$}}
\nc{\gsf}{\mbox{$\mathsf g$}}
\nc{\Asf}{\mbox{$\mathsf A$}}
\nc{\Bsf}{\mbox{$\mathsf B$}}
\nc{\Csf}{\mbox{$\mathsf C$}}
\nc{\Dsf}{\mbox{$\mathsf D$}}

\nc{\repV}{\mathsf{V}}
\nc{\repW}{\mathsf{W}}
\newrgbcolor{darkgreen}{0., 0.733333, 0.0621042}
\definecolor{lightblue}{rgb}{.7,.7,1}
\definecolor{lightestblue}{rgb}{.95,.95,1}
\definecolor{lightlightblue}{rgb}{.93,.93,1}
\definecolor{midblue}{rgb}{.7,.7,1}
\definecolor{lightyellow}{rgb}{1.,.929,.514}
\definecolor{lightorange}{rgb}{.996,.847,.694}
\definecolor{lightpurple}{rgb}{.999,.7,.999}
\definecolor{lightpurple}{rgb}{1,.8,1}
\definecolor{darkpurple}{rgb}{1,.575,1}

\nc{\elegant}{1.5pt}
\nc{\moyen}{1.0pt}
\nc{\mince}{0.5pt}
\def\disp{\displaystyle}
\def\mypmatrix#1{\begin{pmatrix}#1\end{pmatrix}}

\def\ssmat#1{\mbox{\setlength\arraycolsep{2.5pt}\renewcommand*{\arraystretch}{.95} \scriptsize{\mbox{$\mypmatrix{#1}$}}}}
\def\sssmat#1{\mbox{\setlength\arraycolsep{1.5pt}\renewcommand*{\arraystretch}{.95} {\mbox{\scriptsize$\mypmatrix{#1}$}}}}

\def\vvdots{\mathinner{\mkern1mu\raise1pt\vbox{\kern7pt\hbox{.}}\mkern2mu
  \raise4pt\hbox{.}\mkern2mu\raise7pt\hbox{.}\mkern1mu}}

\def\facegrid#1#2{
\psframe[fillstyle=solid,fillcolor=lightlightblue,linewidth=0pt]#1#2
\psgrid[gridlabels=0pt,subgriddiv=1]#1#2}
\def\facegridy#1#2{
\psframe[fillstyle=solid,fillcolor=lightyellow,linewidth=0pt]#1#2
\psgrid[gridlabels=0pt,subgriddiv=1]#1#2}

\def\facegridp#1#2{
\psframe[fillstyle=solid,fillcolor=lightpurple,linewidth=0pt]#1#2
\psgrid[gridlabels=0pt,subgriddiv=1]#1#2}

\def\facegrid#1#2{
\psframe[fillstyle=solid,fillcolor=lightlightblue,linewidth=0pt]#1#2
\psgrid[gridlabels=0pt,subgriddiv=1]#1#2}

\renewcommand{\ge}{\geqslant}
\renewcommand{\le}{\leqslant}

\newcommand{\face}[5]{
\psset{unit=0.8cm}
\begin{pspicture}[shift=-.40](0,0)(1,1)
\facegrid{(0,0)}{(1,1)}
\psarc[linewidth=0.5pt,linecolor=red]{-}(0,0){0.16}{0}{90}
\rput(0.,-.1){\spos{tr}{#1}}
\rput(1.,-.1){\spos{tl}{#2}}
\rput(1.,1.1){\spos{bl}{#3}}
\rput(0.,1.1){\spos{br}{#4}}
\rput(.5,.5){\spos{c}{#5}}
\end{pspicture}}

\nc{\proof}{{\scshape Proof.\ }} 				
\nc{\eproof}{{\hfill \rule{0.5em}{0.5em}\medskip}}		
\def\Exp{{\rm Exp}}
\newcommand{\p}[2]{\makebox(0,0)[#1]{$#2$}}
\newcommand{\pp}[2]{\makebox(0,0)[#1]{$\ss#2$}}
\newcommand{\ade}{\mbox{$A$-$D$-$E$ }}
\renewcommand{\ss}{\scriptstyle}

\def\Wt#1#2#3#4#5{W\Big(\!\begin{array}{cc}#4&#3\\#1&#2\end{array}\!\Big|\,#5\Big)}
\def\Wtt#1#2#3#4#5{#1\Big(\!\begin{array}{cc}#5&#4\\#2&#3\end{array}\Big)}
\nc{\spos}[2]{\makebox(0,0)[#1]{$\sm{#2}$}}
\nc{\sm}[1]{{\scriptstyle #1}}

\def\re{\mathop{\rm re}}

\def\g{\mathsf g}
\def\d{\mathsf d}
\def\vec#1{\boldsymbol{#1}}

\usepackage{todonotes}
\usepackage{braket}

\DeclareMathOperator{\Tr}{Tr}
\usepackage{mathtools}

\DeclarePairedDelimiterX\Gbraket[2]{(}{)}{#1\,\delimsize\vert\,\mathopen{}#2}

\begin{document}

\topmargin -5mm
\oddsidemargin 5mm
\vspace*{-2cm}

\makeatletter 
\newcommand\Larger{\@setfontsize\semiHuge{20.00}{21.00}}
\makeatother 

\setcounter{page}{1}
\mbox{}\vspace{.5cm}\mbox{}
\begin{center}

{\Larger \bf \mbox{
Unitary and nonunitary $A$-$D$-$E$ minimal models:\ \ }
\\[0.18cm] 
{\Larger \bf \mbox{Coset graph fusion algebras, defects}}
\\[0.18cm] 
{\Larger \bf \mbox{and dilogarithm identities }}
\\[.25cm]
}

\end{center}

\vspace{.6cm}
\begin{center}
{\vspace{-2mm}\Large Paul A.~Pearce$^{a,b,c}$, Jared Heymann$^a$, Thomas Quella$^a$}
\\[.5cm]
{\em { }$^a$School of Mathematics and Statistics, University of Melbourne\\
Parkville, Victoria 3010, Australia}
\\[.2cm]
{\em { }$^b$School of Mathematics and Physics, University of Queensland}\\
{\em St Lucia, Brisbane, Queensland 4072, Australia}
\\[.2cm] 
{\em { }$^c$High Energy Physics Research Unit, Faculty of Science}\\
{\em Chulalongkorn University, Bangkok 10330, Thailand}
\\[.2cm] 
\qquad
{\tt papearce@unimelb.edu.au}
\qquad
{\tt jared\!.\!heymann@student.unimelb.edu.au}
\qquad
{\tt Thomas\!.\!Quella@unimelb.edu.au}
\end{center}

\vspace{12mm}
\centerline{{\bf{Abstract}}}
\vskip.3cm
\noindent 
We consider unitary and nonunitary $(A,G)$ coset minimal models on the cylinder with $G=A,D,E$. Elementary topological defects are implemented as non-contractible Verlinde, Pasquier and Ocneanu lines around the cylinder. 
The decomposition of compound defects, formed by fusing elementary defects together, 
is described by the Verlinde, Pasquier and Ocneanu graph fusion algebras. 
The action of the compound defects on the left- or right-vacuum boundary state builds the known conformal cylinder boundary conditions. 
Fusing these defects, in the presence of vacuum boundary conditions, reproduces the known general formulas for the conformal cylinder partition functions. 
We define the coset graph $A\otimes G/\mathbb{Z}_2$, argue that it is a universal object and show that
it encodes (i) the coset graph fusion algebra, (ii) the Affleck-Ludwig boundary $\g$-factors, (iii) the defect $\g$-factors given by quantum dimensions and (iv) the relative Symmetry Resolved Affleck-Ludwig Entropies (SRALEs). 
Additionally, it is shown that the boundary and defect $\g$-factors are related to the asymptotic counting of fusion paths on the coset graph. 
On the lattice, the defects are constructed as Yang-Baxter integrable seams including special braid and graph automorphism transfer matrices. 
Remarkably, many of the boundary CFT discrete structures, such as fusion matrices, modular matrices, quantum dimensions and defects, appear at the level of lattice Yang-Baxter integrable \ade Restricted Solid-On-Solid (RSOS) models and these structures and their properties carry over to the CFT in the continuum scaling limit. 
Importantly, in the continuum scaling limit, the lattice transfer matrix $T$- and $Y$-system functional equations carry over to produce the coset graph fusion algebra for the defect lines. 
Moreover, the effective central charges and conformal weights are expressed in terms of dilogarithms of the braid and bulk asymptotics of the $Y$-system expressed in terms of the quantum dimensions.

\newpage
\tableofcontents

\newpage
\hyphenpenalty=30000

\setcounter{footnote}{0}

%
\section{Introduction}
%

This paper focusses on the discrete structures of Yang-Baxter integrable lattice models~\cite{BaxBook82} and their associated Conformal Field Theories (CFTs)~\cite{FMS97}, realized in the continuum scaling limit. 
We work in the context of the unitary and nonunitary \ade Restricted Solid-On-Solid (RSOS) lattice models~\mbox{\cite{ABF84,FB,Pasquier87a,Pasquier87b,Pasquier87c}} and the minimal CFTs ${\cal M}(m,m')$~\cite{BPZ84}. 
The minimal models are rational CFTs~\cite{MooreSeiberg} with respect to the Virasoro algebra and admit a realization in terms of the Goddard-Kent-Olive (GKO) coset construction~\cite{GKO1985}. 
The modular invariant partition functions of these CFTs on the torus are classified by Cappelli, Itzykson and Zuber~\cite{CIZ87} in terms of the simply-laced \ade Lie algebras and their Dynkin graphs. The conformal partition functions on the cylinder, or equivalently the annulus, for conformal boundary conditions are given  in \cite{BPPZ}.  The twisted partition functions, resulting from the insertion of topological defect lines on the torus, were obtained by Petkova and Zuber~\cite{PetkovaZuber2001,PetkovaZuberFaces2001}.
Recently, it was proposed~\cite{PRasmussen24,PRasmussen25} that these topological defect lines for the \ade minimal CFTs on the torus satisfy the Ocneanu algebra~\cite{Ocneanu}. 

Recent developments across various areas of physics have emphasized that topological defects and interfaces are the natural tool to study global symmetries and dualities \cite{Frohlich:2010maph.conf..608F,Gaiotto:2015JHEP...02..172G,Bhardwaj:2018JHEP...03..189B}. This perspective has been particularly influential and powerful in 2D CFTs which naturally arise as continuum limits of integrable and non-integrable classical and quantum lattice models \cite{Fateev:1985mm,Affleck:1988NuPhB.305..582A,Affleck:1989JPhA...22..511A,BazhResh1989,KP92} and where fusion categories provide a rigorous and well-developed mathematical framework for the study of their symmetries \cite{Fuchs:2002NuPhB.646..353F}. Yet, despite impressive progress there still exist a number of open problems concerning the relation between continuum CFTs and discrete lattice models. Here we explore the interplay between the symmetries and other properties encoded by Dynkin diagrams and their generalizations to fusion graphs.

A primary interest is to study the properties of defect lines of the unitary and nonunitary minimal models in cylindrical geometry in the presence of boundaries. 
The properties of defects for CFTs with diagonal modular invariants are well understood~\cite{STRSaleur2025} in terms of the Verlinde algebra~\cite{Verlinde88} but, for general nonunitary theories with non-diagonal modular invariants, it is necessary to properly incorporate the Pasquier~\cite{PasquierThesis} and Ocneanu~\cite{Ocneanu} graph fusion algebras.
We propose a new and general framework, based on coset graphs and their associated coset graph fusion algebras, to give a concise unified encapsulation of the defects, quantum dimensions, Affleck-Ludwig entropies and dilogarithm identitities. 
Our framework applies to all unitary and nonunitary A-D-E minimal models including nonunitary exceptional $E$ cases and, for example, allows to readily write down the fusion matrices. 
Our central premise is that the quantum dimensions, Affleck-Ludwig entropies and other properties of these theories are encoded through their associated universal {\em coset graphs} (as defined in Appendix~\ref{App:Coset}). This use of the word ``coset'' should not be confused with the distinct use of coset in the GKO construction. 
The insight of our framework will provide systematic tools to apply to more general classes of CFTs.

The layout of the paper is as follows. After the introduction we collate, in Section~\ref{cosetData}, the basic Virasoro minimal model CFT data including (i) the central charges, conformal weights, characters and modular $S$ matrices, (ii)~the coset graph algebras and their graph fusion matrices (nimreps) along with the properties of topological defect lines and (iii) the quantum dimensions, boundary/defect $\g$-factors~\cite{AL91} and (iv)~the relative Symmetry Resolved Affleck-Ludwig Entropies (SRALEs)~\cite{DGMN23,Northe23,KMOP23,DMVSB24,HQ24,CRZ24,SBDSMV24,BGPS25}. In Section~\ref{secExamples}, we present many prototypical examples of unitary and nonunitary Virasoro minimal CFTs alongside their CFT data derived from the coset graph. 
In Section~\ref{secLattice}, we use Yang-Baxter integrability~\cite{BaxBook82} to study the properties of the defect lines in these CFTs via the continuum scaling limit of integrable defect seams of the associated \ade\/ Restricted Solid-On-Solid (RSOS) lattice models~\cite{ABF84,FB,Pasquier87a,Pasquier87b,Pasquier87c,BPO96,BP2001}. More specifically, we (i)~construct integrable defect seams on the lattice, (ii)~consider the associated $T$- and $Y$-systems of functional equations satisfied by the transfer matrices and integrable seams, (iii)~obtain their braid and bulk limits in terms of quantum dimensions and (iv)~exhibit expressions for the central charges and conformal weights in terms of the analytic continuation of dilogarithms with arguments given by the braid and bulk limits.

\section{Data of \ade Minimal CFTs}
\label{cosetData}

\subsection{\ade Lie algebra data}

\begin{figure}[tht]
\begin{center}
\psset{unit=.5cm}
    \setlength{\unitlength}{.5cm}
\begin{equation*}
\mbox{}\hspace{1.1in}\mbox{}
\begin{array}{ccccc}
   \text{\text{Graph $G$\qquad}}& \mbox{}\quad\qquad\text{$m$}\quad\qquad\mbox{}&\text{$\Exp(G)$}&\text{Type/$H$}&\Gamma\\[10pt]
   \begin{pspicture}(6,1)
   	\rput(-3.8,2.2){Lie Algebra}
	\psline[linewidth=.6pt](-6.2,1.5)(26,1.5)
	\psline[linewidth=.6pt](-6.2,-14.6)(26,-14.6)
        \rput(-3.5,0){\p{}{A_{m-1}}}
     \psline[linewidth=.5pt](0,0)(4,0)
        \multirput(1,0)(1,0){2}{\pp{}{\bullet}}
        \rput(4,0){\pp{}{\bullet}}
        \rput(0,0){\pp{}{\bullet}}
       \rput(0,.3){\pp{b}{1}}
        \rput(1,.3){\pp{b}{2}}
        \rput(2,.3){\pp{b}{3}}
        \rput(3,.3){\pp{b}{\cdots}}
        \rput(4,.3){\pp{b}{m-1}}
    \end{pspicture}  & m & 1, 2, \ldots, m-1 &\mbox{I}&{\Bbb Z_2}\\
     \begin{pspicture}(6,2)
        \rput(-3.5,0){\p{}{D_{l +2}\,\mbox{($l$ even)}}}
     \psline[linewidth=.5pt](0,0)(3.5,0)
     \psline[linewidth=.5pt](4.5,1)(3.5,0)
     \psline[linewidth=.5pt](4.5,-1)(3.5,0)
        \multirput(1,0)(1,0){2}{\pp{}{\bullet}}
        \rput(3.5,0){\pp{}{\bullet}}
        \rput(4.5,1){\pp{}{\bullet}}
        \rput(4.5,-1){\pp{}{\bullet}}
        \rput(-.2,-.04){\pp{l}{\bullet}}
        \rput(0,.3){\pp{b}{1}}
        \rput(1,.3){\pp{b}{2}}
        \rput(2,.3){\pp{b}{3}}
        \rput(2.75,.3){\pp{b}{\cdots}}
        \rput(3.5,.3){\pp{b}{l}}
        \rput(4.5,1){\pp{l}{~l+1}}
        \rput(4.5,-.7){\pp{l}{~l+2}}
    \end{pspicture}    & 2l +2 & 1, 3, \ldots, 2l +1, l +1 &\mbox{I}&{\Bbb Z_2} \\[10pt]
 \begin{pspicture}(6,2)
        \rput(-3.5,0){\p{}{D_{l +2}\,\mbox{($l$ odd)}}}
     \psline[linewidth=.5pt](0,0)(3.5,0)
     \psline[linewidth=.5pt](4.5,1)(3.5,0)
     \psline[linewidth=.5pt](4.5,-1)(3.5,0)
        \multirput(0,0)(1,0){3}{\pp{}{\bullet}}
        \rput(3.5,0){\pp{}{\bullet}}
        \rput(4.5,1){\pp{}{\bullet}}
        \rput(0,.3){\pp{b}{1}}
        \rput(1,.3){\pp{b}{2}}
        \rput(2,.3){\pp{b}{3}}
        \rput(2.75,.3){\pp{b}{\cdots}}
        \rput(3.5,.3){\pp{b}{l}}
        \rput(4.5,1){\pp{l}{~l+1}}
        \rput(4.5,-.7){\pp{l}{~l+2}}
        \rput(4.35,-1.05){\pp{l}{\bullet}}
    \end{pspicture}    & 2l +2 & 1, 3, \ldots, 2l +1, l +1 &\mbox{II}/A_{2l+1} &{\Bbb Z_2}\\
   \begin{pspicture}(6,2.5)
        \rput(-3.5,0){\p{}{E_{6}}}
      \psline[linewidth=.5pt](0,0)(4,0)
       \psline[linewidth=.5pt](2,0)(2,1)
        \multirput(1,0)(1,0){4}{\pp{}{\bullet}}
        \rput(2,1){\pp{}{\bullet}}
        \rput(0,0){\pp{}{\bullet}}
        \rput(0,.3){\pp{b}{1}}
        \rput(1,.3){\pp{b}{2}}
        \rput(2,-.3){\pp{t}{3}}
        \rput(3,.3){\pp{b}{4}}
        \rput(4,.3){\pp{b}{5}}
        \rput(2,1.3){\pp{b}{6}}
     \end{pspicture}  & 12 & 1, 4, 5, 7, 8, 11 &\mbox{I}&{\Bbb Z_2} \\
   \begin{pspicture}(6,2.5)
        \rput(-3.5,0){\p{}{E_{7}}}
       \psline[linewidth=.5pt](0,0)(5,0)
       \psline[linewidth=.5pt](3,0)(3,1)
        \multirput(1,0)(1,0){5}{\pp{}{\bullet}}
         \rput(3,1){\pp{}{\bullet}}
        \rput(0,.3){\pp{b}{1}}
        \rput(1,.3){\pp{b}{2}}
        \rput(2,.3){\pp{b}{3}}
        \rput(3,-.3){\pp{t}{4}}
        \rput(4,.3){\pp{b}{5}}
        \rput(5,.3){\pp{b}{6}}
        \rput(3,1.3){\pp{b}{7}}
        \rput(0,0){\pp{}{\bullet}}
     \end{pspicture}  & 18 & 1, 5, 7, 9, 11, 13, 17  &\mbox{II}/D_{10}&1\\
   \begin{pspicture}(6,2.5)
        \rput(-3.5,0){\p{}{E_{8}}}
       \psline[linewidth=.5pt](0,0)(6,0)
       \psline[linewidth=.5pt](4,0)(4,1)
        \multirput(1,0)(1,0){6}{\pp{}{\bullet}}
        \rput(4,1){\pp{}{\bullet}}
        \rput(0,0){\pp{}{\bullet}}
        \rput(0,.3){\pp{b}{1}}
        \rput(1,.3){\pp{b}{2}}
        \rput(2,.3){\pp{b}{3}}
        \rput(3,.3){\pp{b}{4}}
        \rput(4,-.3){\pp{t}{5}}
        \rput(5,.3){\pp{b}{6}}
        \rput(6,.3){\pp{b}{7}}
        \rput(4,1.3){\pp{b}{8}}
     \end{pspicture}  & 30 & 1, 7, 11, 13, 17, 19, 23, 29&\mbox{I}&1\\
\end{array}
\end{equation*}
\end{center}
\caption{Dynkin diagrams of the classical simply-laced \ade 
Lie algebras. The nodes associated with the identity and the fundamental are
labelled by 1 and 2 respectively. We note that the fundamental is the unique neighbour of the identity. Also shown are the Coxeter numbers $m$, exponents $\Exp(G)$, the type I or II, the so-called parent \ade graph $H\ne G$ and the diagram automorphism group $\Gamma$. It is the exponents of the parent graph $H$ that appear in the modular invariant partition functions as in Table~\ref{MIPFs}. 
The $D_4$ graph is exceptional having the noncommutative automorphism group $\Bbb S_3$. The eigenvalues of $G$ are $2\cos \tfrac{\ell\pi}{m}$ with $\ell\in \Exp(G)$. 
By abuse of notation, we use $G$ to denote the graph, its adjacency matrix and its set of vertices with cardinality $|G|$ but the meaning should be clear from context. }
\label{fig:Graphs}
\end{figure}

The \ade minimal CFTs~\cite{FMS97} are coset models built on a pair of Dynkin diagrams of simply-laced \ade Lie algebras with coprime Coxeter numbers $(m,m')$ and data as shown in Figure~\ref{fig:Graphs}. The models can be characterized by the pair $(A,G)$ of \ade diagrams, or more precisely, by the coset graph $A\otimes G/\mathbb{Z}_2$. 
Within graph theory~\cite{AlgGraphTheory}, coset graphs are constructed mathematically by combining the standard graph theory constructs of tensor products and quotients of graphs as described in Appendix~\ref{App:Coset}.
The $(A,A)$ series is associated with the minimal models ${\cal M}(m,m')$ of Belavin, Polyakov and Zamolodchikov~\cite{BPZ84}. 
More generally, these models are in fact unitary ($m'=m\pm 1$) or nonunitary ($m'\neq m\pm 1$) coset models~\cite{GKO1985,FMS97}. 
The unitary \ade minimal models are classified~\cite{CIZ87} into a critical and tricritical series
\be
(A,G)=\begin{cases}
(A_{m-2},D_{(m+2)/2}),\\
(A_{10},E_6),\\
(A_{16},E_7),\\
(A_{28},E_8),
\end{cases}\quad
(A,G)=\begin{cases}
(A_{m},D_{(m+2)/2}),\qquad&m=6,8,10,\ldots\\
(A_{12},E_6),\ \ \  &m=12\\
(A_{18},E_7),\ \ \ &m=18\\
(A_{30},E_8),\ \ \ &m=30
\end{cases}
\ee
where $m$ is the Coxeter number of $G$. 
The coset theories $(A,G)$ and $(G,A)$ are equivalent so we only use the $(A,G)$ pair. There is no distinction between the two $A$-types $(A_{m-1},A_{m})\equiv(A_{m},A_{m-1})$. The critical Ising model is $(A_3,A_4)$ and the tricritical Ising model is $(A_4,A_5)$.  The critical 3-state Potts model is $(A_4,D_4)$ and the tricritical 3-state Potts model is $(A_6,D_4)$. 
As shown in Table~\ref{MIPFs}, the Coxeter exponents $\Exp(A)$ and $\Exp(H)$ appear in the modular invariant partition functions~\cite{CIZ87}. 

Defining q-numbers by $S_a=[a]_{x}=\frac{x^{a}-x^{-a}}{x-x^{-1}}$  with $x=e^{\pi i/m}$, the nondegenerate largest eigenvalue of the adjacency matrix $G$ is $[2]_{x}=x+x^{-1}=2\cos\tfrac{\pi}{m}$ and the associated (unnormalized) Perron-Frobenius eigenvector $\boldsymbol{\psi}$ is 
\be
G\boldsymbol\psi=[2]_x\,\boldsymbol\psi,\qquad
\boldsymbol\psi=(\psi_a)_{1\le a\le |G|}=\begin{cases}
\bigl( [1]_{x},[2]_x,\ldots,[l]_x\bigr),&G=T_l\\[4pt]
\bigl( [1]_{x},[2]_x,\ldots,[l]_x\bigr),&G=A_l\\[4pt]
\bigl( [1]_{x}, [2]_x,\ldots,[l]_x,\frac{[l]_{x}}{[2]_{x}},\frac{[l]_{x}}{[2]_{x}}\bigr),&G=D_{l+2}\\[4pt]
\bigl( [1]_{x}, [2]_{x}, [3]_{x}, [2]_{x}, [1]_{x},\frac{[3]_{x}}{[2]_{x}}\bigr),&G=E_6\\[4pt]
\bigl( [1]_{x}, [2]_{x}, [3]_{x}, [4]_{x}, \frac{[6]_{x}}{[2]_{x}},\frac{[4]_{x}}{[3]_{x}},\frac{[4]_{x}}{[2]_{x}}\bigr),&G=E_7\\[4pt]
\bigl( [1]_{x}, [2]_{x}, [3]_{x}, [4]_{x},  [5]_{x}, \frac{[7]_{x}}{[2]_{x}},\frac{[5]_{x}}{[3]_{x}},\frac{[5]_{x}}{[2]_{x}}\bigr),&G=E_8
\end{cases}
\label{PF}
\ee
where $\psi_a=[a]_x=S_a$ for the $A$ series and $T_l$, with $m=2l+1$, is the tadpole graph with $l$ nodes. To normalize these vectors, we need their norms
\bea
\|\boldsymbol\psi\|=\begin{cases}
\tfrac{\sqrt{2L+1}}2\csc\tfrac{\pi}{2L+1},\ \ \sqrt{\tfrac{L+1}{2}}\csc\tfrac{\pi}{L+1},\ \ \sqrt{\tfrac{L-1}{2}}\csc\tfrac{\pi}{2L-2},\quad&G=T_L, A_L, D_L\\[6pt]
\frac{\sqrt{3-\sqrt{3}}}{\sin\tfrac{\pi}{12}}=\sqrt{\tfrac{24}{3-\sqrt{3}}},\ \ \frac{\sqrt{9/2}}{\sin\tfrac{\pi}{18}},\ \ \frac{\sqrt{15-\sqrt{75+30\sqrt{5}}}}{2\sin\tfrac{\pi}{30}},\ \ \ &G=E_{6,7,8}
\end{cases}
\eea
For unitary models models ($m'\!-\!m=\pm1$), $\boldsymbol{\psi}$ is chosen to be the Perron-Frobenius eigenvector with eigenvalue $2\cos\tfrac{\pi}{m}$. For nonunitary cases, $\boldsymbol{\psi}$ is chosen to be the eigenvector corresponding to the eigenvalue $2\cos \lambda$ with $\lambda=\tfrac{(m'-m)\pi}{m'}$ and $m'\!-\!m\in \Exp(G)$.

\subsection{Central charges, conformal weights and characters}

The  central charges $c$, conformal weights $\Delta$ and Virasoro characters $\chi_{r,s}^{m,m'}(q)$ of the \ade minimal CFTs are
\begin{subequations}
\begin{align}
&c=1-\frac{6(m'\!-\!m)^2}{mm'},\quad \Delta=\Delta_{r,s}^{m,m'}\!=\frac{(rm'\!\!-\!sm)^2\!-\!(m'\!\!-\!m)^2}{4mm'},\quad r=1,2,\ldots,m\!-\!1; s=1,2,\ldots, m'\!\!-\!1\\
&\chi_{r,s}^{m,m'}(q)=\frac{q^{-c/24+\Delta_{r,s}^{m,m'}}}{(q)_\infty} \sum_{k=-\infty}^\infty \big[q^{k(kmm'+rm'-sm)}-q^{(km+r)(km'+s)}\big],\qquad (q)_\infty=\prod_{k=1}^\infty (1-q^k)
\end{align}
\end{subequations}
where $q$ is the modular nome, $(r,s)$ are the Kac labels and $(r,s)=(1,1)$ is the vacuum. 
The Kac symmetry, giving the identification of fields, is given by $\Delta_{r,s}^{m,m'}=\Delta_{m-r,m'-s}^{m,m'}$ with $\chi_{r,s}^{m,m'}(q)=\chi_{m-r,m'-s}^{m,m'}(q)$ so
we can restrict to the set 
\bea
\mathbb{K}=\{\mbox{either $(r,s)$ or $(m\!-\!r,m'\!\!-\!s)$}: 1\le r\le m\!-\!1, 1\le s\le m'\!\!-\!1\},\quad |\mathbb{K}|=\tfrac12(m\!-\!1)(m'\!\!-\!1)
\eea
consisting of one representative from each equivalent pair ${(r,s)\equiv(m\!-\!r,m'\!\!-\!s)}$ of Kac labels. 
We regard this as an ordered set with an arbitrary but fixed choice for the ordering.
Once the ordering of the elements in $\mathbb{K}$ is fixed, the elements can be labelled by $\mu=1,2,\ldots,|\mathbb{K}|$. 
Depending on the parities of $m,m'$, it can be convenient to choose $\mathbb{K}$ by the restriction $r+s$ even or $r\le \tfrac12(m\!-\!1)$ or $s\le\tfrac12(m'\!-\!1)$. 

The effective central charges are
\bea
c_\text{eff}\!=\!c\!-\!24\Delta_\text{min}\!=\!1\!-\!\frac{6}{m m'},\qquad
\Delta_\text{min}=\!\mathop{\mbox{min}}_{(r,s)\in\mathbb{K}}\!\Delta_{r,s}^{m,m'}
\!=\!\Delta_{r_0,s_0}^{m,m'}
\!=\!\frac{1\!-\!(m'\!-\!m)^2}{4mm'}\label{ceff}
\eea
where $(r,s)=(r_0,s_0)$ is the groundstate. Specifically, since $m$ and $m'$ are coprime, $(r_0,s_0)$ is given by the solution of the Diophantine equation
\bea
m'r_0-ms_0=1\label{r0s0}
\eea
which is guaranteed to be unique by the Bezout lemma~\cite{FMS97}. For unitary models with $m'=m\pm 1$, one finds $(r_0,s_0)=(1,1)$, $\Delta_\text{min}=0$ and $c=c_\text{eff}$ whereas, 
for nonunitary models, $\Delta_\text{min}<0$ and $c\neq c_\text{eff}$.

\subsection{Classification of conformal partition functions and nimreps}

\begin{table}[htb]
\begin{center}
\small
\renewcommand{\arraystretch}{1.5}
\begin{tabular}{ll}   \hline Coset $(A,G)$&Modular Invariant Partition Function
\\ \hline \\[-12pt]
$(A_{m-1},A_{m'\!-1})$
&\rule[-15pt]{0pt}{30pt}$\disp{Z=\half\sum_{r=1}^{m-1}\sum_{s=1}^{m'-1}|\chi_{r,s}|^2}$\\
\parbox[t]{0.99in}{
$\disp{(A_{m-1},D_{2\rho+2})}$\par $\quad\scriptstyle{m'=4\rho+2\ge 6}$}
&\rule[-15pt]{0pt}{30pt}$\disp{Z=\half\sum_{r=1}^{m-1}
    \Biggl\{\sum_{s=1\atop s\ {\rm odd}}^{2\rho-1}
    |\chi_{r,s}+\chi_{r,4\rho+2-s}|^2+2|\chi_{r,2\rho+1}|^2\Biggr\}}$\\[16pt]
\parbox[t]{0.99in}{
$\disp{(A_{m-1},D_{2\rho+1})}$\par $\quad\scriptstyle{m'=4\rho\ge 8}$}
&\rule[-15pt]{0pt}{30pt}$\disp{Z=\half\sum_{r=1}^{m-1}
    \Biggl\{\sum_{s=1\atop s\ {\rm odd}}^{4\rho-1}
    |\chi_{r,s}|^2\!+\!|\chi_{r,2\rho}|^2\!+\!
    \sum_{s=2\atop s\ {\rm even}}^{2\rho-2}
(\chi_{r,s}\bar{\chi}_{r,4\rho-s}+\bar{\chi}_{r,s}\chi_{r,4\rho-s})\Biggr\}}
$\\
\parbox[t]{0.99in}{
$(A_{m-1},E_6)$\par $\quad \scriptstyle{m'=12}$}
&\rule[-15pt]{0pt}{30pt}$\disp{Z=\half\sum_{r=1}^{m-1}
     \biggl\{|\chi_{r,1}+\chi_{r,7}|^2+|\chi_{r,4}+\chi_{r,8}|^2+
     |\chi_{r,5}+\chi_{r,11}|^2\biggr\}}$\\
\parbox[t]{0.99in}{
$(A_{m-1},E_7)$\par $\quad \scriptstyle{m'=18}$}
&\rule[-15pt]{0pt}{30pt}$\disp{Z=\half\sum_{r=1}^{m-1}
     \biggl\{|\chi_{r,1}+\chi_{r,17}|^2+|\chi_{r,5}+\chi_{r,13}|^2+
     |\chi_{r,7}+\chi_{r,11}|^2}$\\[-10pt]
     &\qquad\qquad\quad$\disp{\qquad\hbox{}+|\chi_{r,9}|^2
+[(\chi_{r,3}+\chi_{r,15})\bar{\chi}_{r,9}+
     (\bar{\chi}_{r,3}+\bar{\chi}_{r,15})\chi_{r,9}]\biggr\}}$\\[-6pt]
\parbox[t]{0.99in}{
$(A_{m-1},E_8)$\par $\quad \scriptstyle{m'=30}$}
&\rule[-15pt]{0pt}{30pt}$\disp{Z=\half\sum_{r=1}^{m-1}
     \biggl\{|\chi_{r,1}+\chi_{r,11}+\chi_{r,19}+\chi_{r,29}|^2}$\\[-10pt]
&\qquad\qquad\quad$\disp{\qquad\hbox{}+|\chi_{r,7}+\chi_{r,13}+\chi_{r,17}+\chi_{r,23}|
^2
     \biggr\}}$ \vspace{4pt}\\ \hline
\end{tabular}
\renewcommand{\arraystretch}{1.0}
\normalsize
\end{center}
\caption{\ade classification of $(A,G)$ minimal model modular invariant partition functions on the torus. The central charges are
$c=1-{6(m-m')^2\over mm'}$, $\chi_{r,s}=\chi_{r,s}(q)$ are Virasoro
characters and bars denote complex conjugates. 
Here $r,s$ are the Coxeter exponents of $(A,G)$ and 
$(m,m')$ are coprime. The coset theories $(A,G)$ and $(G,A)$ are equivalent. The unitary minimal models have $m'=m\pm1$.}
\label{MIPFs}
\end{table}

The \ade CFTs can be considered in different topologies but they are classified~\cite{CIZ87} by the modular invariant torus partition functions shown in Table~\ref{MIPFs}. 
More generally, the twisted conformal partition functions on the torus are classified in terms of \ade Ocneanu graphs exhibited explicitly in \cite{PetkovaZuber2001}. Our main focus here will be on the cylinder which is equivalent by conformal transformation to the annulus. On the torus, the coset graph fusion algebra is replaced by the coset Ocneanu graph fusion algebra $A\otimes \mbox{Oc}(G)/\mathbb{Z}_2$ as in Appendix~\ref{App:Coset}.

To present the classification of the cylinder partition functions, we first need to define various graph fusion matrices. The fused adjacency matrices (intertwiners) $n_s$ and graph fusion matrices $\widehat{N}_a$ are defined by the finitely truncated recursions
\begin{subequations}
\begin{align}
n_1=&\;I,\quad n_2=G,\quad n_s n_2=n_{s-1}+n_{s+1},\quad n_0=n_{m'}=0;\qquad n_s=U_{n-1}(\tfrac12 G)\label{nRecursion}\\[4pt]
&\qquad\widehat{N}_1=I,\quad \widehat{N}_2=G,\quad G\,\widehat{N}_b=\sum_{c\in G}G_{bc}\,\widehat{N}_c,\qquad a,b,c\in G\label{GRecursion}
\end{align}
\end{subequations}
where $U_n(x)$ are the Chebyshev polynomials of the second kind and the graph fusion matrices $\widehat{N}_{a}$ only exist for graphs $G$ of type~I (see Figure~\ref{fig:Graphs}).
The $A_{m'\!-1}$ fusion matrices, denoted $n_i=N_i$, are the Verlinde fusion matrices~\cite{Verlinde88}. The various fusion matrices form nimreps (nonnegative integer matrix representations) of the commutative associative fusion algebras
\be
n_i\,n_j=\sum_{k=1}^{m'\!-1} N_{ij}{}^k\,n_k,\qquad 
\widehat{N}_a\,\widehat{N}_b=\sum_{c\in G} \widehat{N}_{ab}{}^c\,\widehat{N}_c,\qquad n_s\,\widehat{N}_a=\sum_{b\in G} n_{sa}{}^b\,\widehat{N}_b\label{FusionAlgebras}
\ee
where the nonnegative integer structure constants are
\be
N_{ij}{}^k=(N_i)_j{}^k,\qquad \widehat{N}_{ab}{}^c=(\widehat{N}_a)_b{}^c,\qquad n_{sa}{}^b=(n_s)_a{}^b\ \label{nNcompat}
\ee
The nimreps are all symmetric matrices except for the case $G=D_{4l}$ for which $\widehat{N}_{4l}=\widehat{N}_{4l-1}^T\neq \widehat{N}_{4l}^T$. In these cases, there are some complex eigenvalues. 
If the $\mathbb{Z}_2$ automorphism $\sigma$ is excluded for $D_{2l}$, then all the nimreps are normal, mutually commuting and simultaneously diagonalizable. 
Denoting the complex unitary matrices of (normalized) eigenvectors of $N_2$, $\widehat{N}_2$ by $S_i{}^\ell$ (modular matrix), $\Psi_a{}^\ell$ respectively with $S_1{}^\ell,\Psi_1{}^\ell>0$, the Verlinde and Verlinde-type formulas are
\be
 N_{ij}{}^k=\sum_{\ell=1}^{m'\!-1}{S_i{}^\ell\, S_j{}^\ell\, \overline{S_k{}^\ell}\over S_1{}^\ell},\quad 
 \widehat{N}_{ab}{}^c=\sum_{\ell\in {\rm Exp}(G)} 
{\Psi_a{}^\ell \Psi_b{}^\ell \,\overline{\Psi_c{}^\ell} \over \Psi_1{}^\ell},\quad
 n_{ia}{}^b=\sum_{\ell\in\Exp(G)} {S_i{}^\ell\,\Psi_a{}^\ell\, \overline{\Psi_b{}^\ell}\over S_1{}^\ell}
\ee
where bars denote complex conjugates.

The conformal cylinder partition functions~\cite{BPPZ} are given in terms of the graph fusion matrices $N_i$, $n_s$ and $\widehat{N}_a$ by
\be
Z_{(r'\!,b)|(r''\!,c)}^{m,m'}(q)=\begin{cases}
\disp\sum_{r=1}^{m-1}\sum_{a\in G}N_{rr'}{}^{r''}\!\widehat{N}_{ab}{}^c\, \hat{\chi}_{ra}(q),&\mbox{$G$ is type I}\\[14pt]
\disp\sum_{r=1}^{m-1}\sum_{s=1}^{m'\!-1}\! N_{rr'}{}^{r''}\!n_{sb}{}^{c}\, \chi_{rs}(q),&\mbox{$G$ is type I or II}
\end{cases}
\label{CylPFs}
\ee
where the block characters, associated with extended chiral symmetry in type I cases, are 
\bea
\hat{\chi}_{r,a}=\sum_{s=1}^{m'-1} n_{s1}{}^a \chi_{r,s}(q)
\eea
and the fundamental rectangular intertwiner $C$ has entries
\bea
C_{sa}=n_{s1}{}^a\label{intertwiner}
\eea
For type I theories, these two expressions agree since, by (\ref{nNcompat}),
\bea
\sum_{a\in G} n_{s1}{}^a \hat{N}_a=n_s \hat{N}_1=n_s
\eea
If $G$ is of $A$-type, the fusion matrices $n_s$ and $\widehat{N}_a$ both reduce to the Verlinde matrices $N_s$. The origin of the form of the cylinder partition functions is explained, in terms of propagating defect lines, in Figure~{\ref{PartitionFunctDecomp}. 
In the $G=D_{4l}$ cases, it is understood~\cite{BPZ1998} that the $b$ on the RHS of (\ref{CylPFs}) is replaced with $b^*$ where the star involution is defined by $a^*=a$ for $a=1,2,\ldots, 4l\!-\!2$, $(4l)^*=(4l\!-\!1)$ and $(4l\!-\!1)^*=(4l)$. In all other cases, the star conjugation reduces to the identity. We note that the $\mathbb{Z}_2$ graph automorphism is included as a generator of the graph fusion algebras in all cases except for the $D_{2l}$ cases. In these cases, adding the $\mathbb{Z}_2$ graph automorphism to the graph fusion algebra results in a noncommutative algebra (see for example \cite{PRasmussen25}).

\subsection{Coset graph fusion algebras and polynomial rings}

\begin{subequations}
Symbolically, the $sl(2)$ Verlinde fusion rules for $A_{m-1}$ are
\bea
(r)\times (r')=\sum_{r'\!'=1}^{m-1} N^{(m)}_{rr'}{}^{r'\!'} (r'')\;=\!\!\!\!\!\sum_{{r'\!'=|r-r'|+1}\atop r+r'+r'\!'=1\text{\ mod 2}}^{r_\text{max}} (r''),\qquad 1\le r,r'\le m-1\label{AfusionRules}
\eea
If $r'=1$, then $(r)\times (1)=(r)$ so $(1)$ acts as the identity. If $r'=2$ is the fundamental, then $r''$ are the neighbours of $r$ on $A_{m-1}$, that is, $(r)\times (2)= (r\!-\!1)+(r\!+\!1)$ for $r\le m-1$ and $(m\!-\!1)\times 2=(m\!-\!2)$.

As is well known, the Verlinde fusion rules for the $A$-type ${\cal M}(m,m')$ coset minimal models are given by the tensor product
\begin{align}
&(r,s)\times (r'\!,s') = \sum_{r'\!'\!=1}^{m-1}\sum_{s'\!'\!=1}^{m'\!-1} N^{(m)}_{rr'}{}^{r'\!'} N^{(m')}_{ss'}{}^{s'\!'} (r''\!,s'')=\sum_{{r'\!'=|r-r'|+1}\atop r+r'+r'\!'=1\text{\ mod 2}}^{r_\text{max}}\sum_{{s'\!'=|s-s'|+1}\atop s+s'+s'\!'=1\text{\ mod 2}}^{s_\text{max}} (r''\!,s'')\\[4pt]
&\quad\  r_\text{max}=\mbox{min}[r\!+\!r'\!-\!1,2m\!-\!r\!-\!r'\!-\!1],\qquad s_\text{max}=\mbox{min}[s\!+\!s'\!-\!1,2m'\!-\!s\!-\!s'\!-\!1]
\end{align}
\end{subequations}
subject to application of the ${\Bbb Z}_2$ Kac equivalence $(r,s)\equiv (m\!-\!r,m'\!-\!s)$. For example, restricting $\mathbb{K}$ so that $1\le s\le 3$ for ${\cal M}(5,7)$ and applying the Kac equivalence by hand gives
\begin{align}
(2,2')\times (3,3')&=(2\times 3,2'\!\times 3')=(2+4,2'\!+4')=(2,2')+(2,4')+(4,2')+(4,4')\nonumber\\
&\equiv (2,2')+(3,3')+(4,2')+(1,3')\label{egFusion}
\end{align}
where, for clarity, the $s$ labels in $(r,s)$ are shown with a prime. The $(r,s)$ nodes on the right side of (\ref{egFusion}) are the (diagonal) neighbours of $(3,3')$ in the Kac table. Continuing this process by applying the coset fundamental $(r,s)=(2,2)$ to consecutive neighbours $(r,s)$, with $r+s$ even and $1\le s\le 3$, leads to the connected component coset graph 
$\tilde{G}$ of Figure~\ref{M57CosetGraph}. Since $(3,3')$ is a neighbour of itself, this leads to a loop at this node of $\tilde{G}$. 

Algebraically, fusion rules are formalized as graph fusion algebras. To incorporate the ${\Bbb Z}_2$ Kac table symmetry in coset minimal model fusion graphs, we formalize new general coset graphs $\tilde{G}=(A_{m-1}\otimes A_{m'-1})/\mathbb{Z}_2$ in terms of graph products and quotients of graphs in Appendix~\ref{App:Coset}. 
For $G=A$ or $D$, the Kac symmetry is $(r,a)=(m-r,a)$ leading to the coset graph $\tilde{G}=(A_{m-1}\otimes G)/\mathbb{Z}_2$.
All of the \ade  coset graph fusion rules can be described in this manner giving the coset fusion rules
\bea
(\mu)\times(\mu')=\sum_{\mu''=1}^{|\mathbb{K}|} \tilde{N}_{\mu\mu'}{}^{\mu''} (\mu''),\qquad \mu,\mu'=1,2,\ldots,|\mathbb{K}|\label{cosetFus}
\eea
where $\tilde{N}_{\mu\mu'}{}^{\mu''}=(\tilde{N}_\mu)_{\mu'}{}^{\mu''}$ are the structure constants, $\tilde{N}_\mu$ are the fusion matrices, $\tilde{N}_1=I$ and the fundamental $\tilde{N}_2=\tilde{G}$ is the adjacency matrix of the coset graph. 
The fusion matrices are easily read off from the Cayley table as is illustrated for the trivial case of the critical Ising model $A_3$ 
\bea
\ssmat{\,1\times1\,&\,1\times 2\,&\,1\times3\,\\ \,2\times1\,&\,2\times 2\,&\,2\times3\,\\ \,3\times1\,&\,3\times 2\,&\,3\times3\,}=\!\!\ssmat{1&2&3\\ 2&1\!+\!3&2\\ 3&2&1}
=(1)\!\ssmat{1&0&0\\ 0&1&0\\ 0&0&1}+(2)\!\ssmat{0&1&0\\ 1&0&1\\ 0&1&0}+(3)\!\ssmat{0&0&1\\ 0&1&0\\ 1&0&0}=\sum_{\mu=1}^3 (\mu)\, \tilde{N}_\mu
\eea
The fusion matrices yield nonnegative integer matrix representations (nimreps) of the coset graph fusion algebra
\bea
\tilde{N}_\mu \tilde{N}_{\mu'} =\sum_{\mu''=1}^{|\mathbb{K}|} \tilde{N}_{\mu\mu'}{}^{\mu''} \tilde{N}_{\mu''},\qquad \mu,\mu'=1,2,\ldots,|\mathbb{K}|\label{graphFusAlg}
\eea
The coset graph fusion algebras are satisfied by (i) the nimreps, (ii) the defect lines ${\cal L}_\mu$ and (iii) the defect eigenvalues (quantum dimensions $\tilde{d}_\mu$) including the $\g$-factors corresponding to the Perron-Frobenius eigenvalues. 
General expressions for the coset nimreps as tensor products are given in Appendix~\ref{App:Coset}.

The graph fusion  matrices $\tilde{N}_\mu$ are nonnegative commuting normal matrices. They are therefore simultaneously diagonalizable with the spectral decomposition yielding a Verlinde-like formula. The Perron-Frobenius eigenvalues define the quantum dimensions $\tilde{d}_\mu>0$ leading to a 1-dimensional representation of the coset graph fusion algebra\footnote{The quotient ${\cal S}_{\mu,1}/{\cal S}_{1,1}$ frequently referred to as the quantum dimension in connection with the Verlinde fusion rules is not suitable for nonunitary theories as it may be negative. Working with the Perron-Frobenius eigenvalues 
$\tilde{d}_\mu={\cal S}_{\mu,\mu_0}/{\cal S}_{1,\mu_0}$ where $\mu_0$ denotes the ground state $(r_0,s_0)$ as quantum dimensions guarantees positivity.}
\bea
\tilde{d}_\mu \tilde{d}_{\mu'} =\sum_{\mu''=1}^{|\mathbb{K}|} \tilde{N}_{\mu\mu'}{}^{\mu''} \tilde{d}_{\mu''},\qquad \mu,\mu'=1,2,\ldots,|\mathbb{K}|\label{1dGraphFusAlg}
\eea
Substituting $(\mu)\mapsto \tilde{d}_\mu$ into (\ref{cosetFus}) gives the decomposition of a rank-1 matrix of products of quantum dimensions into a sum of the fusion matrices
\bea
\boldsymbol{\cal D}=\sssmat{
\tilde{d}_{1}^2&\tilde{d}_{1}\tilde{d}_{2}&\cdots&\tilde{d}_{1}\tilde{d}_{n}\\[2pt]
\tilde{d}_{2}\tilde{d}_{1}&\tilde{d}_{2}^2&\cdots&\tilde{d}_{2}\tilde{d}_{n}\\[-4pt]
\vdots&\vdots&\ddots&\vdots\\
\tilde{d}_{n}\tilde{d}_{1}&\tilde{d}_{n}\tilde{d}_{2}&\cdots&\tilde{d}_{n}^2}=\sssmat{\tilde{d}_{1}\\[2pt] \tilde{d}_{2}\\[-4pt]  \vdots\\ \tilde{d}_{n}}\!\!\sssmat{\tilde{d}_{1}&\tilde{d}_{2}&\cdots& \tilde{d}_{n}}
=\sum_{\mu=1}^{n} \tilde{d}_\mu \tilde{N}_\mu,\qquad n=|\mathbb{K}|\label{cosetFusion}
\eea
thus encoding the fusion rules with $\tilde{d}_1=1$ where $\mu=1$ denotes the vacuum $(r,s)=(1,1)$. The total quantum dimension $D$~\cite{KP2006,FNO2017}  is given by
\bea
D^2=\mbox{Tr }\!\boldsymbol{\cal D}=\sum_{\mu=1}^n \tilde{d}_\mu^2,\quad\mbox{with}\quad \frac{\tilde{d}_\mu}{D}={\cal S}_{\mu,\mu_0}
\eea
where $\mu=(r,s)$, $\mu_0=(r_0,s_0)$ and $\cal S$ is the modular ${\cal S}$-matrix defined below in (\ref{SMatrix}). The total quantum dimension $D$ plays a role for CFTs on non-orientable surfaces~\cite{Tu2017}.

The coset graph fusion algebra of ${\cal M}(m,m')$ can also be realized~\cite{FZ93,RasmussenStudent} as a polynomial ring in two indeterminates $\mathbb{Z}[x,y]$ quotiented by an ideal 
${\cal I}=\langle p_1(x),p'_1(y),p_2(x,y)\rangle$
\begin{subequations}
\bea
&\mathbb{Z}[x,y]/\langle p_1(x),p'_1(y),p_2(x,y)\rangle&\\[4pt]
&p_1(x)=U_{m-1}(\tfrac x2),\quad  p'_1(y)=U_{m'-1}(\tfrac y2),\quad  p_2(x,y)=U_{m-2}(\tfrac x2)-U_{m'-2}(\tfrac y2)&
\eea
\end{subequations}
where $U_n(x)$ are Chebyshev polynomials of the second kind. The ideal ${\cal I}$ can always be reduced further, so that it is generated by one or two polynomials, as seen in the examples in Section~\ref{secExamples}. 
The indeterminates are the fundamentals so the polynomials generating the ideal vanish when evaluated with $x\mapsto [2]_x=2\cos\tfrac{\pi}m$ and $y\mapsto [2]_y=2\cos\tfrac{\pi}{m'}$. 
Curiously, we observe that the coset graph fusion algebras of ${\cal M}(m,m')$ coincide with the coset graph fusion algebras of the projective Grothendieck representations for the logarithmic minimal models 
${\cal LM}(m-2,m'-2)$~\cite{PRZ2006,PR2011,LogTY2014}. 
This means, for example, that the critical Ising model ${\cal M}(3,4)$ and dense polymers ${\cal LM}(1,2)$ and similarly the tricritical Ising model ${\cal M}(4,5)$  and critical percolation ${\cal LM}(2,3)$ are described by the same coset graph fusion algebras. 

\subsection{Modular matrices and modular transformations}

For general $(A_{m-1},A_{m'\!-1})$ theories, the fusion rules and adjacency matrix $\tilde A$ of the coset graph \mbox{$(A_{m-1}\otimes A_{m'-1})/\mathbb{Z}_2$} are diagonalized by the modular $\cal S$-matrix~\cite{FMS97}
\bea
  \label{SMatrix}
{\cal S}_{rs,r'\!s'}:={\cal S}_{(r,s),(r'\!,s')}=\sqrt{\frac{8}{mm'}}(-1)^{1+r s'+s r'}\sin\tfrac{\pi rr'm'}{m}\sin\tfrac{\pi ss'm}{m'},\quad (r,s),(r'\!,s')\in\mathbb{K}
\eea
In this form, the formula applies to unitary and nonunitary theories and is invariant under the Kac table symmetry $(r,s)\equiv (m\!-\!r,m'\!-\!s)$. 
Other common forms of this equation apply only to unitary theories. 
The $\cal S$-matrix is a real orthogonal matrix satisfying ${\cal S}^2=I$ and is unique up to a reordering of the elements of $\mathbb{K}$. The unique row/column with all positive entries coincides with the groundstate $(r,s)=(r_0,s_0)$. 
For unitary theories, the groundstate $(r_0,s_0)$ coincides with the vacuum $(1,1)$ but this is not the case for nonunitary theories. More generally, for $(A_{m-1},G)$ theories, the coset graph fusion nimreps are simultaneously diagonalized by the matrix $\Psi$.

Under modular transformation ${\cal S}$, the characters transform as
\bea
\chi_{r,s}(q)=\sum_{(r',s')\in\mathbb{K}}\!\! {\cal S}_{(r,s),(r',s')}\,\chi_{r',s'}(\tilde{q}),\qquad q=e^{2\pi i\tau},\ \ \tilde{q}=e^{-2\pi i/\tau},\ \ \mbox{Im}(\tau)>0
\eea
where $\tau$ is the modular parameter and $\tilde{q}$ is the conjugate modular nome.

\subsection{Defect lines and their properties}

\begin{figure}[htb]
\begin{align}
Z^{m,m'}_{(r'\!,b)|(r'\!'\!,c)}(q)\;&=\;\!
\begin{pspicture}[shift=-.4](-.15,.3)(2.6,1.7)
\psframe[linewidth=0pt,fillstyle=solid,fillcolor=lightlightblue](0,0)(2.5,1.5)
\psline[linewidth=1.5pt](0,0)(0,1.5)
\psline[linewidth=1.5pt](2.5,0)(2.5,1.5)
\rput(0,1.7){\tiny $(r'\!\!,\!b)$}
\rput(2.5,1.7){\tiny $(r'\!'\!\!\!,c)$}
\end{pspicture}
\;=\;\!\begin{pspicture}[shift=-.4](-.15,.3)(2.6,1.7)
\psframe[linewidth=0pt,fillstyle=solid,fillcolor=lightlightblue](0,0)(2.5,1.5)
\psline[linewidth=1.5pt](.8,0)(.8,1.5)
\psline[linewidth=1.5pt](1.7,0)(1.7,1.5)
\psline[linewidth=.75pt](0,0)(0,1.5)
\psline[linewidth=.75pt](2.5,0)(2.5,1.5)
\rput(.8,1.7){\scriptsize $\widehat{\cal L}_{(r'\!\!,b)}$}
\rput(1.7,1.7){\scriptsize $\widehat{\cal L}_{(r'\!'\!\!,c)}$}
\rput(1.25,.75){\scriptsize $\times$}
\rput(0,1.7){\tiny $(1,\!1)$}
\rput(2.5,1.7){\tiny $(1,\!1)$}
\end{pspicture}
\;=\;\!\sum_{r=1}^{m-1}\sum_{a\in G}N_{rr'}{}^{r''}\widehat{N}_{ab}{}^{c}\begin{pspicture}[shift=-.4](-.15,.3)(2.6,1.7)
\psframe[linewidth=0pt,fillstyle=solid,fillcolor=lightlightblue](0,0)(2.5,1.5)
\psline[linewidth=1.5pt](1.25,0)(1.25,1.5)
\psline[linewidth=.75pt](0,0)(0,1.5)
\psline[linewidth=.75pt](2.5,0)(2.5,1.5)
\rput(1.25,1.7){\scriptsize $\widehat{\cal L}_{(r,a)}$}
\rput(0,1.7){\tiny $(1,\!1)$}
\rput(2.5,1.7){\tiny $(1,\!1)$}
\end{pspicture}\nonumber\\[22pt]
&=\;\!\sum_{r=1}^{m-1}\sum_{a\in G}N_{rr'}{}^{r''}\widehat{N}_{ab}{}^{c}\begin{pspicture}[shift=-.4](-.15,.3)(2.6,1.7)
\psframe[linewidth=0pt,fillstyle=solid,fillcolor=lightlightblue](0,0)(2.5,1.5)
\psline[linewidth=1.5pt](2.5,0)(2.5,1.5)
\psline[linewidth=.75pt](0,0)(0,1.5)
\psline[linewidth=.75pt](2.5,0)(2.5,1.5)
\rput(2.5,1.7){\tiny $(r,\!a)$}
\rput(0,1.7){\tiny $(1,\!1)$}
\end{pspicture}
=\;\!\sum_{r=1}^{m-1}\sum_{a\in G}N_{rr'}{}^{r''}\widehat{N}_{ab}{}^{c}\;Z^{m,m'}_{(1,1)|(r\!,a)}(q),\ \ \mbox{$G$ is type I}
\nonumber\\[24pt]
Z^{m,m'}_{(r'\!,b)|(r'\!'\!,c)}(q)\;&=\;\!
\begin{pspicture}[shift=-.4](-.15,.3)(2.6,1.7)
\psframe[linewidth=0pt,fillstyle=solid,fillcolor=lightlightblue](0,0)(2.5,1.5)
\psline[linewidth=1.5pt](0,0)(0,1.5)
\psline[linewidth=1.5pt](2.5,0)(2.5,1.5)
\rput(0,1.7){\tiny $(r'\!\!,\!b)$}
\rput(2.5,1.7){\tiny $(r'\!'\!\!\!,c)$}
\end{pspicture}
\;=\;\!\begin{pspicture}[shift=-.4](-.15,.3)(2.6,1.7)
\psframe[linewidth=0pt,fillstyle=solid,fillcolor=lightlightblue](0,0)(2.5,1.5)
\psline[linewidth=1.5pt](.8,0)(.8,1.5)
\psline[linewidth=1.5pt](1.7,0)(1.7,1.5)
\psline[linewidth=.75pt](0,0)(0,1.5)
\psline[linewidth=.75pt](2.5,0)(2.5,1.5)
\rput(.8,1.7){\scriptsize $\widehat{\cal L}_{(r'\!\!,b)}$}
\rput(1.7,1.7){\scriptsize $\widehat{\cal L}_{(r'\!'\!\!,c)}$}
\rput(1.25,.75){\scriptsize $\times$}
\rput(0,1.7){\tiny $(1,\!1)$}
\rput(2.5,1.7){\tiny $(1,\!1)$}
\end{pspicture}
\;=\;\!\sum_{r=1}^{m-1}\sum_{s=1}^{m'\!-1}N_{rr'}{}^{r''}\!n_{sb}{}^{c}\begin{pspicture}[shift=-.4](-.15,.3)(2.6,1.7)
\psframe[linewidth=0pt,fillstyle=solid,fillcolor=lightlightblue](0,0)(2.5,1.5)
\psline[linewidth=1.5pt](1.25,0)(1.25,1.5)
\psline[linewidth=.75pt](0,0)(0,1.5)
\psline[linewidth=.75pt](2.5,0)(2.5,1.5)
\rput(1.25,1.7){\scriptsize ${\cal L}_{(r,s)}$}
\rput(0,1.7){\tiny $(1,\!1)$}
\rput(2.5,1.7){\tiny $(1,\!1)$}
\end{pspicture}\nonumber\\[22pt]
&=\;\!\sum_{r=1}^{m-1}\sum_{s=1}^{m'\!-1}N_{rr'}{}^{r''}\!n_{sb}{}^{c}\begin{pspicture}[shift=-.4](-.15,.3)(2.6,1.7)
\psframe[linewidth=0pt,fillstyle=solid,fillcolor=lightlightblue](0,0)(2.5,1.5)
\psline[linewidth=1.5pt](2.5,0)(2.5,1.5)
\psline[linewidth=.75pt](0,0)(0,1.5)
\psline[linewidth=.75pt](2.5,0)(2.5,1.5)
\rput(2.5,1.7){\tiny $(r,\!s)$}
\rput(0,1.7){\tiny $(1,\!1)$}
\end{pspicture}
=\;\!\sum_{r=1}^{m-1}\sum_{s=1}^{m'\!-1}N_{rr'}{}^{r''}\!n_{sb}{}^{c}\,Z^{m,m'}_{(1,1)|(r\!,s)}(q),\ \ \mbox{$G$ is type I/II}\
\nonumber\\[20pt]
&\hspace{1.8cm} Z^{m,m'}_{(1,1)|(r\!,a)}(q)\,=\,\hat{\chi}_{ra}(q),\qquad Z^{m,m'}_{(1,1)|(r\!,s)}(q)\,=\,\chi_{rs}(q)\nonumber
\end{align}
\caption{The conformal cylinder partition functions (\ref{CylPFs}) are generated by propagating the defect lines ${\cal L}_\mu$ (with $\mu=(r,a)$ or $\mu=(r,s)$) glued to the two boundaries to the center and fusing them. If $G$ is of $A$-type, then $n_{sb}{}^c$ reduces to the Verlinde structure constants $N_{ss'}{}^{s'\!'}$ and the cylinder partition functions are then compatible with the Kac symmetry since 
${\cal L}_{(r,s)}={\cal L}_{(m-r,m'\!-s)}$. Similar arguments apply for more general topological defects.
\label{PartitionFunctDecomp}}
\end{figure}

The conformal defect lines are labelled by fusions $\mu$ which encode their internal structure (charges). The defect lines are in fact operators. They satisfy the same fusion relations as their fusion labels and possess a number of properties. In particular, the defect lines ${\cal L}_\mu$: 
\begin{enumerate}
\item[(i)] are topological in the sense that they freely propagate,\\[-20pt]
\item[(ii)] are mutually commuting (Abelian) so they pass through one another,\\[-20pt]
\item[(iii)] satisfy the Kac symmetry ${\cal L}_{(r,s)}={\cal L}_{(m-r,m'\!-s)}$,\\[-20pt]
\item[(iv)] factorize as ${\cal L}_{(r,s)}={\cal L}_r {\cal L}_s$, ${\cal L}_{(r,a)}={\cal L}_r {\cal L}_a$ with ${\cal L}_r={\cal L}_{(r,1)}$, ${\cal L}_s={\cal L}_{(1,s)}$ 
and ${\cal L}_a={\cal L}_{(1,a)}$,\\[-20pt]
\item[(v)] satisfy the coset graph fusion algebra (\ref{cosetFus}),\\[-20pt]
\item[(vi)] carry a defect entropy $S^\text{defect}_{(r,a)}=\log\hat{d}_{(r,a)}$ with $\hat{d}_{(r,a)}=\sum_{s=1}^{m'\!-1} n_{s1}{}^a\,\tilde{d}_{(r,s)}$,\\[-20pt]
\item[(vii)] exhibit an eigenvalue spectrum of quantum dimensions given by the solutions $\tilde{d}_\mu$ of (\ref{cosetFusion}).
\end{enumerate}
The Abelian property (ii) holds generally since we do not add the $\mathbb{Z}_2$ automorphism in the $D_{2l}$ cases.

Within the context of CFT, the properties of defect lines are initially posited but are ultimately confirmed to be consistent and formalized as the axioms of various kinds of fusion categories. Alternatively, as we pursue here, the properties of the defects can be established in the context of integrable (defect) seams for the associated Yang-Baxter integrable \ade RSOS lattice models as in Section~\ref{secLattice}. The properties (i)--(vii) are then inherited by the defect lines in the continuum scaling limit. This is the approach that we adopt here.

\subsection{Boundary and defect $\g$-factors and coset quantum dimensions}

For the $A$-type ${\cal M}(m,m')$ minimal models, the 1-boundary Affleck-Ludwig $\g$-factors~\cite{AL91} are
\bea
\tilde{\g}_{(r,s)}\!=\!\big({\tfrac{8}{m m'}}\big)^{1/4} \frac{\sin\tfrac{r\pi}{m}\sin\tfrac{s\pi}{m'}}{\sqrt{\sin\tfrac{\pi}{m}\sin\tfrac{\pi}{m'}}}
=\tilde{\g}_{(m-r,s)}=\tilde{\g}_{(r,m'\!-s)}=\tilde{\g}_{(1,1)}\tilde{d}_{(r,s)}
\eea
where
\bea
\tilde{\g}_{(1,1)}\!=\!\big({\tfrac{8}{m m'}}\big)^{1/4} {\sqrt{\sin\tfrac{\pi}{m}\sin\tfrac{\pi}{m'}}},\qquad 
\tilde{d}_{(r,s)}\!=\!\frac{\sin\tfrac{r\pi}{m}\sin\tfrac{s\pi}{m'}}{\sin\tfrac{\pi}{m}\sin\tfrac{\pi}{m'}}=[r]_{e^{\pi i/m}} [s]_{e^{\pi i/m'}}
\eea
and the coset quantum dimension $\tilde{d}_{(r,s)}=\tilde{d}_{(r,1)}\tilde{d}_{(1,s)}$ gives the $\g$-factor associated with the $(r,s)$ defect.
The 1-boundary $\g$-factors are not directly measurable so we focus instead on 2-boundary $\g$-factors $\tilde{\g}_{(r,s)|(r',s')}=\tilde{\g}_{(r,s)}\tilde{\g}_{(r',s')}$ with
\bea
\tilde{\g}_{(1,1)|(r,s)}=\tilde{\g}_{(1,1)}\tilde{\g}_{(r,s)}=\tilde{\g}_{(1,1)|(1,1)}\tilde{d}_{(r,s)}=\sqrt{\tfrac{8}{mm'}}\sin\tfrac{r\pi}{m}\sin\tfrac{s\pi}{m'}={\cal S}_{(r_0,s_0),(r,s)} 
\eea
The last equality follows straightforwardly using the relation (\ref{r0s0}) between $r_0$ and $s_0$.

More generally, for the $(A,G)$ minimal models, the 2-boundary $\g$-factors are given by~\cite{BPPZ}
\bea
\hat{\g}_{(1,1)|(r,a)}=\hat{\g}_{(1,1)}\hat{\g}_{(r,a)}=\tilde{\g}_{(1,1)|(1,1)} \hat{d}_{(r,a)}\label{g11ra}=\sqrt{\tfrac{2m'}m}\,\frac{\sin\tfrac{r\pi}m}{\sin\tfrac{\pi}{m'}}\,\Psi_1{}^1\Psi_a{}^1,\qquad \hat{d}_{(r,a)}=\sum_{s=1}^{m'\!-1} n_{s1}{}^a\,\tilde{d}_{(r,s)}
\eea
where $\Psi_{1}{}^{1}=\|\boldsymbol\psi\|^{-1}$ and $\hat{\g}_{(1,1)|(1,1)}=\sqrt{\tfrac{2m'}m}\,\frac{\sin\tfrac{\pi}m}{\sin\tfrac{\pi}{m'}}\,\|\boldsymbol\psi\|^{-2}$ with $\boldsymbol\psi$ given by (\ref{PF}). The quantum dimension $\hat{d}_{(r,a)}$ gives the \mbox{$\g$-factor} associated with the $(r,a)$ defect.
These defect $\g$-factors give a 1-dimensional representation of the coset graph fusion algebra. The boundary and defect entropies are
\bea
S^\text{bdy}_{(r,a)}=\log\hat{\g}_{(r,a)},\qquad S^\text{defect}_{(r,a)}=\log\hat{d}_{(r,a)}
\eea
The $\g$-factors are multiplicative while the entropies are additive.

Numerical $\g$-factors can be calculated from the limits of conformal tower degeneracies. 
The $q$-series for the Virasoro characters are
\bea
\chi_{r,s}^{m,m'}(q)=q^{-c/24+\Delta_{r,s}^{m,m'}}\!\!\! \sum_{n=0,1,2,\ldots}^\infty\!\!\! \d_n^{\,(r,s)} q^n
\eea
where $n=E\in{\Bbb N}_0$ is an energy level in the conformal tower with degeneracy $\d_n^{\,(r,s)} \in{\Bbb N}_0$. Remarkably, the degeneracies $\d_n^{\,(r,s)} $ allow for the numerical calculation of the effective central charge $c_\text{eff}$ (\ref{ceff}) and the 2-boundary $\g$-factors $\tilde{\g}_{(1,1)|(r,a)}$ (\ref{g11ra})
\begin{subequations}
\label{sequences}
\begin{align}
c_\text{eff}&=c\!-\!24\Delta_\text{min}=1\!-\!\frac{6}{m m'}=\lim_{n\to\infty} \frac{3}{2\pi^2 n} (\log \d_n^{\,(r,s)})^2,\quad\mbox{independent of $(r,s)$}\\[4pt] 
\tilde{\g}_{(1,1)|(r,s)}&=\lim_{n\to\infty} 2\Big(\frac{6n^3}{c_\text{eff}}\Big)^{\!1/4}\!\exp\!\Big(\!\!-\!2\pi\sqrt{\frac{nc_\text{eff}}{6}}\Big)\,\d_n^{\,(r,s)},\qquad
\hat{\g}_{(1,1)|(r,a)}=\sum_{s=1}^{m'\!-1} n_{s1}{}^a\,\tilde{\g}_{(1,1)|(r,s)}
\end{align}
\end{subequations}
These two formulas generalize relations in \cite{NahmRT93,AL91} respectively by replacing $c$ with $c_\text{eff}$ to allow for nonunitary theories. For unitary theories, $m'=m\!\pm\!1$, $\Delta_\text{min}=0$ and $c_\text{eff}=c$. These limits are checked numerically against the analytic expressions for each theory that we study. 
The extrapolation of some typical sequences are shown in Figure~\ref{vBStables}.

\begin{figure}[htb]
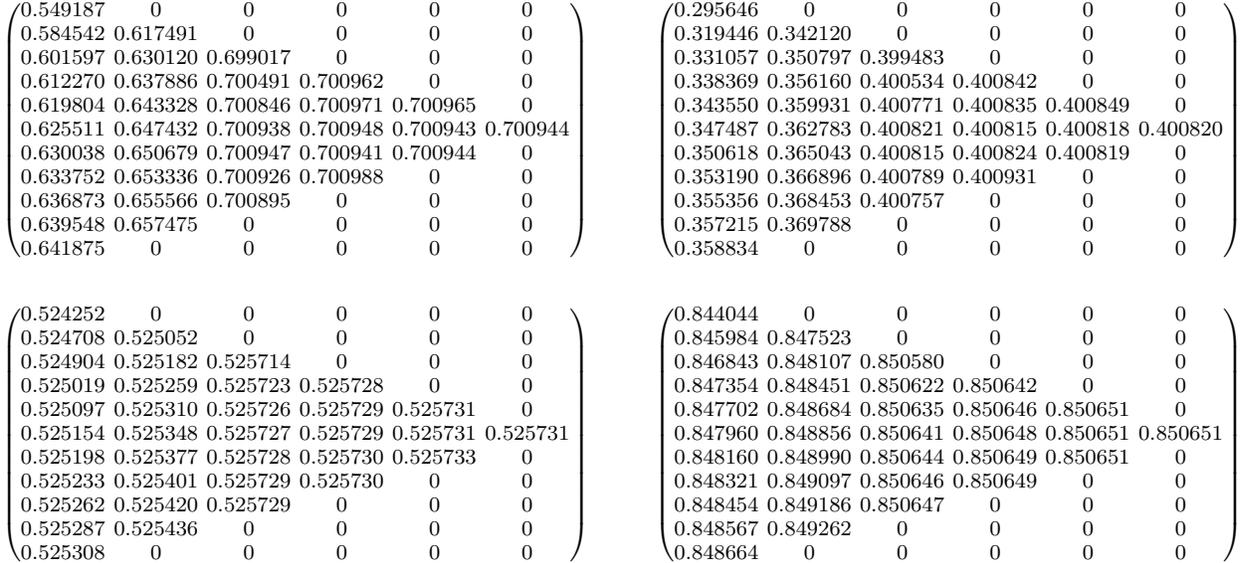

\begin{align}
&\sssmat{
  0.549187 & 0 & 0 & 0 & 0 & 0 \\
 0.584542 & 0.617491 & 0 & 0 & 0 & 0 \\
 0.601597 & 0.630120 & 0.699017 & 0 & 0 & 0 \\
 0.612270 & 0.637886 & 0.700491 & 0.700962 & 0 & 0 \\
 0.619804 & 0.643328 & 0.700846 & 0.700971 & 0.700965 & 0 \\
 0.625511 & 0.647432 & 0.700938 & 0.700948 & 0.700943 & 0.700944 \\
 0.630038 & 0.650679 & 0.700947 & 0.700941 & 0.700944 & 0 \\
 0.633752 & 0.653336 & 0.700926 & 0.700988 & 0 & 0 \\
 0.636873 & 0.655566 & 0.700895 & 0 & 0 & 0 \\
 0.639548 & 0.657475 & 0 & 0 & 0 & 0 \\
 0.641875 & 0 & 0 & 0 & 0 & 0}\quad
&\sssmat{
 0.295646 & 0 & 0 & 0 & 0 & 0 \\
 0.319446 & 0.342120 & 0 & 0 & 0 & 0 \\
 0.331057 & 0.350797 & 0.399483 & 0 & 0 & 0 \\
 0.338369 & 0.356160 & 0.400534 & 0.400842 & 0 & 0 \\
 0.343550 & 0.359931 & 0.400771 & 0.400835 & 0.400849 & 0 \\
 0.347487 & 0.362783 & 0.400821 & 0.400815 & 0.400818 & 0.400820 \\
 0.350618 & 0.365043 & 0.400815 & 0.400824 & 0.400819 & 0 \\
 0.353190 & 0.366896 & 0.400789 & 0.400931 & 0 & 0 \\
 0.355356 & 0.368453 & 0.400757 & 0 & 0 & 0 \\
 0.357215 & 0.369788 & 0 & 0 & 0 & 0 \\
 0.358834 & 0 & 0 & 0 & 0 & 0}
 &\nonumber\\[12pt]
 &\sssmat{
 0.524252 & 0 & 0 & 0 & 0 & 0 \\
 0.524708 & 0.525052 & 0 & 0 & 0 & 0 \\
 0.524904 & 0.525182 & 0.525714 & 0 & 0 & 0 \\
 0.525019 & 0.525259 & 0.525723 & 0.525728 & 0 & 0 \\
 0.525097 & 0.525310 & 0.525726 & 0.525729 & 0.525731 & 0 \\
 0.525154 & 0.525348 & 0.525727 & 0.525729 & 0.525731 & 0.525731 \\
 0.525198 & 0.525377 & 0.525728 & 0.525730 & 0.525733 & 0 \\
 0.525233 & 0.525401 & 0.525729 & 0.525730 & 0 & 0 \\
 0.525262 & 0.525420 & 0.525729 & 0 & 0 & 0 \\
 0.525287 & 0.525436 & 0 & 0 & 0 & 0 \\
 0.525308 & 0 & 0 & 0 & 0 & 0}\quad
 &\sssmat{
 0.844044 & 0 & 0 & 0 & 0 & 0 \\
 0.845984 & 0.847523 & 0 & 0 & 0 & 0 \\
 0.846843 & 0.848107 & 0.850580 & 0 & 0 & 0 \\
 0.847354 & 0.848451 & 0.850622 & 0.850642 & 0 & 0 \\
 0.847702 & 0.848684 & 0.850635 & 0.850646 & 0.850651 & 0 \\
 0.847960 & 0.848856 & 0.850641 & 0.850648 & 0.850651 & 0.850651 \\
 0.848160 & 0.848990 & 0.850644 & 0.850649 & 0.850651 & 0 \\
 0.848321 & 0.849097 & 0.850646 & 0.850649 & 0 & 0 \\
 0.848454 & 0.849186 & 0.850647 & 0 & 0 & 0 \\
 0.848567 & 0.849262 & 0 & 0 & 0 & 0 \\
 0.848664 & 0 & 0 & 0 & 0 & 0}&\nonumber
\end{align}
\caption{Vanden Broeck-Schwartz \cite{vBS} extrapolations of degeneracy sequences for $c_\text{eff}$ (top row) and 2-boundary $\g$-factors (bottom row).
The top row relates to (i) the central charge $c=c_\text{eff}=\tfrac{7}{10}$ of the tricritical Ising model ${\cal M}(4,5)$, (ii) the effective central charge $c_\text{eff}=\tfrac25$ of the Lee-Yang model ${\cal M}(2,5)$. The bottom row relates to (iii-iv) the 2-boundary $\g$-factors $\tilde{\g}_{(1,1)|(1,1)}=0.525731\ldots$ and $\tilde{\g}_{(1,1)|(1,2)}=0.850651\ldots$ of the Lee-Yang model ${\cal M}(2,5)$. 
The values of $n$ in the sequences (\ref{sequences}) range from $n=1000$ to $n=11000$ in increments of $\Delta n=1000$. 
The convergence of the $\g$-factors is faster than the convergence of the effective central charges.
\label{vBStables}}
\end{figure}

\subsection{Asymptotic counting of coset fusion paths}

The fundamental fusion matrix $\tilde{N}_2$ acts to move from one node of the coset fusion graph $\tilde{G}$ to a neighbour. So the application of $N$ fundamental fusions corresponds to an $N$-step fusion path on $\tilde{G}$. For large $N$, the number of such $N$-step fusion paths grows as $\tilde{d}_2^N$. It is therefore natural to look at the large-$N$ asymptotics $\lim_{N\to \infty}(\tilde{G}/\tilde{d}_2)^N$ of the dimensions $\mbox{dim}\,{\cal V}^{(N)}_{\mu,\mu'}$ of the vector spaces spanned by such $N$-step fusion paths from $\mu$ to $\mu'$. The details depend (i) on the parities of $m$ and $m'$, (ii) on whether $\tilde{G}$ is bipartite or not and (iii) on the number of connected components of $\tilde{G}$ but, ultimately, a rank-1 matrix emerges
\bea
\lim_{N\to \infty}\!\Big(\frac{\tilde{G}}{\tilde{d}_2}\Big)^N\!\propto \!\!\ssmat{
\tilde{\g}_{1|1}&\tilde{\g}_{1|2}&\cdots&\tilde{\g}_{1|n}\\ 
\tilde{\g}_{2|1}&\tilde{\g}_{2|2}&\cdots&\tilde{\g}_{2|n}\\
\vdots&\vdots&\ddots&\vdots\\
\tilde{\g}_{n|1}&\tilde{\g}_{n|2}&\cdots&\tilde{\g}_{n|n}}\!
=\tilde{\g}_{1|1}\!\!\sssmat{1\\ \tilde{d}_{2}\\ \vdots\\ \tilde{d}_{n}}\!\!\sssmat{1&\tilde{d}_{2}&\cdots&\tilde{d}_{n}}=\tilde{\g}_{1|1}\sum_{\mu=1}^n \tilde{d}_\mu \tilde{N}_\mu=\tilde{\g}_{1|1} \boldsymbol{\cal D},\quad n=|\mathbb{K}|
\label{limFusPaths}
\eea
where $\tilde{d}_1=1$ and $\mu=2$ denotes the fundamental. If the coset graph $\tilde{G}$ is bipartite, we combine the limits for the odd and even sublattices. 
Both the 2-boundary $\g$-factors $\tilde{\g}_{\mu|\mu'}$ and the quantum dimensions $\tilde{d}_\mu$ can be obtained from (\ref{limFusPaths}). 
Indeed, the quantum dimensions $\tilde{d}_\mu$ can be obtained by factoring out $\tilde{\g}_{1|1}$ and solving the equations
\bea
\lim_{N\to \infty}\frac{\tilde{G}^N}{(\tilde{G}^N)_{1,1}}=\sssmat{1\\ \tilde{d}_{2}\\ \vdots\\ \tilde{d}_{n}}\!\!\sssmat{1&\tilde{d}_{2}&\cdots&\tilde{d}_{n}}=\sum_{\mu=1}^n \tilde{d}_\mu \tilde{N}_\mu
\eea

The physical interpretation (see Figure~\ref{PartitionFunctDecomp}) of these equations is that $\tilde{\g}_{\mu|\mu'}$ gives the $\g$-factor for the system with boundary condition $\mu$ on the left and $\mu'$ on the right. The boundaries $\mu$, $\mu'$ are implemented by the action of the defects ${\cal L}_\mu$ on the left vacuum boundary and ${\cal L}_{\mu'}$ on the right vacuum boundary. The defects carry the defect $\g$-factors given by the quantum dimensions $\tilde{d}_\mu$ and $\tilde{d}_{\mu'}$. Finally, leaving the vacuum contribution $\tilde{\g}_{1|1}$ behind, the defects can propagate to the center of the cylinder where the fusion product is decomposed in accord with the fusion rules. A number of prototypical examples of this procedure are given in the examples in Section~\ref{secProtoExamples}.

\subsection{Relative symmetry resolved Affleck-Ludwig entropies}}

Symmetry Resolved Entanglement Entropies (SREEs) have garnered interest recently due to the boundary CFT approach to entanglement entropy~\cite{DGMN23,Northe23,KMOP23,DMVSB24,HQ24,CRZ24,SBDSMV24,L13,OT15,CT16}.
In this section, we consider relative Symmetry Resolved Affleck-Ludwig Entropies (SRALEs). 
These quantities have some similarities to SREEs but they do not involve partitioning of space and are not genuine entanglement entropies. 
They are much simpler objects describing the decomposition or partitioning of Affleck-Ludwig $\g$-factors into sectors labelled by CFT quantum numbers such as $(r,s)$ or $(r,a)$ according to the fusion rules. In this sense, the symmetry being resolved is quantum symmetry.

The vector spaces of quantum Hamiltonians, corresponding to the cylinder partition functions (\ref{CylPFs}), decompose into sectors according to the fusion rules 
\bea
{\cal V}_{\mu'|\mu''}=\bigoplus_{\mu\in\mathbb{K}} \tilde{N}_{\mu\mu'}{}^{\mu''} {\cal V}_{1|\mu}
\eea
where the fusion labels $\mu$ are quantum numbers. In unitary cases, the vector spaces are Hilbert spaces. 
The reduced density matrix is a block-diagonal sum over sectors
\bea
\rho_{\mu'|\mu''}=\bigoplus_{\mu\in\mathbb{K}} p_{\mu}\,\tilde{N}_{\mu\mu'}{}^{\mu''}\rho_{1|\mu}
\eea
where $p_{\mu}$ are probabilities. 
In terms of the cylinder partition functions (\ref{CylPFs}), the R\'enyi entropies with $n$ replicas are defined by
\bea
    S^n_{\mu'|\mu''} := \frac{1}{1-n} \log \Tr \rho_{\mu'|\mu''}^n(q)  = \frac{1}{1-n} \log{\frac{Z_{\mu'|\mu''}(q^n)}{[Z_{\mu'|\mu''}(q)]^n}}
\eea
Although originating as an integer, $n$ is treated as a continuous variable. Indeed, the von Neumann entanglement entropies are recovered as $\displaystyle\lim_{n\to 1} S^n_{\mu'|\mu''}$.

We are interested in relative SRALEs $\Delta S^n_{\mu,\mu'\!,\mu''}$, defined as 
\begin{align}
\Delta S^n_{\mu,\mu'\!,\mu'\!'}&:=\lim_{q\to 1}(S^n_{1|\mu}-S^n_{\mu'|\mu'\!'})
=\lim_{q\to 1}\frac{1}{1\!-\!n}\bigg[\log\frac{Z_{1,\mu}(q^n)}{\big[Z_{1,\mu}(q)\big]^n}-\log\frac{Z_{\mu'\!,\mu''}(q^n)}{\big[Z_{\mu'\!,\mu''}(q)\big]^n}\bigg]\nonumber\\
&\phantom{:}=\,\lim_{q\to 1}\frac{1}{1\!-\!n} \bigg[\log\frac{Z_{1,\mu}(q^n)}{Z_{\mu'\!,\mu''}(q^n)}-n\log\frac{Z_{1,\mu}(q)}{Z_{\mu'\!,\mu''}(q)}\bigg]=\lim_{q\to 1}\log\frac{Z_{1,\mu}(q^n)}{Z_{\mu'\!,\mu''}(q^n)}\label{relSREEs}
\end{align}
where $\mu$ appears in the decomposition of the fusion product $\mu'\times\mu''$. The relative SRALEs $\Delta S^n_{\mu,\mu'\!,\mu''}$ measure the contribution of the sector $\mu$ to the entanglement entropy $\disp\lim_{q\to 1}S^n_{\mu'|\mu''}$. Using the asymptotic result of Affleck and Ludwig~\cite{AL91} (see also (4.5) of \cite{BPPZ}) in (\ref{relSREEs}) 
\bea
\log Z_{\mu|\mu'}\sim -\log\tilde{q}^{\frac{c_{\text{eff}}}{24}}+\log\tilde{\g}_{(1,1)|(1,1)}+\log \hat{d}_\mu \hat{d}_{\mu'},\qquad \tilde{q}\to 0\ \ \mbox{or $q\to1$}
\eea
now yields the desired result independent of $n$
\bea
\Delta S^n_{\mu,\mu'\!,\mu'\!'}=\log\frac{\hat{d}_\mu}{\hat{d}_{\mu'}\hat{d}_{\mu''}}
\eea
Since this result is independent of $n$, no ambiguity can arise from the analytic continuation in $n$. From (\ref{1dGraphFusAlg}), it follows that
\bea
1=\sum_{\mu=1}^\mathbb{|K|} \tilde{N}_{\mu'\mu''}{}^{\mu} \frac{\hat{d}_\mu}{\hat{d}_{\mu'}\hat{d}_{\mu''}}=\sum_{\mu=1}^\mathbb{|K|} \tilde{N}_{\mu'\mu''}{}^{\mu}\,  \exp(\Delta S^n_{\mu,\mu'\!,\mu'\!'})
\eea
so the exponentials of the relative SRALEs, given by the ratios of quantum dimensions, are probabilities.

\bigskip
\section{Prototypical CFTs}\label{secExamples}

In this section we provide unitary and nonunitary examples showcasing the central role of the coset graphs. 
Simple examples, such as the Ising model, 3-state Potts model and Lee-Yang model are well understood but are included for illustrative purposes and completeness. 

Except for special cases, such as $m=2$ or $m=3$, the tensor product graphs $A_{m-1}\otimes G$ decompose into two disconnected graphs equivalent to the coset graph $\tilde{G}=A\otimes G/\mathbb{Z}_2$. 
The coset graph $\tilde{G}$ may be bipartite or not bipartitite depending on the absence or presence of loops. For ${\cal M}(m,m')$, we find
\bea
\tilde{G}=\begin{cases} 
\mbox{single non-bipartite component},&\mbox{$m=2$ and $m'$ odd}\\[-3pt]
\mbox{single bipartite component},&\mbox{$m=3$ and $m'$ odd or even}\\[-3pt]
\mbox{single bipartite component},&\mbox{$m\ge4$ and $mm'$ even}\\[-3pt]
\mbox{direct sum of two non-bipartite components},\quad&\mbox{$m\ge5$ odd and $m'$ odd}
\end{cases}
\eea
For ${\cal M}(2,m')$, the coset graph $A_{1}\otimes A_{m'-1}/\mathbb{Z}_2$ is the tadpole graph $\tilde{G}=T_{(m'-1)/2}$.
For ${\cal M}(3,m')$, the coset graph $A_{2}\otimes A_{m'-1}/\mathbb{Z}_2$ is the graph $\tilde{G}=A_{m'-1}$. The general expressions for the coset graphs are given in Appendix~\ref{App:Coset}.

\subsection{Prototypical unitary CFTs}

\subsubsection{Critical Ising model ${\cal M}(3,4)=(A_2,A_3)$}

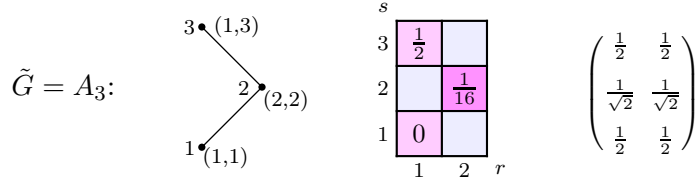
\begin{figure}[h]
\scriptsize
\begin{center}
\qquad\qquad\begin{pspicture}(0,.3)(1,1.6)
\psset{unit=.8cm}
\psline[linewidth=.5pt](0,0)(1,1)(0,2)
\pscircle*(0,0){.06}
\pscircle*(1,1){.06}
\pscircle*(0,2){.06}
\rput(-2.3,1.05){\normalsize $\tilde{G}=A_3$:}
\rput(.4,0){\pp{t}{$(1,1)$}}
\rput(1.4,.95){\pp{t}{$(2,2)$}}
\rput(.2,2){\pp{l}{$(1,3)$}}
\rput(-.2,.1){\pp{t}{$1$}}
\rput(.7,1.1){\pp{t}{$2$}}
\rput(-.3,2){\pp{l}{$3$}}
\end{pspicture} \qquad\qquad\quad
\psset{unit=.6cm}
\begin{pspicture}(0,.7)(2,3)
\psframe[linewidth=0pt,fillstyle=solid,fillcolor=lightlightblue](0,0)(2,3)
\multirput(0,0)(0,2){2}{\multirput[bl](0,0)(2,0){1}{\psframe[linewidth=0pt,fillstyle=solid,fillcolor=lightpurple](0,0)(1,1)}}
\multirput[bl](1,1)(0,2){1}{\psframe[linewidth=0pt,fillstyle=solid,fillcolor=darkpurple](0,0)(1,1)}
\psgrid[gridlabels=0pt,subgriddiv=1](0,0)(2,3)
\rput(.5,.5){\small $0$}
\rput(1.5,1.5){\small $\tfrac{1}{16}$}
\rput(.5,2.5){\small $\tfrac12$}
\rput(.5,-.3){$\small 1$}
\rput(1.5,-.3){$\small 2$}
\rput(2.3,-.3){$\small r$}
\rput(-.3,.5){$\small 1$}
\rput(-.3,1.5){$\small 2$}
\rput(-.3,2.5){$\small 3$}
\rput(-.3,3.3){$\small s$}
\end{pspicture}\qquad\qquad
\raisebox{10pt}{\ssmat{\frac{1}{2} & \frac{1}{2} \\[8pt]
 \frac{1}{\sqrt{2}} & \frac{1}{\sqrt{2}} \\[8pt]
 \frac{1}{2} & \frac{1}{2}}}
\end{center}

\caption{The bipartite coset graph $\tilde{G}=A_2\otimes A_3/\mathbb{Z}_2=A_3$ and Kac tables of conformal weights and 2-boundary $\g$-factors $\g_{(1,1)|(r,s)}$  for the critical Ising model ${\cal M}(3,4)$ with $c=\tfrac12$.  
Under the Kac table symmetry $(1,2)\equiv (2,2)$, so the nodes $(r,s)=(1,1),(2,2),(1,3)$ are simply labelled by $s=1,2,3$.
\label{IsingCosetGraph}}
\end{figure}

The coset graph $\tilde{G}$ of the Ising model ${\cal M}(3,4)$ is $A_2\otimes A_3/\mathbb{Z}_2=A_3$ as shown in Figure~\ref{IsingCosetGraph}. 
Explicitly, the nimrep fusion matrices are
\bea
\tilde{N}_1=I=\!\ssmat{1&0&0\\ 0&1&0\\ 0&0&1},\quad \tilde{N}_{2}=\tilde{G}=\!\ssmat{0&1&0\\ 1&0&1\\ 0&1&0},\quad 
\tilde{N}_{3}=\sigma=\!\ssmat{0&0&1\\ 0&1&0\\ 1&0&0}
\eea
with $\tilde{N}_2^2=I+\tilde{N}_3$, $\tilde{N}_2\tilde{N}_1=\tilde{N}_2\tilde{N}_3=\tilde{N}_2$, $\tilde{N}_3^2=I$ and quantum dimensions $\tilde{d}_{s}=[s]_y$ where $s=1,2,3$, $y=e^{\pi i/4}$ and $\sigma=\tilde{N}_3$ is the spin-reversal operator. 
The coset graph fusion algebra is realized as the polynomial ring
\bea
\mathbb{Z}[y]/\langle y^2\!-\!2\rangle
\eea
The conformal cylinder partition functions $Z_{s|s'}(q)=Z_{s'|s}(q)$ are
\bea
Z_{1|s}(q)=\chi_{1,s}^{3,4}(q),\quad Z_{2|2}(q)=\chi_{1,1}^{3,4}(q)+\chi_{1,3}^{3,4}(q),\quad Z_{2|3}(q)=\chi_{1,2}^{3,4}(q),\quad Z_{3|3}(q)=\chi_{1,1}^{3,4}(q)
\eea
The unitary matrix ${\cal S}$ that diagonalizes $\tilde{G}$ is the modular matrix
\bea
{\cal S}={\cal S}^T=\tfrac12\!\!\sssmat{1&\sqrt{2}&1\\ \sqrt{2}&0&-\sqrt{2}\\ 1&-\sqrt{2}&1},\qquad 
{\cal S}^{-1} \tilde{N}_2{\cal S}=\!\ssmat{\sqrt{2}&0&0\\ 0&0&0\\ 0&0&-\sqrt{2}},\qquad {\cal S}^2=I
\eea
The 2-boundary $\g$-factors $\tilde{\g}_{s|s'}=\tilde{\g}_s\tilde{\g}_{s'}$ are 
\bea
\tilde{\g}_{1|1}\!=\!\tilde{\g}_{1|3}\!=\!{\cal S}_{1,1}\!=\!\tfrac12,\qquad 
\tilde{\g}_{1|2}\!=\!{\cal S}_{1,2}\!=\!\tfrac{1}{\sqrt{2}},\qquad \tilde{\g}_{2|2}=\frac{\tilde{\g}_{1|2}^2}{\tilde{\g}_{1|1}}=1
\eea
The multiplicative defect $\g$-factors $\tilde{d}_s$  
\bea
\tilde{d}_s=\frac{\tilde{\g}_{1|s}}{\tilde{\g}_{1|1}},\quad \tilde{d}_{1}=\tilde{d}_{3}=1,\quad \tilde{d}_{2}=2\cos\tfrac{\pi}{4}=\sqrt{2}=1.414214\ldots,\quad \tilde{d}^{\,2}_2=1+ \tilde{d}_3
\eea
give a 1-dimensional representation of the coset graph fusion algebra.

In terms of the asymptotics of counting fusion paths, we find
\bea
\lim_{\substack{N\to\infty\\[1pt] \text{$N\!=\!\kappa$\,mod\,2}}} (\tfrac{1}{\sqrt{2}})^{N}\tilde{G}^N\!\!=\lim_{\substack{N\to\infty\\[1pt] \text{$N\!=\!\kappa$\,mod\,2}}} (\tfrac{1}{\sqrt{2}})^{N}\!\!\ssmat{0&1&0\\ 1&0&1\\ 0&1&0}^{\!N}\!\!={\cal S}\!\!\ssmat{1&0&0\\ 0&0&0\\ 0&0&(-1)^\kappa}{\cal S}^{-1}=\!
\begin{cases}\vec D_+,&\mbox{$N$ even}\\\vec D_-,&\mbox{$N$ odd}\end{cases}
\eea
with
\bea 
\vec D_+=\tfrac12\!\!\sssmat{1&0&1\\ 0&2&0\\ 1&0&1}=\tfrac12 (\tilde{N}_1+\tilde{N}_3),\qquad
\vec D_-=\tfrac12\!\!\sssmat{0&\sqrt{2}&0\\ \sqrt{2}&0&\sqrt{2}\\ 0&\sqrt{2}&0}=\tfrac{1}{\sqrt{2}}\tilde{N}_2
\eea
Combining the even and odd matrices, gives the rank-1 matrix
\bea
\vec D\!=\!\tfrac12(\vec D_+\!+\!\vec D_-)\!=\!\tfrac14\!\!\sssmat{1&\sqrt{2}&1\\ \sqrt{2}&2&\sqrt{2}\\ 1&\sqrt{2}&1}\!
=\!\tilde{\g}_{1|1}\!\!\ssmat{\tilde{\g}_{1|1}&\tilde{\g}_{1|2}&\tilde{\g}_{1|3}\\ 
\tilde{\g}_{2|1}&\tilde{\g}_{2|2}&\tilde{\g}_{2|3}\\
\tilde{\g}_{3|1}&\tilde{\g}_{3|2}&\tilde{\g}_{3|3}}\!
=\tilde{\g}_{1|1}^2\!\!\ssmat{1&\tilde{d}_{2}&\tilde{d}_{3}\\ \tilde{d}_{2}&\tilde{d}_{2}^2&\tilde{d}_{2}\tilde{d}_{3}\\
\tilde{d}_{3}&\tilde{d}_{3}\tilde{d}_{2}&\tilde{d}^2_{3}}
\!=\!\tilde{\g}^2_{1|1}\!\!\sssmat{1\\ \tilde{d}_{2}\\ \tilde{d}_{3}}\!\!\sssmat{1&\tilde{d}_{2}&\tilde{d}_{3}}
\eea
The boundary and defect $\g$-factors are simply obtained by solving these equations. Notice that
\bea
\vec D=\tilde{\g}^2_{1|1} \sum_{s=1}^3 \tilde{d}_s \tilde{N}_s
\eea

\subsubsection{Tricritical Ising model ${\cal M}(4,5)=(A_3,A_4)$}

\begin{figure}[h]
\scriptsize
\begin{center}
\qquad\qquad\quad\begin{pspicture}[shift=2pt](0,0)(2,3)
\psset{unit=.8cm}
\psline[linewidth=.5pt](0,0)(1,1)(2,0)
\psline[linewidth=.5pt](1,3)(0,2)(1,1)(2,2)(1,3)
\pscircle*(0,0){.06}
\pscircle*(1,1){.06}
\pscircle*(2,0){.06}
\pscircle*(1,3){.06}
\pscircle*(0,2){.06}
\pscircle*(2,2){.06}
\rput(-2.1,1.5){\normalsize $\tilde{G}$:}
\rput(.4,0){\pp{t}{$(1,1)$}}
\rput(2.4,0){\pp{t}{$(3,1)$}}
\rput(1.2,1){\pp{l}{$(2,2)$}}
\rput(.2,2){\pp{l}{$(1,3)$}}
\rput(2.1,2){\pp{l}{$(3,3)$}}
\rput(1.2,3){\pp{l}{$(2,4)$}}
\rput(-.2,-.05){\pp{t}{$1$}}
\rput(.6,1){\pp{l}{$2$}}
\rput(1.6,2){\pp{l}{$3$}}
\rput(.6,3){\pp{l}{$4$}}
\rput(-.3,2){\pp{l}{$5$}}
\rput(1.8,-.05){\pp{t}{$6$}}
\end{pspicture} \qquad\qquad\qquad
\psset{unit=.6cm}
\begin{pspicture}(0,0)(3,4)
\psframe[linewidth=0pt,fillstyle=solid,fillcolor=lightlightblue](0,0)(3,4)
\multirput(0,0)(0,2){2}{\multirput[bl](0,0)(2,0){2}{\psframe[linewidth=0pt,fillstyle=solid,fillcolor=lightpurple](0,0)(1,1)}}
\multirput[bl](1,1)(0,2){2}{\psframe[linewidth=0pt,fillstyle=solid,fillcolor=darkpurple](0,0)(1,1)}
\psgrid[gridlabels=0pt,subgriddiv=1](0,0)(3,4)
\rput(.5,.5){\small $0$}
\rput(1.5,1.5){\small $\tfrac{3}{80}$}
\rput(2.5,2.5){\small $\tfrac{1}{10}$}
\rput(1.5,3.5){\small $\tfrac{7}{16}$}
\rput(.5,2.5){\small $\tfrac35$}
\rput(2.5,.5){\small $\tfrac32$}
\rput(.5,-.3){$\small 1$}
\rput(1.5,-.3){$\small 2$}
\rput(2.5,-.3){$\small 3$}
\rput(3.3,-.3){$\small r$}
\rput(-.3,.5){$\small 1$}
\rput(-.3,1.5){$\small 2$}
\rput(-.3,2.5){$\small 3$}
\rput(-.3,3.5){$\small 4$}
\rput(-.3,4.3){$\small s$}
\end{pspicture}\qquad\qquad
\raisebox{30pt}{\ssmat{
\sqrt{\tfrac{(5-\sqrt{5})}{40}}&\sqrt{\tfrac{(5-\sqrt{5})}{20}}&\sqrt{\tfrac{(5-\sqrt{5})}{40}}\\[4pt]
\sqrt{\tfrac{(5+\sqrt{5})}{40}}&\sqrt{\tfrac{(5+\sqrt{5})}{20}}&\sqrt{\tfrac{(5+\sqrt{5})}{40}}\\[4pt]
\sqrt{\tfrac{(5+\sqrt{5})}{40}}&\sqrt{\tfrac{(5+\sqrt{5})}{20}}&\sqrt{\tfrac{(5+\sqrt{5})}{40}} \\[4pt]
\sqrt{\tfrac{(5-\sqrt{5})}{40}}&\sqrt{\tfrac{(5-\sqrt{5})}{20}}&\sqrt{\tfrac{(5-\sqrt{5})}{40}} }}
\end{center}
\caption{The bipartite coset graph $\tilde{G}=\tilde{N}_2=A_3\otimes A_4/\mathbb{Z}_2$ and Kac tables of conformal weights and 2-boundary $\g$-factors $\g_{(1,1)|(r,s)}$  for the tricritical Ising model ${\cal M}(4,5)$ with $c=\tfrac7{10}$. 
Choosing the $r\!+\!s$ even sublattice, the nodes $(r,s)\in\mathbb{K}=\{(1,1),(2,2),(3,3),(2,4),(1,3),(3,1)\}$ are labelled by $\mu=1,2,\ldots,6$. The fundamental $(2,2)$ is labelled by $\mu=2$.
\label{TricritIsingCosetGraph}}
\end{figure}
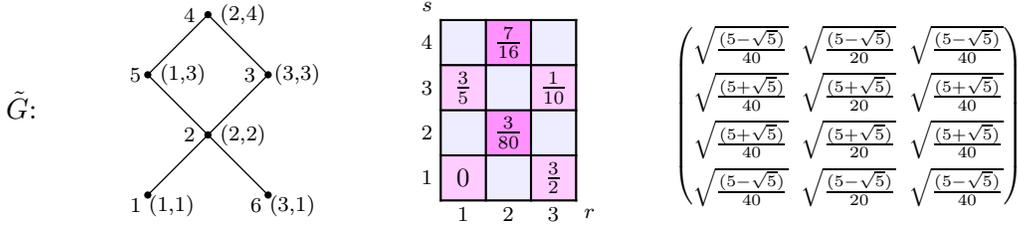

The coset graph $\tilde{G}$ of the tricritical Ising model~\cite{PK91,KP91,LMC91} ${\cal M}(4,5)$ is $\tilde{G}=A_3\otimes A_4/\mathbb{Z}_2$ as shown in Figure~\ref{TricritIsingCosetGraph}. 
With these choices, the explicit nimrep fusion matrices are
\bea
\tilde{N}_1=\!\ssmat{
 1 & 0 & 0 & 0 & 0 & 0 \\
 0 & 1 & 0 & 0 & 0 & 0 \\
 0 & 0 & 1 & 0 & 0 & 0 \\
 0 & 0 & 0 & 1 & 0 & 0 \\
 0 & 0 & 0 & 0 & 1 & 0 \\
 0 & 0 & 0 & 0 & 0 & 1},\quad 
 \tilde{N}_{2}=\!\ssmat{
 0 & 1 & 0 & 0 & 0 & 0 \\
 1 & 0 & 1 & 0 & 1 & 1 \\
 0 & 1 & 0 & 1 & 0 & 0 \\
 0 & 0 & 1 & 0 & 1 & 0 \\
 0 & 1 & 0 & 1 & 0 & 0 \\
 0 & 1 & 0 & 0 & 0 & 0},\quad 
 \tilde{N}_{3}=\!\ssmat{
 0 & 0 & 1 & 0 & 0 & 0 \\
 0 & 1 & 0 & 1 & 0 & 0 \\
 1 & 0 & 0 & 0 & 1 & 0 \\
 0 & 1 & 0 & 0 & 0 & 0 \\
 0 & 0 & 1 & 0 & 0 & 1 \\
 0 & 0 & 0 & 0 & 1 & 0}\nonumber\\
 \tilde{N}_4=\!\ssmat{
 0 & 0 & 0 & 1 & 0 & 0 \\
 0 & 0 & 1 & 0 & 1 & 0 \\
 0 & 1 & 0 & 0 & 0 & 0 \\
 1 & 0 & 0 & 0 & 0 & 1 \\
 0 & 1 & 0 & 0 & 0 & 0 \\
 0 & 0 & 0 & 1 & 0 & 0},\quad 
 \tilde{N}_{5}=\!\ssmat{
 0 & 0 & 0 & 0 & 1 & 0 \\
 0 & 1 & 0 & 1 & 0 & 0 \\
 0 & 0 & 1 & 0 & 0 & 1 \\
 0 & 1 & 0 & 0 & 0 & 0 \\
 1 & 0 & 0 & 0 & 1 & 0 \\
 0 & 0 & 1 & 0 & 0 & 0},\quad 
 \tilde{N}_{6}=\!\ssmat{
 0 & 0 & 0 & 0 & 0 & 1 \\
 0 & 1 & 0 & 0 & 0 & 0 \\
 0 & 0 & 0 & 0 & 1 & 0 \\
 0 & 0 & 0 & 1 & 0 & 0 \\
 0 & 0 & 1 & 0 & 0 & 0 \\
 1 & 0 & 0 & 0 & 0 & 0}
\eea
Since $\tilde{N}_{\mu 1}{}^{\mu'}=\delta_{\mu,\mu'}$, the fusion matrices are linearly independent and $\mu$ labels the position of the single 1 in the first row of $\tilde{N}_\mu$.
The fusion matrices, along with the quantum dimensions $\tilde{d}_{(r,s)}=[r]_x[s]_y$ for $(r,s)\in\mathbb{K}$, $x=e^{\pi i/4}$ and $y=e^{\pi i/5}$, satisfy the coset graph fusion algebra. 
The coset graph fusion algebra is realized as the polynomial ring
\bea
\mathbb{Z}[x,y]/\langle x^3\!-\!2x,y^2\!-\!x^2 y\!+\!y\!-\!1\rangle
\eea
The conformal cylinder partition functions $Z_{s|s'}(q)=Z_{s'|s}(q)$ are given by (\ref{CylPFs}).
The unitary matrix ${\cal S}$ that diagonalizes the fusion matrices is the modular matrix
\bea
{\cal S}\!=\!{\cal S}^T\!\!=\!\tfrac12\!\!\sssmat{
s_1 & \sqrt{2} s_2 & s_2 & \sqrt{2} s_1 & s_2 & s_1 \\
 \sqrt{2} s_2 & 0 & \sqrt{2} s_1 & 0 & -\sqrt{2} s_1 & -\sqrt{2} s_2 \\
 s_2 & \sqrt{2} s_1 & -s_1 & -\sqrt{2} s_2 & -s_1 & s_2 \\
 \sqrt{2} s_1 & 0 & -\sqrt{2} s_2 & 0 & \sqrt{2} s_2 & -\sqrt{2} s_1 \\
 s_2 & -\sqrt{2} s_1 & -s_1 & \sqrt{2} s_2 & -s_1 & s_2 \\
 s_1 & -\sqrt{2} s_2 & s_2 & -\sqrt{2} s_1 & s_2 & s_1
   },\ \  
{\cal S}^{-1}\!\tilde{N}_2\,{\cal S}=\!\!\sssmat{\lambda_+&0&0&0&0&0\\ 0&0&0&0&0&0\\ 0&0&\lambda_-&0&0&0\\ 0&0&0&0&0&0\\ 0&0&0&0&\!-\lambda_-\!&0\\ 0&0&0&0&0&\!\!-\lambda_+},\ \  {\cal S}^2\!=\!I
\eea
where $s_1\!=\!\tfrac{2}{\sqrt{5}}\sin\tfrac{\pi}{5}$, $s_2\!=\!\tfrac{2}{\sqrt{5}}\sin\tfrac{2\pi}{5}$ and $\lambda_\pm\!=\!\sqrt{3\!\pm\!\sqrt{5}}$. 
The multiplicative boundary $\g$-factors $\tilde{\g}_{s|s'}\!=\!\tilde{\g}_s\tilde{\g}_{s'}$ are 
\bea
\tilde{\g}_{1|1}\!=\!\tilde{\g}_{1|6}\!=\!{\cal S}_{1,1}\!=\!\tfrac12 s_1,\quad 
\tilde{\g}_{1|2}\!=\!{\cal S}_{1,2}\!=\!\tfrac{1}{\sqrt{2}} s_2,\quad 
\tilde{\g}_{1|3}\!=\!\tilde{\g}_{1|5}\!=\!{\cal S}_{1,3}\!=\!\tfrac12 s_2,\quad
\tilde{\g}_{1|4}\!=\!{\cal S}_{1,4}\!=\!\tfrac{1}{\sqrt{2}} s_1
\eea
The multiplicative defect $\g$-factors are
\begin{subequations}
\begin{align}
&\qquad\quad\tilde{d}_\mu=\frac{\tilde{\g}_{1|\mu}}{\tilde{\g}_{1|1}}=\frac{{\cal S}_{1,\mu}}{{\cal S}_{1,1}}=\Big\{1,\sqrt{3\!+\!\sqrt{5}},\tfrac12(1\!+\!\sqrt{5}),\sqrt{2},\tfrac12(1\!+\!\sqrt{5}),1\Big\}\\
&\tilde{d}_{(1,1)}=\tilde{d}_{(3,1)}=1,\quad \tilde{d}_{(2,2)}=\sqrt{3\!+\!\sqrt{5}},\quad \tilde{d}_{(3,3)}=\tilde{d}_{(1,3)}=\tfrac12(1\!+\!\sqrt{5}),\quad \tilde{d}_{(2,4)}=\sqrt{2}
\end{align}
\end{subequations}

In terms of the asymptotics of counting fusion paths, we find
\bea
\lim_{\substack{N\to\infty\\[1pt] \text{$N\!=\!\kappa$\,mod\,2}}} (\tfrac{1}{\lambda_+})^{N}\tilde{G}^{N}\!\!={\cal S}\,\text{diag}\{1,0,0,0,0,(-1)^\kappa\}{\cal S}^{-1}=\!
\begin{cases}\vec D_+,&\mbox{$N$ even}\\\vec D_-,&\mbox{$N$ odd}\end{cases}
\eea
Combining the even and odd matrices, gives the rank-1 matrix
\bea
\vec D\!=\!\tfrac12\,(\vec D_+\!+\!\vec D_-)\!=\!\frac12\!\!\ssmat{
\tfrac{1}{20}(5\!-\!\sqrt{5})&\tfrac{1}{\sqrt{10}}&\tfrac{1}{2\sqrt{5}}&\sqrt{\tfrac{1}{20}(3\!-\!\sqrt{5})}&\tfrac{1}{2\sqrt{5}}&\tfrac{1}{20}(5\!-\!\sqrt{5})\\[4pt]
\frac{1}{\sqrt{10}}&\tfrac{1}{10}(5\!+\!\sqrt{5})&\sqrt{\tfrac{1}{20}(3\!+\!\sqrt{5})}&\tfrac{1}{\sqrt{5}}&\sqrt{\tfrac{1}{20}(3\!+\!\sqrt{5})}&\tfrac{1}{\sqrt{10}}\\[4pt]
\tfrac{1}{2\sqrt{5}}&\sqrt{\tfrac{1}{20}(3\!+\!\sqrt{5})}&\tfrac{1}{20}(5\!+\!\sqrt{5})&\tfrac{1}{\sqrt{10}}&\tfrac{1}{20}(5\!+\!\sqrt{5})&\tfrac{1}{2\sqrt{5}}\\[4pt]
\sqrt{\tfrac{1}{20}(3\!-\!\sqrt{5})}&\tfrac{1}{\sqrt{5}}&\tfrac{1}{\sqrt{10}}&\tfrac{1}{10}(5\!-\!\sqrt{5})&\tfrac{1}{\sqrt{10}}&\sqrt{\tfrac{1}{20}(3\!-\!\sqrt{5})}\\[4pt]
\tfrac{1}{2\sqrt{5}}&\sqrt{\tfrac{1}{20}(3\!+\!\sqrt{5})}&\tfrac{1}{20}(5\!+\!\sqrt{5})&\tfrac{1}{\sqrt{10}}&\tfrac{1}{20}(5\!+\!\sqrt{5})&\tfrac{1}{2\sqrt{5}}\\[4pt]
\tfrac{1}{20}(5\!-\!\sqrt{5})&\tfrac{1}{\sqrt{10}}&\tfrac{1}{2\sqrt{5}}&\sqrt{\tfrac{1}{20}(3\!-\!\sqrt{5})}&\tfrac{1}{2\sqrt{5}}&\tfrac{1}{20}(5\!-\!\sqrt{5})
}
\eea
The boundary and defect $\g$-factors are obtained by solving 
\bea
\vec D=\hat{\g}_{1|1}\!\!\!\sssmat{\hat{\g}_{1|1}&\hat{\g}_{1|2}&\ldots&\hat{\g}_{1|6}\\
\hat{\g}_{2|1}&\hat{\g}_{2|2}&\ldots&\hat{\g}_{2|6}\\
\vdots&\vdots&\ddots&\vdots\\
\hat{\g}_{6|1}&\hat{\g}_{6|2}&\ldots&\hat{\g}_{6|6}}
=\hat{\g}^2_{1|1}\!\sssmat{1&\hat{d}_{2}&\ldots&\hat{d}_{6}\\ 
\hat{d}_{2}&\hat{d}_{2}^2&\ldots&\hat{d}_{2}\hat{d}_{6}\\[-4pt]
\vdots&\vdots&\ddots&\vdots\\
\hat{d}_{6}&\hat{d}_{2}\hat{d}_{6}&\ldots&\hat{d}^2_{6}}\!
=\hat{\g}^2_{1|1}\!\sssmat{1\\ \hat{d}_{2}\\[-4pt] \vdots\\ \hat{d}_{6}}\!\sssmat{1&\hat{d}_{2}&\ldots&\hat{d}_{6}}
\eea
Notice that 
\bea
\vec D=\hat{\g}^2_{1|1} \sum_{\mu=1}^6 \hat{d}_\mu\hat{N}_\mu
\eea

Alternatively, using $\tilde{G}=A_3\otimes A_4/\mathbb{Z}_2= A_3\otimes T_2$ with $(r,s)$ representatives restricted by $1\le s\le 2$, we could have chosen $\mathbb{K}=\{(1,1),(1,2),(2,1),(2,2),(3,1),(3,2)\}$.
The nimrep fusion matrices would then be $\tilde{N}_{r,s}=N_r^{(A_3)}\!\otimes\! N_s^{(T_2)}$ with
\bea
\tilde{N}_{1,s}\!=\!\!\!\sssmat{{N}^{(T_2)}_s&0&0\\ 0&{N}^{(T_2)}_s&0\\0&0&{N}^{(T_2)}_s},\ 
\tilde{N}_{2,s}\!=\!\!\!\sssmat{0&{N}^{(T_2)}_s&0\\ {N}^{(T_2)}_s&0&{N}^{(T_2)}_s\\0&{N}^{(T_2)}_s&0},\ 
\tilde{N}_{3,s}\!=\!\!\!\sssmat{0&0&{N}^{(T_2)}_s\\ 0&{N}^{(T_2)}_s&0\\ {N}^{(T_2)}_s&0&0},
\quad s=1,2
\eea
Allowing for a simultaneous permutation of the rows and columns of $\tilde{N}_{r,s}$, ${\cal S}$, $\vec D$ and an associated reordering of the elements of $\mathbb{K}$, the results are identical.

\subsubsection{Critical 3-state Potts model $(A_4,D_4)$}

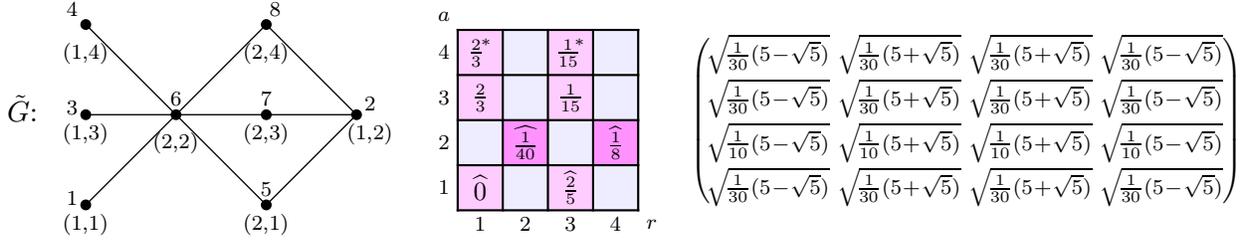
\begin{figure}[htb]
\scriptsize
\begin{center}
\qquad\quad\begin{pspicture}(0,0)(3.75,2.3)
\psset{unit=1.2cm}
\psline[linewidth=.5pt](0,0)(1,1)(2,0)(3,1)
\psline[linewidth=.5pt](0,2)(1,1)(2,2)(3,1)
\psline[linewidth=.5pt](0,1)(3,1)
\pscircle*(0,0){.06}
\pscircle*(0,1){.06}
\pscircle*(0,2){.06}
\pscircle*(1,1){.06}
\pscircle*(2,0){.06}
\pscircle*(2,1){.06}
\pscircle*(2,2){.06}
\pscircle*(3,1){.06}
\rput(-.7,1.07){\normalsize $\tilde{G}$:}
\rput(0,-.1){\pp{t}{$(1,1)$}}
\rput(0,.9){\pp{t}{$(1,3)$}}
\rput(0,1.8){\pp{t}{$(1,4)$}}
\rput(1,.8){\pp{t}{$(2,2)$}}
\rput(2,-.1){\pp{t}{$(2,1)$}}
\rput(2,.9){\pp{t}{$(2,3)$}}
\rput(2,1.8){\pp{t}{$(2,4)$}}
\rput(3.15,.9){\pp{t}{$(1,2)$}}
\rput(-.15,.15){\pp{t}{$1$}}
\rput(3.15,1.2){\pp{t}{$2$}}
\rput(-.15,1.15){\pp{t}{$3$}}
\rput(-.15,2.25){\pp{t}{$4$}}
\rput(2,.25){\pp{t}{$5$}}
\rput(1,1.25){\pp{t}{$6$}}
\rput(2,1.25){\pp{t}{$7$}}
\rput(2.1,2.25){\pp{t}{$8$}}
\end{pspicture}\qquad\qquad
\psset{unit=.6cm}
\begin{pspicture}[shift=-2pt](0,0)(4,4)
\psframe[linewidth=0pt,fillstyle=solid,fillcolor=lightlightblue](0,0)(4,4)
\multirput(0,0)(0,2){2}{\multirput[bl](0,0)(2,0){2}{\psframe[linewidth=0pt,fillstyle=solid,fillcolor=lightpurple](0,0)(1,1)}}
\multirput(0,3)(0,2){1}{\multirput[bl](0,0)(2,0){2}{\psframe[linewidth=0pt,fillstyle=solid,fillcolor=lightpurple](0,0)(1,1)}}
\multirput(0,0)(2,0){2}{\multirput[bl](1,1)(0,2){1}{\psframe[linewidth=0pt,fillstyle=solid,fillcolor=darkpurple](0,0)(1,1)}}
\psgrid[gridlabels=0pt,subgriddiv=1](0,0)(4,4)
\rput(.5,.5){\small $\widehat 0$}
\rput(1.5,1.5){$\widehat{\small \tfrac{1}{40}}$}
\rput(2.5,2.5){$\small \tfrac{1}{15}$}
\rput(2.5,3.5){$\small \tfrac{1}{15}^{\!\!*}$}
\rput(.5,2.5){$\small \tfrac{2}{3}$}
\rput(.5,3.5){$\small \tfrac{2}{3}^{\!*}$}
\rput(3.5,1.5){$\small \widehat{\tfrac18}$}
\rput(2.5,.5){$\small \widehat{\tfrac25}$}
\rput(.5,-.3){$\small 1$}
\rput(1.5,-.3){$\small 2$}
\rput(2.5,-.3){$\small 3$}
\rput(3.5,-.3){$\small 4$}
\rput(4.3,-.3){$\small r$}
\rput(-.3,.5){$\small 1$}
\rput(-.3,1.5){$\small 2$}
\rput(-.3,2.5){$\small 3$}
\rput(-.3,3.5){$\small 4$}
\rput(-.3,4.3){$\small a$}
\end{pspicture}\qquad
\raisebox{30pt}{\sssmat{
\!\sqrt{\!\tfrac{1}{30}(5\!-\!\sqrt{5})}&\sqrt{\!\tfrac{1}{30}(5\!+\!\sqrt{5})}&\sqrt{\!\tfrac{1}{30}(5\!+\!\sqrt{5})}&\sqrt{\!\tfrac{1}{30}(5\!-\!\sqrt{5})}\\[4pt]
\!\sqrt{\!\tfrac{1}{30}(5\!-\!\sqrt{5})}&\sqrt{\!\tfrac{1}{30}(5\!+\!\sqrt{5})}&\sqrt{\!\tfrac{1}{30}(5\!+\!\sqrt{5})}&\sqrt{\!\tfrac{1}{30}(5\!-\!\sqrt{5})}\\[4pt]
\!\sqrt{\!\tfrac{1}{10}(5\!-\!\sqrt{5})}&\sqrt{\!\tfrac{1}{10}(5\!+\!\sqrt{5})}&\sqrt{\!\tfrac{1}{10}(5\!+\!\sqrt{5})}&\sqrt{\!\tfrac{1}{10}(5\!-\!\sqrt{5})}\\[4pt]
\!\sqrt{\!\tfrac{1}{30}(5\!-\!\sqrt{5})}&\sqrt{\!\tfrac{1}{30}(5\!+\!\sqrt{5})}&\sqrt{\!\tfrac{1}{30}(5\!+\!\sqrt{5})}&\sqrt{\!\tfrac{1}{30}(5\!-\!\sqrt{5})}
}}
\end{center}
\caption{The bipartite coset graph $\tilde{G}=A_4\otimes D_4/\mathbb{Z}_2=T_2\otimes D_4$ and Kac tables of conformal weights and 2-boundary $\g$-factors $\hat{\g}_{(1,1)|(r,a)}$ of the critical 3-state Potts model $(A_4,D_4)$ with $c=\tfrac45$ where $T_2$ is the tadpole on 2 nodes. 
The nodes are $(r,a)$ with $r=1,2$ and $a=1,2,3,4$. Alternatively, the ordered nodes $(r,a)=(1,1),(1,2),(1,3),(1,4),(2,1),(2,2),(2,3),(2,4)$ are labelled by $\mu=1,2,\ldots,8$. 
The label $a=4$ can be replaced with $a=3^*$ reflecting the fact that $a=3$ and $4$ are on the same sublattice and rows 3 and 4 are duplicates. 
The $D_4$ graph arises as the $\mathbb{Z}_2$ orbifold of the $A_5$ diagram. 
Under the $(A_4,A_5)$ Kac table symmetry, $(2,1)\equiv (3,1)$, $(2,3) \equiv (3,3)$, $(2,4) \equiv (3,4)$ and $(1,2) \equiv (4,2)$. 
The extended conformal weights are 
$\widehat{0}=0+3$, $\small \widehat{\tfrac{1}{40}}=\tfrac{1}{40}+\tfrac{21}{40}$, $\small \widehat{\tfrac18}=\tfrac18+\tfrac{13}{8}$, $\widehat{\tfrac25}=\tfrac25+\tfrac75$.
\label{3PottsCosetGraph}}
\end{figure}

The coset graph of the 3-state Potts model~\cite{BPZ1998,DKMM94,SaleurEtAl2024} is $\tilde{G}=A_4\otimes D_4/\mathbb{Z}=T_2\otimes D_4$ as shown in Figure~\ref{3PottsCosetGraph}.
Explicitly, the nimrep fusion matrices are
\bea
\tilde{N}_{r,a}=N_r^{(T_2)}\!\otimes\! \hat{N}_a^{(D_4)},\  \tilde{N}_{1,a}\!=\!\!\!\ssmat{\hat{N}^{(D_4)}_a&0\\ 0&\hat{N}^{(D_4)}_a},\ 
\tilde{N}_{2,a}\!=\!\!\!\ssmat{0&\hat{N}^{(D_4)}_a\\ \hat{N}^{(D_4)}_a&\hat{N}^{(D_4)}_a},\quad r=1,2;\ a=1,2,3,4
\eea
The nimreps of $T_2$ and $D_4$ and the adjacency matrix (fundamental nimrep) $\tilde{G}=\tilde{N}_{2,2}$ are
\begin{align}
N_r^{(T_2)}=\!\!\sssmat{1&0\\ 0&1}, \!\!\sssmat{0&1\\ 1&1};\ \ 
\hat{N}^{(D_4)}_a=
\sssmat{1&0&0&0\\ 0&1&0&0\\ 0&0&1&0\\ 0&0&0&1}, \!\!
\sssmat{0&1&0&0\\ 1&0&1&1\\ 0&1&0&0\\ 0&1&0&0},  \!\!
\sssmat{0&0&1&0\\ 0&1&0&0\\ 0&0&0&1\\ 1&0&0&0},  \!\!
\sssmat{0&0&0&1\\ 0&1&0&0\\ 1&0&0&0\\ 0&0&1&0};\ \ 
&\tilde{G}=\!\!
\sssmat{
0&0&0&0&0&1&0&0\\ 
0&0&0&0&1&0&1&1\\ 
0&0&0&0&0&1&0&0\\ 
0&0&0&0&0&1&0&0\\
0&1&0&0&0&1&0&0\\ 
1&0&1&1&1&0&1&1\\ 
0&1&0&0&0&1&0&0\\ 
0&1&0&0&0&1&0&0}
\end{align}
\begin{subequations}
The polynomial ring for this case will be discussed elsewhere. The Cayley table of the $T_2\otimes D_4$ graph fusion algebra is shown in Figure~\ref{3PottsCayley}. 
It is possible to add the $\mathbb{Z}_2$ diagram automorphism $\sigma$ to the coset graph fusion algebra. In this case, the enlarged graph fusion algebra becomes noncommutative because it contains within it the noncommutative symmetric group $\mathbb{S}_3$ but we do not do this here.

\begin{figure}[htb]
\begin{center}
\footnotesize
\mbox{}\hspace{-0pt}\mbox{}
\arraycolsep=1.0pt
$\begin{array}{c||c|c|c|c||c|c|c|c|}
&\ 11\ &\ 12\ &\ 13\ &\ 14\ &\ 21\ &\ 22\ &\ 23\ &\ 24\  \rule{0pt}{6pt}\\
\hline\hline
11&11&12&13&14&21&22&23&24\rule{0pt}{6pt}\\
\hline
12&12&11\!+\!13\!+\!14&12&12&22&21\!+\!23\!+\!24&22&22\rule{0pt}{6pt}\rule{0pt}{6pt}\\
\hline
13&13&12&14&11&23&22&24&21\rule{0pt}{6pt}\rule{0pt}{6pt}\\
\hline
14&14&12&11&13&24&22&21&23\rule{0pt}{6pt}\rule{0pt}{6pt}\\
\hline\hline
21&21&22&23&24&11\!+\!21&12\!+\!22&13\!+\!23&14\!+\!24 \rule{0pt}{6pt}\\
\hline
22&22&21\!+\!23\!+\!24&22&22&12\!+\!22&11\!+\!13\!+\!14\!+21\!+\!23\!+\!24&12\!+\!22&12\!+\!22\rule{0pt}{6pt}\\
\hline
23&23&22&24&21&13\!+\!23&12\!+\!22&14\!+\!24&11\!+\!21\rule{0pt}{6pt}\\
\hline
24&24&22&21&23&14\!+\!24&12\!+\!22&11\!+\!21&13\!+\!23\rule{0pt}{6pt}\\
\hline
\end{array}\nonumber$
\end{center}
\caption{Cayley table of the coset graph fusion algebra of the critical 3-state Potts model. We use the compact notations $ra=\tilde{N}_{r,a}$, $r=1,2$; $a=1,2,3,4$ with the order
$(r,a)=(1,1),(1,2),(1,3),(1,4),(2,1),(2,2),(2,3),(2,4)$ labelled by $\mu=1,2,\ldots,8$.
\label{3PottsCayley}}
\end{figure}

The unitary matrix $\boldsymbol\Psi$ that diagonalizes $\tilde{G}$ and the coset fusion matrices is~\cite{BPZ1998}
\begin{align}
&\qquad \boldsymbol\Psi={\cal S}\otimes \Psi, \qquad {\cal S}=\tfrac{2}{\sqrt{5}}\!\sssmat{\sin\tfrac{\pi}5&\sin\tfrac{2\pi}5\\[4pt] \sin\tfrac{2\pi}5&-\sin\tfrac{\pi}5},\qquad 
\Psi=\tfrac{1}{\sqrt{3}}\!
\sssmat{
\frac{1}{\sqrt{2}}&\frac{1}{\sqrt{2}}&1&1\\[4pt]
\sqrt{\frac32}&-\sqrt{\frac32}&0&0\\[4pt]
\frac{1}{\sqrt{2}}&\frac{1}{\sqrt{2}}&\omega&\omega^2\\[4pt]
\frac1{\sqrt{2}}&\frac1{\sqrt{2}}&\omega^2&\omega}\\
&\boldsymbol\Psi^{-1}\tilde{G}\,\boldsymbol\Psi=
\mbox{diag}\Big\{\tfrac{\sqrt{3}}{2}(1\!+\!\sqrt{5}),-\tfrac{\sqrt{3}}{2}(1\!+\!\sqrt{5}),0,0,-\tfrac{\sqrt{3}}{2}(\sqrt{5}\!-\!1),\tfrac{\sqrt{3}}{2}(\sqrt{5}\!-\!1),0,0\Big\}
\end{align}
\end{subequations}
where $\omega=\exp(2\pi i/3)$. The 2-boundary $\g$-factors $\hat{\g}_{(1,1)|(r,a)}$ and defect $\g$-factors $\hat{d}_{(r,a)}$ are
\begin{align}
\hat{\g}_{(1,1)|(r,a)}\!=\!\tfrac{4}{\sqrt{15}}\sin\tfrac{\pi}{5}\sin\tfrac{\pi}6\,\tilde{d}_{(r,a)},\quad
\hat{d}_{(r,a)}&=\{1,\sqrt{3},1,1,\tfrac12(1\!+\!\sqrt{5}),\tfrac{\sqrt{3}}{2}(1\!+\!\sqrt{5}),\tfrac12(1\!+\!\sqrt{5}),\tfrac12(1\!+\!\sqrt{5})\}\qquad
\end{align}

In terms of the asymptotics of counting fusion paths, we find
\bea
\lim_{\substack{N\to\infty\\[1pt] \text{$N\!=\!\kappa$\,mod\,2}}} (\tfrac{1}{\lambda_+})^{N}\tilde{G}^{N}\!\!=\boldsymbol\Psi\,\text{diag}\{1,(-1)^\kappa,0,0,0,0,0,0\}\boldsymbol\Psi^{-1}=\!
\begin{cases}\vec D_+,&\mbox{$N$ even}\\ \vec D_-,&\mbox{$N$ odd}\end{cases}
\eea
with $\lambda_+=\tfrac{\sqrt{3}}{2}(1\!+\!\sqrt{5})$. Combining the even and odd matrices, gives the rank-1 matrix $\vec D\!=\!\vec D_+\!+\!\vec D_-$\\
\begin{align}
\vec D\!&=\!\!\!\sssmat{\tfrac{1}{30}(5\!-\!\sqrt{5})&\tfrac{1}{\sqrt{45}}\\ \tfrac{1}{\sqrt{45}}&\tfrac{1}{30}(5\!+\!\sqrt{5})}\otimes 
\!\!\!\sssmat{1&\sqrt{3}&1&1\\ \sqrt{3}&3&\sqrt{3}&\sqrt{3}\\ 1&\sqrt{3}&1&1\\ 1&\sqrt{3}&1&1\\ }
\end{align}
Using the fact that $\hat{\g}_{1|1}=\tfrac{4}{\sqrt{15}}\sin\tfrac{\pi}{5}\sin\tfrac{\pi}6$ and $\hat{\g}_{1|1}^2=\tfrac{1}{30}(5\!-\!\sqrt{5})$, the 2-boundary $\g$-factors $\hat{\g}_{\mu|\mu'}$ and defect $\g$-factors 
$\hat{d}_\mu$ are obtained by solving 
\bea
\vec D=\hat{\g}_{1|1}\!\!\!\sssmat{\hat{\g}_{1|1}&\hat{\g}_{1|2}&\ldots&\hat{\g}_{1|8}\\
\hat{\g}_{2|1}&\hat{\g}_{2|2}&\ldots&\hat{\g}_{2|8}\\
\vdots&\vdots&\ddots&\vdots\\
\hat{\g}_{8|1}&\hat{\g}_{8|2}&\ldots&\hat{\g}_{8|8}}
=\hat{\g}^2_{1|1}\!\sssmat{1\\ \hat{d}_{2}\\[-4pt] \vdots\\ \hat{d}_{8}}\!\sssmat{1&\hat{d}_{2}&\ldots&\hat{d}_{8}}
=\hat{\g}^2_{1|1} \sum_{\mu=1}^8 \hat{d}_\mu\hat{N}_\mu
\eea

\subsubsection{Tricritical 3-state Potts model $(A_6,D_4)$}

\nobreak

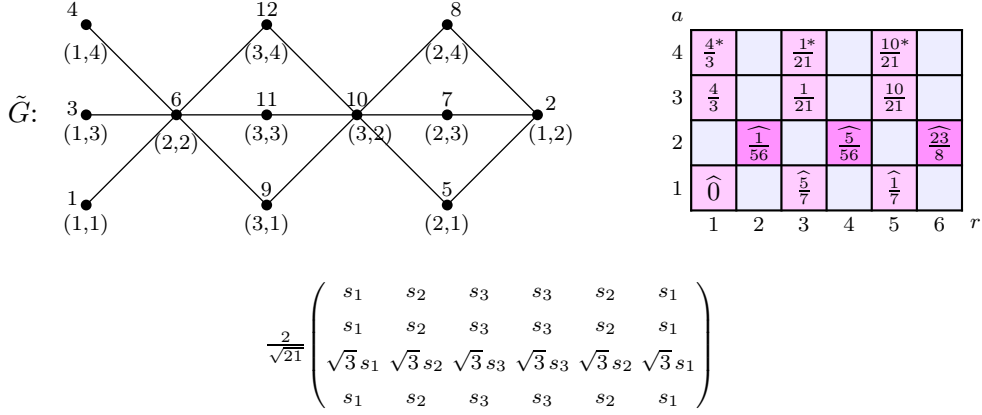
\begin{figure}[h]
\scriptsize
\begin{center}
\qquad\quad\begin{pspicture}(0,0)(6.25,2.5)
\psset{unit=1.2cm}
\psline[linewidth=.5pt](0,0)(1,1)(2,0)(3,1)(4,0)(5,1)
\psline[linewidth=.5pt](0,2)(1,1)(2,2)(3,1)(4,2)(5,1)
\psline[linewidth=.5pt](0,1)(5,1)
\pscircle*(0,0){.06}
\pscircle*(0,1){.06}
\pscircle*(0,2){.06}
\pscircle*(1,1){.06}
\pscircle*(2,0){.06}
\pscircle*(2,1){.06}
\pscircle*(2,2){.06}
\pscircle*(3,1){.06}
\pscircle*(4,0){.06}
\pscircle*(4,1){.06}
\pscircle*(4,2){.06}
\pscircle*(5,1){.06}
\rput(-.7,1.07){\normalsize $\tilde{G}$:}
\rput(0,-.1){\pp{t}{$(1,1)$}}
\rput(0,.9){\pp{t}{$(1,3)$}}
\rput(0,1.8){\pp{t}{$(1,4)$}}
\rput(1,.8){\pp{t}{$(2,2)$}}
\rput(2,-.1){\pp{t}{$(3,1)$}}
\rput(4,-.1){\pp{t}{$(2,1)$}}
\rput(2,.9){\pp{t}{$(3,3)$}}
\rput(4,.9){\pp{t}{$(2,3)$}}
\rput(2,1.8){\pp{t}{$(3,4)$}}
\rput(4,1.8){\pp{t}{$(2,4)$}}
\rput(3.15,.9){\pp{t}{$(3,2)$}}
\rput(5.15,.9){\pp{t}{$(1,2)$}}
\rput(-.15,.15){\pp{t}{$1$}}
\rput(5.15,1.2){\pp{t}{$2$}}
\rput(3.,1.25){\pp{t}{$10$}}
\rput(-.15,1.15){\pp{t}{$3$}}
\rput(-.15,2.25){\pp{t}{$4$}}
\rput(4,.25){\pp{t}{$5$}}
\rput(2,.25){\pp{t}{$9$}}
\rput(1,1.25){\pp{t}{$6$}}
\rput(4,1.25){\pp{t}{$7$}}
\rput(2,1.25){\pp{t}{$11$}}
\rput(4.1,2.25){\pp{t}{$8$}}
\rput(2.,2.25){\pp{t}{$12$}}
\end{pspicture}\qquad\qquad\qquad
\psset{unit=.6cm}
\begin{pspicture}[shift=-2pt](0,0)(6,4.4)
\psframe[linewidth=0pt,fillstyle=solid,fillcolor=lightlightblue](0,0)(6,4)
\multirput(0,0)(0,2){2}{\multirput[bl](0,0)(2,0){3}{\psframe[linewidth=0pt,fillstyle=solid,fillcolor=lightpurple](0,0)(1,1)}}
\multirput(0,3)(0,2){1}{\multirput[bl](0,0)(2,0){3}{\psframe[linewidth=0pt,fillstyle=solid,fillcolor=lightpurple](0,0)(1,1)}}
\multirput(0,0)(2,0){3}{\multirput[bl](1,1)(0,2){1}{\psframe[linewidth=0pt,fillstyle=solid,fillcolor=darkpurple](0,0)(1,1)}}
\psgrid[gridlabels=0pt,subgriddiv=1](0,0)(6,4)
\rput(.5,.5){\small $\widehat 0$}
\rput(1.5,1.5){$\widehat{\small \tfrac{1}{56}}$}
\rput(2.5,2.5){$\small \tfrac{1}{21}$}
\rput(2.5,3.5){$\small \tfrac{1}{21}^{\!\!\!*}$}
\rput(4.5,2.5){$\small \tfrac{10}{21}$}
\rput(4.5,3.5){$\small \tfrac{10}{21}^{\!*}$}
\rput(.5,2.5){$\small \tfrac{4}{3}$}
\rput(.5,3.5){$\small \tfrac{4}{3}^{\!*}$}
\rput(3.5,1.5){$\small \widehat{\tfrac{5}{56}}$}
\rput(5.5,1.5){$\small \widehat{\tfrac{23}{8}}$}
\rput(2.5,.5){$\small \widehat{\tfrac57}$}
\rput(4.5,.5){$\small \widehat{\tfrac{1}7}$}
\rput(.5,-.3){$\small 1$}
\rput(1.5,-.3){$\small 2$}
\rput(2.5,-.3){$\small 3$}
\rput(3.5,-.3){$\small 4$}
\rput(4.5,-.3){$\small 5$}
\rput(5.5,-.3){$\small 6$}
\rput(6.3,-.3){$\small r$}
\rput(-.3,.5){$\small 1$}
\rput(-.3,1.5){$\small 2$}
\rput(-.3,2.5){$\small 3$}
\rput(-.3,3.5){$\small 4$}
\rput(-.3,4.3){$\small a$}
\end{pspicture}
\raisebox{-1.6cm}
{$\frac{2}{\sqrt{21}}$\!\!\sssmat{
s_1&s_2&s_3&s_3&s_2&s_1\\[4pt]
s_1&s_2&s_3&s_3&s_2&s_1\\[4pt]
\sqrt{3}\,s_1&\sqrt{3}\,s_2&\sqrt{3}\,s_3&\sqrt{3}\,s_3&\sqrt{3}\,s_2&\sqrt{3}\,s_1\\[4pt]
s_1&s_2&s_3&s_3&s_2&s_1
}}
\end{center}
\caption{The bipartite coset graph $\tilde{G}=A_6\otimes D_4/\mathbb{Z}_2=T_3\otimes D_4$ and Kac tables of conformal weights and 2-boundary $\g$-factors $\hat{\g}_{(1,1)|(r,a)}$ of the tricritical 3-state Potts model $(A_6,D_4)$ with $c=\tfrac67$ where $T_3$ is the tadpole on 3 nodes and $s_r=\sin \tfrac{r\pi}{7}$. 
The nodes are $(r,a)$ with $r=1,2,3$ and $a=1,2,3,4$. Alternatively, the ordered nodes $(r,a)=(1,1),(1,2),(1,3),(1,4),(2,1),(2,2),(2,3),(2,4),(3,1),(3,2),(3,3),(3,4)$ are labelled by $\mu=1,2,\ldots,12$. 
The label $a=4$ can be replaced with $a=3^*$ reflecting the fact that $a=3$ and $4$ are on the same sublattice and rows 3 and 4 are duplicates. 
The $D_4$ graph arises as the $\mathbb{Z}_2$ orbifold of the $A_5$ diagram. 
Under the $(A_6,A_5)$ Kac table symmetry, $(3,2)\equiv (4,2)$, $(2,1) \equiv (5,1)$, $(2,3) \equiv (5,3)$, $(2,4) \equiv (5,4)$ and $(1,2) \equiv (6,2)$.
The extended conformal weights are 
$\widehat{0}=0+5$, $\small \widehat{\tfrac{1}{56}}=\tfrac{1}{56}+\tfrac{85}{56}$, $\small \widehat{\tfrac5{56}}=\tfrac5{56}+\tfrac{33}{56}$, $\widehat{\tfrac17}=\tfrac17+\tfrac{22}7$, $\widehat{\tfrac38}=\tfrac38+\tfrac{23}8$, $\widehat{\tfrac57}=\tfrac57+\tfrac{12}7$.
\label{Tri3PottsCosetGraph}}
\end{figure}

The coset graph of the tricritical 3-state Potts model~\cite{OBP95,Iino21} is $\tilde{G}=A_6\otimes D_4/\mathbb{Z}=T_3\otimes D_4$ as shown in Figure~\ref{Tri3PottsCosetGraph}.
Explicitly, the nimrep fusion matrices $\tilde{N}_{r,a}=N_r^{(T_3)}\!\otimes\! \hat{N}_a^{(D_4)}$ are
\bea
\mbox{}\hspace{-16pt}\tilde{N}_{1,a}\!=\!\!\!\sssmat{\hat{N}^{(D_4)}_a&0&0\\ 0&\hat{N}^{(D_4)}_a&0\\0&0&\hat{N}^{(D_4)}_a}, 
\tilde{N}_{2,a}\!=\!\!\!\sssmat{0&\hat{N}^{(D_4)}_a&0\\ \hat{N}^{(D_4)}_a&0&\hat{N}^{(D_4)}_a\\ 0&\hat{N}^{(D_4)}_a&\hat{N}^{(D_4)}_a},
\tilde{N}_{3,a}\!=\!\!\!\sssmat{0&0&\hat{N}^{(D_4)}_a\\ 0&\hat{N}^{(D_4)}_a&\hat{N}^{(D_4)}_a\\\hat{N}^{(D_4)}_a&\hat{N}^{(D_4)}_a&\hat{N}^{(D_4)}_a},\ r\le 3;\ a\le 4
\eea
The adjacency matrix (fundamental nimrep) is $\tilde{G}=\tilde{N}_{2,2}$ and the nimreps of $T_3$ and $D_4$ are
\begin{align}
N_r^{(T_3)}=\!\!
\sssmat{1&0&0\\ 0&1&0\\ 0&0&1}, 
\!\!\sssmat{0&1&0\\ 1&0&1\\ 0&1&1},
\!\!\sssmat{0&0&1\\ 0&1&1\\ 1&1&1};\quad
\hat{N}^{(D_4)}_a=
\sssmat{1&0&0&0\\ 0&1&0&0\\ 0&0&1&0\\ 0&0&0&1}, \!\!
\sssmat{0&1&0&0\\ 1&0&1&1\\ 0&1&0&0\\ 0&1&0&0},  \!\!
\sssmat{0&0&1&0\\ 0&1&0&0\\ 0&0&0&1\\ 1&0&0&0},  \!\!
\sssmat{0&0&0&1\\ 0&1&0&0\\ 1&0&0&0\\ 0&0&1&0}
\end{align}
\begin{subequations}
for $r=1,2,3$ and $a=1,2,3,4$. The polynomial ring for this case will be discussed elsewhere.

The unitary matrix $\boldsymbol\Psi$ that diagonalizes $\tilde{G}$ and the coset fusion matrices is
\begin{align}
&\qquad \boldsymbol\Psi={\cal S}\otimes \Psi, \quad {\cal S}={\cal S}^T=
\tfrac{2}{\sqrt{7}}\!\sssmat{
\sin\tfrac{2\pi}7&-\sin\tfrac{3\pi}7&\sin\tfrac{\pi}7\\[3pt] 
-\sin\tfrac{3\pi}7&-\sin\tfrac{\pi}7&\sin\tfrac{2\pi}7\\[3pt] 
\sin\tfrac{\pi}7&\sin\tfrac{2\pi}7&\sin\tfrac{3\pi}7},\quad 
\Psi=\tfrac{1}{\sqrt{3}}\!
\sssmat{
\frac{1}{\sqrt{2}}&\frac{1}{\sqrt{2}}&1&1\\[4pt]
\sqrt{\frac32}&-\sqrt{\frac32}&0&0\\[4pt]
\frac{1}{\sqrt{2}}&\frac{1}{\sqrt{2}}&\omega&\omega^2\\[4pt]
\frac1{\sqrt{2}}&\frac1{\sqrt{2}}&\omega^2&\omega}\\
&\boldsymbol\Psi^{-1}\tilde{G}\,\boldsymbol\Psi=
2\sqrt{3}\;\mbox{diag}\Big\{\!\!-\!\cos\tfrac{2\pi}7,\cos\tfrac{2\pi}7,0,0,\cos\tfrac{3\pi}7,\!-\!\cos\tfrac{3\pi}7,0,0,\cos\tfrac{\pi}7,\!-\!\cos\tfrac{\pi}7,0,0\Big\}
\end{align}
\end{subequations}
where $\omega=\exp(2\pi i/3)$. The 2-boundary $\g$-factors $\hat{\g}_{(1,1)|(r,a)}$ and defect $\g$-factors $\hat{d}_{(r,a)}$ are
\begin{align}
\hat{\g}_{(1,1)|(r,a)}\!=\!\tfrac{2}{\sqrt{21}}\sin\tfrac{\pi}{6}\sin\tfrac{\pi}7\,\tilde{d}_{(r,a)},\quad
\hat{d}_{(r,a)}&\!=\!s_1^{-1}\{s_1,\!\sqrt{3}\,s_1,\!s_1,\!s_1,\!s_2,\!\sqrt{3}\,s_2,\!s_2,\!s_2,\!s_3,\!\sqrt{3}\,s_3,\!s_3,\!s_3\}\qquad
\end{align}

In terms of the asymptotic counting of fusion paths, we find
\bea
\lim_{\substack{N\to\infty\\[1pt] \text{$N\!=\!\kappa$\,mod\,2}}} (\tfrac{1}{\lambda_+})^{N}\tilde{G}^{N}\!\!=\boldsymbol\Psi\,\text{diag}\{0,0,0,0,0,0,0,0,1,(-1)^\kappa,0,0\}\boldsymbol\Psi^{-1}=\!
\begin{cases}\vec D_+,&\mbox{$N$ even}\\ \vec D_-,&\mbox{$N$ odd}\end{cases}
\eea
with $\lambda_+=\cos\tfrac{\pi}7$. Combining the even and odd matrices, gives the rank-1 matrix $\vec D\!=\!\vec D_+\!+\!\vec D_-$
\bea
\vec D\!=\!\hat{\g}_{1|1}^2\!
\sssmat{1\\[1pt] \tfrac{s_2}{s_1}\\[4pt] \tfrac{s_3}{s_1}}\!\!\sssmat{1& \tfrac{s_2}{s_1}& \tfrac{s_3}{s_1}}\,\otimes
\sssmat{1\\ \sqrt{3}\\ 1\\ 1\\}\!\!\sssmat{1&\sqrt{3}&1&1}
\eea
Using the fact that $\hat{\g}_{1|1}=\tfrac{2}{\sqrt{21}}s_1$ and $\hat{\g}_{1|1}^2=\tfrac{4}{21}s_1^2$, the 2-boundary $\g$-factors $\hat{\g}_{\mu|\mu'}$ and defect $\g$-factors 
$\hat{d}_\mu$ are obtained from
\bea
\vec D=\hat{\g}_{1|1}\!\!\!\sssmat{\hat{\g}_{1|1}&\hat{\g}_{1|2}&\ldots&\hat{\g}_{1|12}\\
\hat{\g}_{2|1}&\hat{\g}_{2|2}&\ldots&\hat{\g}_{2|12}\\
\vdots&\vdots&\ddots&\vdots\\
\hat{\g}_{12|1}&\hat{\g}_{12|2}&\ldots&\hat{\g}_{12|12}}
=\hat{\g}^2_{1|1}\!\sssmat{1\\ \hat{d}_{2}\\[-4pt] \vdots\\ \hat{d}_{12}}\!\sssmat{1&\hat{d}_{2}&\ldots&\hat{d}_{12}}
=\hat{\g}^2_{1|1} \sum_{\mu=1}^{12} \hat{d}_\mu\hat{N}_\mu
\eea

It is possible to add the $\mathbb{Z}_2$ diagram automorphism $\sigma$ to the coset graph fusion algebra. In this case, the enlarged graph fusion algebra becomes noncommutative because it contains within it the noncommutative symmetric group $\mathbb{S}_3$ but we do not do this here.

\subsection{Prototypical nonunitary CFTs}
\label{secProtoExamples}

\subsubsection{Critical Lee-Yang model ${\cal M}(2,5)=(A_1,A_4)$}

\begin{figure}[h]
\scriptsize
\begin{center}
\qquad\qquad\begin{pspicture}(0,.3)(1,1.3)
\psset{unit=.9cm}
\psline[linewidth=.5pt](0,0)(0,1)
\pscircle*(0,0){.06}
\pscircle*(0,1){.06}
\pscircle[linewidth=.5pt](0,1.3){.3}
\rput(-2.5,.8){\normalsize $\tilde{G}=T_2$:}
\rput(.4,0){\pp{t}{$(1,1)$}}
\rput(.4,.95){\pp{t}{$(1,2)$}}
\rput(-.2,-.025){\pp{t}{$1$}}
\rput(-.2,.925){\pp{t}{$2$}}
\end{pspicture}\qquad\qquad
\psset{unit=.6cm}
\begin{pspicture}[shift=-1cm](0,0)(1,4)
\psframe[linewidth=0pt,fillstyle=solid,fillcolor=lightlightblue](0,0)(1,4)
\multirput(0,0)(2,0){1}{\multirput[bl](0,0)(0,2){2}{\psframe[linewidth=0pt,fillstyle=solid,fillcolor=lightpurple](0,0)(1,1)}}
\psgrid[gridlabels=0pt,subgriddiv=1](0,0)(1,4)
\rput(.5,.5){\small $0$}
\rput(.5,1.5){${\small -\tfrac{1}{5}}$}
\rput(.5,2.5){$\small -\tfrac{1}{5}$}
\rput(.5,3.5){\small $0$}
\rput(.5,-.3){$\small 1$}
\rput(1.3,-.3){$\small r$}
\rput(-.3,.5){$\small 1$}
\rput(-.3,1.5){$\small 2$}
\rput(-.3,2.5){$\small 3$}
\rput(-.3,3.5){$\small 4$}
\rput(-.3,4.3){$\small s$}
\end{pspicture} 
\qquad\qquad
\raisebox{2pt}{\ssmat{
\sqrt{\tfrac{(5-\sqrt{5})}{10}}\\[4pt]
\sqrt{\tfrac{(5+\sqrt{5})}{10}}\\[4pt]
\sqrt{\tfrac{(5+\sqrt{5})}{10}}\\[4pt]
\sqrt{\tfrac{(5-\sqrt{5})}{10}}
}}
\end{center}
\caption{The non-bipartite coset graph $\tilde{G}=A_1\otimes A_4/\mathbb{Z}_2=T_2$ and Kac tables of conformal weights and 2-boundary $\g$-factors $\g_{(1,1)|(r,s)}$ for the Lee-Yang model ${\cal M}(2,5)$ with $c=-\tfrac{22}5$ and $c_\text{eff}=\tfrac25$. 
$T_2$ is the tadpole on 2 nodes resulting from the $\mathbb{Z}_2$ folding of $A_4$. The nodes $(r,s)=(1,1),(1,2)$ are labelled by $\mu=1,2$. The groundstate is $(r_0,s_0)=(1,2)$.
\label{LeeYangCosetGraph}}
\end{figure}

The coset graph $\tilde{G}$ of the Lee-Yang model~\cite{LeeYang,BLZ97,DPTW,DRTW,BDP2015} ${\cal M}(2,5)$ is $A_1\otimes A_4/\mathbb{Z}_2=T_2$ as shown in Figure~\ref{LeeYangCosetGraph}. 
Explicitly, the nimrep fusion matrices are
\bea
\tilde{N}_1=I=\!\ssmat{1&0\\ 0&1},\quad \tilde{N}_{2}=\tilde{G}=\!\ssmat{0&1\\ 1&1}\label{T2nimreps}
\eea
with $\tilde{N}_2^2=I+\tilde{N}_2$ and quantum dimensions $\tilde{d}_1=1$, $\tilde{d}_2=2\cos\tfrac{\pi}{5}=[2]_x$ where $x=e^{\pi i/5}$. 
The coset graph fusion algebra is realized as the polynomial ring
\bea
\mathbb{Z}[y]/\langle y^2\!-\!y\!-\!1\rangle
\eea
The conformal cylinder partition functions $Z_{s|s'}(q)=Z_{(1,s)|(1,s')}(q)$ are
\bea
Z_{1|1}(q)=\chi_{1,1}^{2,5}(q),\qquad  Z_{1|2}(q)=\chi_{1,2}^{2,5}(q),\qquad  Z_{2|2}(q)=\chi_{1,1}^{2,5}(q)+\chi_{1,2}^{2,5}(q)
\eea
The unitary matrix ${\cal S}$ that diagonalizes $\tilde{G}$ is the modular matrix
\bea
{\cal S}={\cal S}^T\!=\!\tfrac{2}{\sqrt{5}}\!\!\sssmat{\sin\tfrac{\pi}5&\sin\tfrac{2\pi}5\\[4pt] \sin\tfrac{2\pi}5&-\sin\tfrac{\pi}5},\quad 
{\cal S}^{-1} \tilde{N}_2\,{\cal S}=\!\!\!\ssmat{2\cos\tfrac{\pi}{5}&0\\ 0&2\cos\tfrac{2\pi}{5}}=\!\!\!\sssmat{\tfrac12(1\!+\!\sqrt{5})&0\\ 0&\tfrac12(1\!-\!\sqrt{5})},
\quad {\cal S}^2=I
\eea
The 2-boundary $\g$-factors $\tilde{\g}_{s|s'}=\tilde{\g}_s\tilde{\g}_{s'}$ are 
\bea
\tilde{\g}_{1|1}\!=\!{\cal S}_{1,1}\!=\!\sqrt{\tfrac{5-\sqrt{5}}{10}}\!=\!\tfrac{2}{\sqrt{5}}\sin\tfrac{\pi}{5}\!=\!0.525731\ldots,\qquad 
\tilde{\g}_{1|2}\!=\!{\cal S}_{1,2}\!=\!\sqrt{\tfrac{5+\sqrt{5}}{10}}\!=\!\tfrac{2}{\sqrt{5}}\sin\tfrac{2\pi}{5}\!=\!0.850651\ldots
\eea
The defect $\g$-factors $\tilde{d}_s$ are 
\bea
\tilde{d}_1=1,\qquad \tilde{d}_2=\frac{\tilde{\g}_{1|2}}{\tilde{\g}_{1|1}}=2\cos\tfrac{\pi}{5}=1.61803\ldots,\qquad \tilde{d}_2^2=1+ \tilde{d}_2
\eea

In terms of the asymptotics of counting fusion paths, we find the rank-1 matrix
\bea
\vec D=\lim_{N\to\infty} (2\cos\tfrac{\pi}{5})^{-N}\!\!\sssmat{0&1\\ 1&1}^{\!N}\!\!\!={\cal S}\!\!\sssmat{1&0\\ 0&0}{\cal S}^{-1}=\!\!\!\sssmat{\tfrac{1}{10}(5\!-\!\sqrt{5})&\tfrac{1}{\sqrt{5}}\\ \tfrac{1}{\sqrt{5}}&\tfrac{1}{10}(5\!+\!\sqrt{5})}\!
=\!\tilde{\g}_{1|1}\!\!\ssmat{\tilde{\g}_{1|1}&\tilde{\g}_{1|2}\\ \tilde{\g}_{2|1}&\tilde{\g}_{2|2}}
=\tilde{\g}^2_{1|1}\!\ssmat{1&\tilde{d}_2\\ \tilde{d}_2&\tilde{d}_2^2}
\eea
This is just the asymptotics of Fibonacci numbers. The boundary and defect $\g$-factors are simply obtained by solving these equations. 
Notice that
\bea
\vec D=\tilde{\g}^2_{1|1} \sum_{s=1}^2 \tilde{d}_s \tilde{N}_s
\eea

\subsubsection{Critical ${\cal M}(2,7)=(A_1,A_6)$ model}

\begin{figure}[h]
\normalsize\begin{center}
\qquad\qquad\begin{pspicture}(0,.3)(1,2.3)
\psset{unit=.7cm}
\psline[linewidth=.5pt](0,0)(0,2)
\pscircle*(0,0){.06}
\pscircle*(0,1){.06}
\pscircle*(0,2){.06}
\pscircle[linewidth=.5pt](0,2.3){.3}
\rput(-3.,1.2){\normalsize $\tilde{G}=T_3$:}
\rput(.6,0){\pp{t}{(1,1)}}
\rput(.6,1){\pp{t}{(1,2)}}
\rput(.6,1.95){\pp{t}{(1,3)}}
\rput(-.2,0){\pp{t}{1}}
\rput(-.2,1){\pp{t}{2}}
\rput(-.2,1.95){\pp{t}{3}}
\end{pspicture} \qquad\qquad
\small
\psset{unit=.6cm}
\begin{pspicture}[shift=-1.3cm](0,0)(1,6)
\psframe[linewidth=0pt,fillstyle=solid,fillcolor=lightlightblue](0,0)(1,6)
\multirput(0,0)(2,0){1}{\multirput[bl](0,0)(0,2){3}{\psframe[linewidth=0pt,fillstyle=solid,fillcolor=lightpurple](0,0)(1,1)}}
\psgrid[gridlabels=0pt,subgriddiv=1](0,0)(1,6)
\rput(.5,.5){\small $0$}
\rput(.5,1.5){${\small -\tfrac{2}{7}}$}
\rput(.5,2.5){$\small -\tfrac{3}{7}$}
\rput(.5,3.5){$\small -\tfrac{3}{7}$}
\rput(.5,4.5){${\small -\tfrac{2}{7}}$}
\rput(.5,5.5){$\small 0$}
\rput(.5,-.3){$\small 1$}
\rput(1.3,-.3){$\small r$}
\rput(-.3,.5){$\small 1$}
\rput(-.3,1.5){$\small 2$}
\rput(-.3,2.5){$\small 3$}
\rput(-.3,3.5){$\small 4$}
\rput(-.3,4.5){$\small 5$}
\rput(-.3,5.5){$\small 6$}
\rput(-.3,6.3){$\small s$}
\end{pspicture}\qquad\qquad
\raisebox{10pt}{
\ssmat{
\tfrac{2}{\sqrt{7}} \sin\tfrac{\pi}{7}\\[8pt]
\tfrac{2}{\sqrt{7}} \sin\tfrac{2\pi}{7}\\[8pt]
\tfrac{2}{\sqrt{7}} \sin\tfrac{3\pi}{7}\\[8pt]
\tfrac{2}{\sqrt{7}} \sin\tfrac{3\pi}{7}\\[8pt]
\tfrac{2}{\sqrt{7}} \sin\tfrac{2\pi}{7}\\[8pt]
\tfrac{2}{\sqrt{7}} \sin\tfrac{\pi}{7}}
}
\end{center}
\caption{The non-bipartite coset graph $\tilde{G}=A_1\otimes A_6/\mathbb{Z}_2=T_3$ and Kac tables of conformal weights and 2-boundary $\g$-factors $\g_{(1,1)|(r,s)}$ for the ${\cal M}(2,7)$ model with $c=-\tfrac{68}7$ and $c_\text{eff}=\tfrac47$.  $T_3$ is the tadpole on 3 nodes. 
The tadpole results from the $\mathbb{Z}_2$ folding of $A_6$. The nodes $(r,s)=(1,1),(1,2),(1,3)$ are labelled by $s=1,2,3$. The groundstate is $(r_0,s_0)=(1,3)$.
\label{M27CosetGraph}}
\end{figure}

The coset graph $\tilde{G}$ of the ${\cal M}(2,7)$ model is $A_1\otimes A_6/\mathbb{Z}_2=T_3$ as shown in Figure~\ref{M27CosetGraph}. 
Explicitly, the nimrep fusion matrices are
\bea
\tilde{N}_1=I=\!\ssmat{1&0&0\\ 0&1&0\\ 0&0&1},\quad \tilde{N}_{2}=\tilde{G}=\!\ssmat{0&1&0\\ 1&0&1\\ 0&1&1},\quad \tilde{N}_{3}=\!\ssmat{0&0&1\\ 0&1&1\\ 1&1&1}
\eea
with $\tilde{N}_2^2=I+\tilde{N}_3$, $\tilde{N}_2\tilde{N}_3=\tilde{N}_2+\tilde{N}_3$, , $\tilde{N}_3^2=I+\tilde{N}_2+\tilde{N}_3$ and quantum dimensions 
$\tilde{d}_1=1$, $\tilde{d}_2=2\cos\tfrac{\pi}{7}=[2]_x$ , $\tilde{d}_3=1+2\cos\tfrac{2\pi}{7}=[3]_x$ where $x=e^{\pi i/7}$. 
The coset graph fusion algebra is realized as the polynomial ring
\bea
\mathbb{Z}[y]/\langle y^3\!-\!y^2\!-\!2y\!+\!1\rangle
\eea
The conformal cylinder partition functions $Z_{s|s'}(q)=Z_{(1,s)|(1,s')}(q)$ are
\begin{align}
&Z_{1|1}(q)=\chi_{1,1}^{2,7}(q),\quad  Z_{1|2}(q)=\chi_{1,2}^{2,7}(q),\quad  Z_{1|3}(q)=\chi_{1,3}^{2,7}(q),\quad Z_{2|2}(q)=\chi_{1,1}^{2,7}(q)+\chi_{1,3}^{2,7}(q)\nonumber\\ 
&\qquad\qquad Z_{2|3}(q)=\chi_{1,2}^{2,7}(q)+\chi_{1,3}^{2,7}(q),\quad Z_{3|3}(q)=\chi_{1,1}^{2,7}(q)+\chi_{1,2}^{2,7}(q)+\chi_{1,3}^{2,7}(q) 
\end{align}
The unitary matrix ${\cal S}$ that diagonalizes $\tilde{G}$ is the modular matrix
\bea
{\cal S}={\cal S}^T\!=\!\tfrac{2}{\sqrt{7}}\!\!\sssmat{
\sin\tfrac{2\pi}7&-\sin\tfrac{3\pi}7&\sin\tfrac{\pi}7\\[3pt] 
-\sin\tfrac{3\pi}7&-\sin\tfrac{\pi}7&\sin\tfrac{2\pi}7\\[3pt] 
\sin\tfrac{\pi}7&\sin\tfrac{2\pi}7&\sin\tfrac{3\pi}7},\quad 
{\cal S}^{-1} \tilde{N}_2\,{\cal S}=\!\!\!\ssmat{2\cos\tfrac{5\pi}{7}&0&0\\ 0&2\cos\tfrac{3\pi}{7}&0\\ 0&0&2\cos\tfrac{\pi}{7}},
\quad {\cal S}^2=I
\eea
Since $s=3$ is the groundstate, the 2-boundary $\g$-factors $\tilde{\g}_{s|s'}=\tilde{\g}_s\tilde{\g}_{s'}$ are 
\bea
\tilde{\g}_{3|s}\!={\cal S}_{3,s}\!=\!\tfrac{2}{\sqrt{7}}\sin\tfrac{s\pi}{7}=0.327985.., 0.591009.., 0.736976..
\eea
The defect $\g$-factors $\tilde{d}_s$ 
\bea
\tilde{d}_1=1,\quad 
\tilde{d}_2=\frac{\tilde{\g}_{1|2}}{\tilde{\g}_{1|1}}=2\cos\tfrac{\pi}{7}=1.80194..,
\quad \tilde{d}_3=\frac{\tilde{\g}_{1|3}}{\tilde{\g}_{1|1}}=1\!+\!2\cos\tfrac{2\pi}{7}=2.24698..
\eea
give a 1-dimensional representation of the coset fusion algebra corresponding to the largest eigenvalues of the fusion matrices. 

In terms of the asymptotic counting of fusion paths, we find the rank-1 matrix
\bea
\vec D=\lim_{N\to\infty}\!(2\cos\tfrac{\pi}{7})^{-N}\!\!\sssmat{0&1&0\\ 1&0&1\\ 0&1&1}^{\!N}\!\!\!={\cal S}\!\!\sssmat{1&0&0\\ 0&0&0\\ 0&0&0}{\cal S}^{-1}\!
=\tilde{\g}_{1|1}\!\!\ssmat{\tilde{\g}_{1|1}&\tilde{\g}_{1|2}&\tilde{\g}_{1|3}\\ 
\tilde{\g}_{2|1}&\tilde{\g}_{2|2}&\tilde{\g}_{2|3}\\
\tilde{\g}_{3|1}&\tilde{\g}_{3|2}&\tilde{\g}_{3|3}}\!
\!=\tilde{\g}^2_{1|1}\!\!\sssmat{1\\ \tilde{d}_{2}\\ \tilde{d}_{3}}\!\!\sssmat{1&\tilde{d}_{2}&\tilde{d}_{3}}
\eea
Notice that
\bea
\vec D=\tilde{\g}^2_{1|1}\sum_{s=1}^3 \tilde{d}_s \tilde{N}_s
\eea

\subsubsection{Critical ${\cal M}(3,5)=(A_2,A_4)$ model}

\begin{figure}[h]
\normalsize\begin{center}
\qquad\qquad
\begin{pspicture}[shift=-.7cm](0,0)(.5,2)
\psset{unit=.7cm}
\psline[linewidth=.5pt](0,0)(0,3)
\pscircle*(0,0){.06}
\pscircle*(0,1){.06}
\pscircle*(0,2){.06}
\pscircle*(0,3){.06}
\rput(-2.5,1.45){\normalsize $\tilde{G}$:}
\rput(.6,0){\pp{t}{(1,1)}}
\rput(.6,1){\pp{t}{(1,2)}}
\rput(.6,2){\pp{t}{(1,3)}}
\rput(.6,3){\pp{t}{(1,4)}}
\rput(-.2,0){\pp{t}{1}}
\rput(-.2,1){\pp{t}{2}}
\rput(-.2,2){\pp{t}{3}}
\rput(-.2,3){\pp{t}{4}}
\end{pspicture}\qquad\qquad\quad
\psset{unit=.65cm}
\small
\begin{pspicture}[shift=-1cm](0,0)(2,4)
\psframe[linewidth=0pt,fillstyle=solid,fillcolor=lightlightblue](0,0)(2,4)
\multirput(0,0)(1,1){2}{\multirput[bl](0,0)(0,2){2}{\psframe[linewidth=0pt,fillstyle=solid,fillcolor=lightpurple](0,0)(1,1)}}
\psgrid[gridlabels=0pt,subgriddiv=1](0,0)(2,4)
\rput(.5,.5){\small $0$}
\rput(.5,1.5){${\small -\!\tfrac{1}{20}}$}
\rput(.5,2.5){$\small \tfrac{1}{5}$}
\rput(.5,3.5){\small $\tfrac34$}
\rput(1.5,.5){\small $\tfrac34$}
\rput(1.5,1.5){$\small \tfrac{1}{5}$}
\rput(1.5,2.5){${\small -\!\tfrac{1}{20}}$}
\rput(1.5,3.5){\small $0$}
\rput(.5,-.3){$\small 1$}
\rput(1.5,-.3){$\small 2$}
\rput(2.3,-.3){$\small r$}
\rput(-.3,.5){$\small 1$}
\rput(-.3,1.5){$\small 2$}
\rput(-.3,2.5){$\small 3$}
\rput(-.3,3.5){$\small 4$}
\rput(-.3,4.3){$\small s$}
\end{pspicture}\qquad\qquad
\raisebox{6pt}{
\ssmat{
\sqrt{\tfrac{5-\sqrt{5}}{20}}&\sqrt{\tfrac{5-\sqrt{5}}{20}}\\[6pt]
\sqrt{\tfrac{5+\sqrt{5}}{20}}&\sqrt{\tfrac{5+\sqrt{5}}{20}}\\[6pt]
\sqrt{\tfrac{5+\sqrt{5}}{20}}&\sqrt{\tfrac{5+\sqrt{5}}{20}}\\[6pt]
\sqrt{\tfrac{5-\sqrt{5}}{20}}&\sqrt{\tfrac{5-\sqrt{5}}{20}}
}}
\end{center}
\caption{The bipartite coset graph $\tilde{G}=A_4$ and Kac tables of conformal weights and 2-boundary $\tilde{\g}$-factors $\g_{(1,1)|(r,s)}$ 
for the ${\cal M}(3,5)$ model with $c=-\tfrac35$ and $c_\text{eff}=\tfrac35$.  
The nodes $(r,s)=(1,1),(1,2),(1,3),(1,4)$ are labelled by $s=1,2,3,4$. The groundstate is $(r_0,s_0)=(1,2)$.
\label{M35CosetGraph}}
\end{figure}
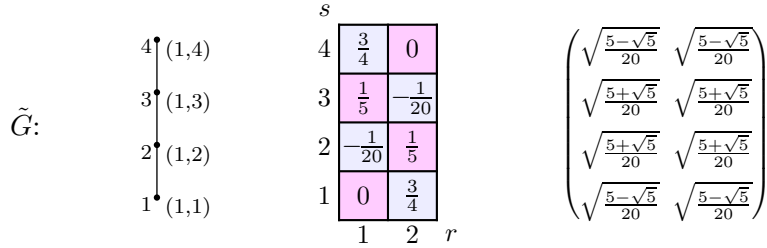

The coset graph $\tilde{G}$ of the ${\cal M}(3,5)$ model is $A_4$ as shown in Figure~\ref{M35CosetGraph}. 
Explicitly, the nimrep fusion matrices are
\bea
\tilde{N}_1\!=\!I,\quad
\tilde{N}_{2}\!=\!\!\!\!\sssmat{0&1&0&0\\ 1&0&1&0\\ 0&1&0&1\\ 0&0&1&0},\quad 
\tilde{N}_{3}\!=\!\!\!\!\sssmat{0&0&1&0\\ 0&1&0&1\\ 1&0&1&0\\ 0&1&0&0},\quad
\tilde{N}_4\!=\!\!\!\!\sssmat{0&0&0&1\\ 0&0&1&0\\ 0&1&0&0\\ 1&0&0&0}
\eea
with $\tilde{N}_2^2=I\!+\!\tilde{N}_3$, $\tilde{N}_2\tilde{N}_3=\tilde{N}_2\!+\!\tilde{N}_4$, $\tilde{N}_2\tilde{N}_4=\tilde{N}_3$, $\tilde{N}_3^2=\tilde{N}_1\!+\!\tilde{N}_3$, $\tilde{N}_3\tilde{N}_4=\tilde{N}_2$, $\tilde{N}_4^2=I$. The fusion matrices, along with the quantum dimensions $\tilde{d}_{(r,s)}=[r]_x[s]_y$ for $(r,s)\in\mathbb{K}$, $x=e^{\pi i/3}$ and $y=e^{\pi i/5}$, satisfy the coset graph fusion algebra. 
The coset graph fusion algebra is realized as the polynomial ring
\bea
\mathbb{Z}[x,y]/\langle x^2\!-\!1,y^2\!-\!xy\!-\!1\rangle
\eea
The conformal cylinder partition functions $Z_{s|s'}(q)=Z_{s'|s}(q)=Z_{(1,s)|(1,s')}(q)$ are
\begin{align}
&Z_{1|s}(q)=\chi_{1,s}^{3,5}(q),\quad  Z_{2|2}(q)=\chi_{1,1}^{3,5}(q)+\chi_{1,3}^{3,5}(q),\quad  Z_{2|3}(q)=\chi_{1,2}^{3,5}(q)+\chi_{1,4}^{3,5}(q),\nonumber\\ 
&Z_{2|4}(q)=\chi_{1,3}^{3,5}(q),\quad Z_{3|3}(q)=\chi_{1,1}^{3,5}(q)+\chi_{1,3}^{3,5}(q),\quad Z_{3|4}(q)=\chi_{1,2}^{3,5}(q),\quad Z_{4|4}(q)=\chi_{1,1}^{3,5}(q)
\end{align}
The unitary matrix ${\cal S}$ that diagonalizes $\tilde{G}$ is the modular matrix
\bea
{\cal S}\!=\!{\cal S}^T\!\!=\!\sqrt{\tfrac25}\!
\sssmat{
\sin\tfrac{2\pi}5&\sin\tfrac{\pi}5&-\sin\tfrac{\pi}5&&-\sin\tfrac{2\pi}5\\[3pt] 
\sin\tfrac{\pi}5&\sin\tfrac{2\pi}5&\sin\tfrac{2\pi}5&&\sin\tfrac{\pi}5\\[3pt] 
-\sin\tfrac{\pi}5&\sin\tfrac{2\pi}5&-\sin\tfrac{2\pi}5&&\sin\tfrac{\pi}5\\[3pt]
-\sin\tfrac{2\pi}5&\sin\tfrac{\pi}5&\sin\tfrac{\pi}5&&-\sin\tfrac{2\pi}5},\quad
{\cal S}^{-1} \tilde{N}_2\,{\cal S}=\!\!\!
\ssmat{2\cos\tfrac{2\pi}{5}&0&0&0\\ 0&2\cos\tfrac{\pi}{5}&0&0\\ 0&0&-2\cos\tfrac{\pi}{5}&0\\ 0&0&0&-2\cos\tfrac{2\pi}{5}}
\eea
with ${\cal S}^2=I$. 
Since $s=2$ is the groundstate, the 2-boundary $\g$-factors $\tilde{\g}_{s|s'}=\tilde{\g}_s\tilde{\g}_{s'}$ are 
\bea
\tilde{\g}_{1|s}\!={\cal S}_{2,s}\!=\!\sqrt{\tfrac25}\sin\tfrac{s\pi}{5}=0.371748.., 0.601501.., 0.601501..,0.371748..
\eea
The defect $\g$-factors $\tilde{d}_\mu$ giving a 1-dimensional representation of the coset graph fusion algebra are 
\bea
\tilde{d}_1=\tilde{d}_4=1,\quad 
\tilde{d}_2=\tilde{d}_3=\frac{\tilde{\g}_{1|2}}{\tilde{\g}_{1|1}}=2\cos\tfrac{\pi}{5}=1.61803..,
\eea

In terms of the asymptotics of counting fusion paths, we find
\bea
\lim_{\substack{N\to\infty\\[1pt] \text{$N\!=\!\kappa$\,mod\,2}}} (2\cos\tfrac{\pi}{5})^{-N}\!\!\sssmat{0&1&0&0\\ 1&0&1&0\\ 0&1&0&1\\ 0&0&1&0}^{\!N}\!\!\!={\cal S}\!\!\sssmat{0&0&0&0\\ 0&1&0&0\\ 0&0&(-1)^\kappa&0\\ 0&0&0&0}{\cal S}^{-1}=\!
\begin{cases}\vec D_+,&\mbox{$N$ even}\\ \vec D_-,&\mbox{$N$ odd}\end{cases}
\eea
with
\bea 
\!\!\!\!\!\!\vec D_+\!=\!\!\!\sssmat{\tfrac{5-\sqrt{5}}{10}\!\!&0&\tfrac{1}{\sqrt{5}}&0\\ 0&\tfrac{5-\sqrt{5}}{10}\!\!&0&\tfrac{1}{\sqrt{5}}\\ \tfrac{1}{\sqrt{5}}&0&\tfrac{5+\sqrt{5}}{10}\!\!&0\\ 0&\tfrac{1}{\sqrt{5}}&0&\tfrac{5+\sqrt{5}}{10}}
\!=\!\tfrac{5-\sqrt{5}}{10}\tilde{N}_1\!+\!\tfrac{1}{\sqrt{5}}\tilde{N}_3,\quad
\vec D_-\!=\!\!\!\sssmat{0&\tfrac{1}{\sqrt{5}}&0&\tfrac{5-\sqrt{5}}{10}\!\!\\ \tfrac{1}{\sqrt{5}}&0&\tfrac{5+\sqrt{5}}{10}\!\!&0\\ 0&\tfrac{5+\sqrt{5}}{10}\!\!&0&\tfrac{1}{\sqrt{5}}\\ \tfrac{5-\sqrt{5}}{10}\!\!&0&\tfrac{1}{\sqrt{5}}&0}
\!=\!\tfrac{5-\sqrt{5}}{10}\tilde{N}_4\!+\!\tfrac{1}{\sqrt{5}}\tilde{N}_2\hfill
\eea
Combining the even and odd matrices, gives the rank-1 matrix
\bea
\vec D=\tfrac12(\vec D_+\!+\!\vec D_-)
=\tilde{\g}_{1|1}\!\!\ssmat{
\tilde{\g}_{1|1}&\tilde{\g}_{1|2}&\tilde{\g}_{1|3}&\tilde{\g}_{1|4}\\ 
\tilde{\g}_{2|1}&\tilde{\g}_{2|2}&\tilde{\g}_{2|3}&\tilde{\g}_{2|4}\\
\tilde{\g}_{3|1}&\tilde{\g}_{3|2}&\tilde{\g}_{3|3}&\tilde{\g}_{3|4}\\
\tilde{\g}_{4|1}&\tilde{\g}_{4|2}&\tilde{\g}_{4|3}&\tilde{\g}_{4|4}}
=\tilde{\g}^2_{1|1}\!\!\sssmat{1\\ \tilde{d}_{2}\\ \tilde{d}_{3}\\  \tilde{d}_{4}}\!\!\sssmat{1&\tilde{d}_{2}&\tilde{d}_{3}& \tilde{d}_{4}}
\eea
The boundary and defect $\g$-factors are simply obtained by solving these equations. Notice that
\bea
\vec D=\tilde{\g}^2_{1|1}\sum_{s=1}^4 \tilde{d}_s \tilde{N}_s
\eea

\subsubsection{Critical ${\cal M}(5,7)=(A_4,A_6)$ model}

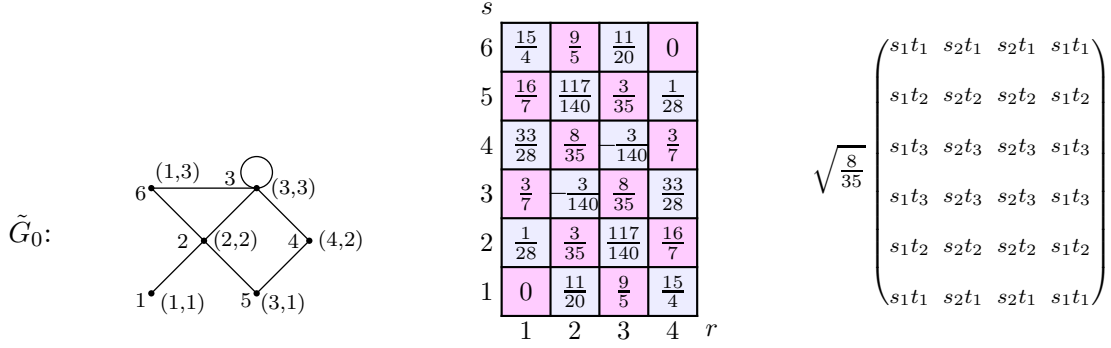
\begin{figure}[h]
\normalsize\begin{center}
\qquad\qquad
\begin{pspicture}[shift=-.7cm](0,0)(3.5,2)
\psset{unit=.7cm}
\psline[linewidth=.5pt](0,0)(2,2)(3,1)(2,0)(0,2)(2,2)
\pscircle*(0,0){.06}
\pscircle*(1,1){.06}
\pscircle*(2,2){.06}
\pscircle*(3,1){.06}
\pscircle*(2,0){.06}
\pscircle*(0,2){.06}
\pscircle[linewidth=.5pt](2,2.3){.3}
\rput(-2.3,1.2){\normalsize $\tilde{G}_0$:}
\rput(.6,0){\pp{t}{(1,1)}}
\rput(1.6,1){\pp{c}{(2,2)}}
\rput(2.7,2){\pp{c}{(3,3)}}
\rput(.5,2.3){\pp{c}{(1,3)}}
\rput(3.6,1.0){\pp{c}{(4,2)}}
\rput(2.5,0){\pp{t}{(3,1)}}
\rput(-.2,0){\pp{t}{1}}
\rput(.6,1){\pp{c}{2}}
\rput(1.5,2.2){\pp{c}{3}}
\rput(2.7,1.0){\pp{c}{4}}
\rput(1.8,0){\pp{t}{5}}
\rput(-.2,1.9){\pp{c}{6}}
\end{pspicture}\qquad\quad
\psset{unit=.65cm}
\small
\begin{pspicture}[shift=-1cm](0,0)(4,6)
\psframe[linewidth=0pt,fillstyle=solid,fillcolor=lightlightblue](0,0)(4,6)
\multirput(0,0)(1,1){2}{\multirput[bl](0,0)(0,2){3}{\psframe[linewidth=0pt,fillstyle=solid,fillcolor=lightpurple](0,0)(1,1)}}
\multirput(2,0)(1,1){2}{\multirput[bl](0,0)(0,2){3}{\psframe[linewidth=0pt,fillstyle=solid,fillcolor=lightpurple](0,0)(1,1)}}
\psgrid[gridlabels=0pt,subgriddiv=1](0,0)(4,6)
\rput(.5,.5){\small $0$}
\rput(.5,1.5){${\small \tfrac{1}{28}}$}
\rput(.5,2.5){$\small \tfrac{3}{7}$}
\rput(.5,3.5){\small $\tfrac{33}{28}$}
\rput(.5,4.5){$\small \tfrac{16}{7}$}
\rput(.5,5.5){\small $\tfrac{15}{4}$}
\rput(1.5,.5){\small $\tfrac{11}{20}$}
\rput(1.5,1.5){$\small \tfrac{3}{35}$}
\rput(1.5,2.5){${\small -\!\tfrac{3}{140}}$}
\rput(1.5,3.5){\small $\tfrac{8}{35}$}
\rput(1.5,4.5){$\small \tfrac{117}{140}$}
\rput(1.5,5.5){${\small \tfrac{9}{5}}$}
\rput(2.5,5.5){\small $\tfrac{11}{20}$}
\rput(2.5,4.5){$\small \tfrac{3}{35}$}
\rput(2.5,3.5){${\small -\!\tfrac{3}{140}}$}
\rput(2.5,2.5){\small $\tfrac{8}{35}$}
\rput(2.5,1.5){$\small \tfrac{117}{140}$}
\rput(2.5,.5){${\small \tfrac{9}{5}}$}
\rput(3.5,5.5){\small $0$}
\rput(3.5,4.5){${\small \tfrac{1}{28}}$}
\rput(3.5,3.5){$\small \tfrac{3}{7}$}
\rput(3.5,2.5){\small $\tfrac{33}{28}$}
\rput(3.5,1.5){$\small \tfrac{16}{7}$}
\rput(3.5,.5){\small $\tfrac{15}{4}$}
\rput(.5,-.3){$\small 1$}
\rput(1.5,-.3){$\small 2$}
\rput(2.5,-.3){$\small 3$}
\rput(3.5,-.3){$\small 4$}
\rput(4.3,-.3){$\small r$}
\rput(-.3,.5){$\small 1$}
\rput(-.3,1.5){$\small 2$}
\rput(-.3,2.5){$\small 3$}
\rput(-.3,3.5){$\small 4$}
\rput(-.3,4.5){$\small 5$}
\rput(-.3,5.5){$\small 6$}
\rput(-.3,6.3){$\small s$}
\end{pspicture}\qquad\qquad
\raisebox{24pt}{
$\sqrt{\tfrac{8}{35}}$\!\!\!
\ssmat{
s_1t_1&s_2t_1&s_2t_1&s_1t_1\\[10pt]
s_1t_2&s_2t_2&s_2t_2&s_1t_2\\[10pt]
s_1t_3&s_2t_3&s_2t_3&s_1t_3\\[10pt]
s_1t_3&s_2t_3&s_2t_3&s_1t_3\\[10pt]
s_1t_2&s_2t_2&s_2t_2&s_1t_2\\[10pt]
s_1t_1&s_2t_1&s_2t_1&s_1t_1
}}
\end{center}
\caption{The non-bipartite coset graph is $\tilde{G}=\tilde{G}_0\oplus\tilde{G}_0$ where $\tilde{G}_0$ is shown. Also shown are the Kac tables of conformal weights and 2-boundary $\tilde{\g}$-factors $\g_{(2,3)|(r,s)}$ 
for the ${\cal M}(5,7)$ model with $c=\tfrac{11}{35}$ and $c_\text{eff}=\tfrac{29}{35}$. Restricting to the lower-half of the Kac table,  $s_r=\sin\tfrac{r\pi}{5}$, $r=1,2,3,4$ and $t_s=\sin\tfrac{s\pi}{7}$, $s=1,2,3$. 
The nodes \mbox{$(r,s)=(1,1),(2,2),(3,3),(4,2),(3,1),(1,3),(4,1),(3,2),(2,3),(1,2),(2,1),(4,3)\in\mathbb{K}$} are labelled by $\mu=1,2,\ldots,12$. 
The groundstate $\mu=9$ is $(r_0,s_0)=(2,3)$.
\label{M57CosetGraph}}
\end{figure}

The coset graph of the ${\cal M}(5,7)$ model is $\tilde{G}=\tilde{G}_0\oplus\tilde{G}_0$ where $\tilde{G}_0$ is shown in Figure~\ref{M57CosetGraph}. 
Explicitly, the nimrep fusion matrices are
\bea
\tilde{\cal N}_\mu=\begin{cases}
\sssmat{\tilde{N}_\mu&0\\ 0&\tilde{N}_\mu},&\mu=1,2,\ldots,6\ \  \mbox{($r+s$ even)}\\[6pt]
\sssmat{0&\tilde{N}_\mu\\ \tilde{N}_\mu&0},&\mu=7,8,\ldots,12\ \  \mbox{($r+s$ odd)}
\end{cases}
\eea
where, with the basis ordering of $\mathbb{K}$ as in Figure~\ref{M57CosetGraph}, 
\bea
\tilde{N}_\mu\big|_{\mu=1,2,\ldots,6}\!=\!\!\!
\sssmat{1 & 0 & 0 & 0 & 0 & 0 \\
 0 & 1 & 0 & 0 & 0 & 0 \\
 0 & 0 & 1 & 0 & 0 & 0 \\
 0 & 0 & 0 & 1 & 0 & 0 \\
 0 & 0 & 0 & 0 & 1 & 0 \\
 0 & 0 & 0 & 0 & 0 & 1},
\sssmat{0 & 1 & 0 & 0 & 0 & 0 \\
 1 & 0 & 1 & 0 & 1 & 1 \\
 0 & 1 & 1 & 1 & 0 & 1 \\
 0 & 0 & 1 & 0 & 1 & 0 \\
 0 & 1 & 0 & 1 & 0 & 0 \\
 0 & 1 & 1 & 0 & 0 & 0 },
\sssmat{0 & 0 & 1 & 0 & 0 & 0 \\
 0 & 1 & 1 & 1 & 0 & 1 \\
 1 & 1 & 1 & 1 & 1 & 1 \\
 0 & 1 & 1 & 0 & 0 & 0 \\
 0 & 0 & 1 & 0 & 0 & 1 \\
 0 & 1 & 1 & 0 & 1 & 0},
\sssmat{ 0 & 0 & 0 & 1 & 0 & 0 \\
 0 & 0 & 1 & 0 & 1 & 0 \\
 0 & 1 & 1 & 0 & 0 & 0 \\
 1 & 0 & 0 & 0 & 0 & 1 \\
 0 & 1 & 0 & 0 & 0 & 0 \\
 0 & 0 & 0 & 1 & 0 & 1 },
\sssmat{0 & 0 & 0 & 0 & 1 & 0 \\
 0 & 1 & 0 & 1 & 0 & 0 \\
 0 & 0 & 1 & 0 & 0 & 1 \\
 0 & 1 & 0 & 0 & 0 & 0 \\
 1 & 0 & 0 & 0 & 1 & 0 \\
 0 & 0 & 1 & 0 & 0 & 0},
\sssmat{0 & 0 & 0 & 0 & 0 & 1 \\
 0 & 1 & 1 & 0 & 0 & 0 \\
 0 & 1 & 1 & 0 & 1 & 0 \\
 0 & 0 & 0 & 1 & 0 & 1 \\
 0 & 0 & 1 & 0 & 0 & 0 \\
 1 & 0 & 0 & 1 & 0 & 1}
\eea
The fusion matrices, along with the quantum dimensions $\tilde{d}_{(r,s)}=[r]_x[s]_y$ for $(r,s)\in\mathbb{K}$, $x=e^{\pi i/5}$ and $y=e^{\pi i/7}$, satisfy the coset graph fusion algebra. 
The coset graph fusion algebra is realized as the polynomial ring
\bea
\mathbb{Z}[x,y]/\langle x^4\!-\!3x^2\!+\!1,y^3\!-\!x^3y^2\!+\!2xy^2\!-\!2y\!+\!x^3\!-\!2x\rangle
\eea
The ${\cal M}(5,7)$ modular matrix that diagonalizes the fusion rules is
\begin{subequations}
\begin{align}
&\qquad\qquad{\cal S}\!=\!{\cal S}^T\!=\!\!\sssmat{-S_0&S_0\\ S_0&S_0},\qquad 
S_0\!=\!\sqrt{\tfrac8{35}}\!
\sssmat{-s_2 t_2 & -s_1 t_3 & s_1 t_1 & s_2 t_3 & s_1 t_2 & -s_2 t_1 \\
 -s_1 t_3 & -s_2 t_1 & s_2 t_2 & -s_1 t_1 & -s_2 t_3 & s_1 t_2 \\
 s_1 t_1 & s_2 t_2 & s_2 t_3 & s_1 t_2 & s_2 t_1 & s_1 t_3 \\
 s_2 t_3 & -s_1 t_1 & s_1 t_2 & s_2 t_1 & -s_1 t_3 & -s_2 t_2 \\
 s_1 t_2 & -s_2 t_3 & s_2 t_1 & -s_1 t_3 & s_2 t_2 & s_1 t_1 \\
 -s_2 t_1 & s_1 t_2 & s_1 t_3 & -s_2 t_2 & s_1 t_1 & -s_2 t_3
}\\
&\ \ \sssmat{-S_0&S_0\\S_0&S_0}\!\!\sssmat{\tilde{N}_2&0\\ 0&\tilde{N}_2}\!\!\sssmat{-S_0&S_0\\S_0&S_0}\!=\!\!\!\sssmat{2S_0\tilde{N}_2S_0&0\\0&2S_0\tilde{N}_2S_0},\qquad
2S_0 \tilde{N}_2\,S_0=\mbox{diag}\{\lambda_\mu\}\big|_{\mu=1,2,\ldots,6}\\[6pt]
&s_r=\sin\tfrac{r\pi}{5},\ \ \  t_s=\sin\tfrac{s\pi}{7},\ \ \  S_0^2=\tfrac12 I,\ \ \  {\cal S}^2=I,\ \ \  \lambda_\mu=\lambda_{\mu+6}=\lambda_{r,s}=4\cos\tfrac{2\pi r}{5}\cos\tfrac{2\pi s}{7}
\end{align}
\end{subequations}
Since the groundstate is $\mu=9$ corresponding to $(r_0,s_0)=(2,3)$, the 2-boundary $\g$-factors $\tilde{\g}_{\mu|\mu'}=\tilde{\g}_\mu\tilde{\g}_{\mu'}$ are 
\bea
\tilde{\g}_{\mu|\mu'}\!=\!\frac{\tilde{\g}_{1|\mu}\tilde{\g}_{1|\mu'}}{\tilde{\g}_{1|1}},\ \  \mu,\mu'\!=\!1,2,\ldots,12;\qquad
\tilde{\g}_{1|\mu}\!=\!{\cal S}_{9,\mu}\!=\!\sqrt{\tfrac8{35}}\sin\tfrac{r\pi}{5}\sin\tfrac{s\pi}{7},\ \  (r,s)\in\mathbb{K}
\eea
The defect $\g$-factors $\tilde{d}_\mu$ giving a 1-dimensional representation of the coset graph fusion algebra are 
\bea
\tilde{d}_\mu=\frac{\tilde{\g}_{1|\mu}}{\tilde{\g}_{1|1}}=\!\tfrac{\sin\tfrac{r\pi}{5}\sin\tfrac{s\pi}{7}}{\sin\tfrac{\pi}{5}\sin\tfrac{\pi}{7}},\qquad \mu=1,2,\ldots,12
\eea

In terms of the asymptotics of counting fusion paths, we find the rank-1 matrix
\bea
\vec D\!=
\!\lim_{N\to\infty} \lambda_{2,3}^{-N}\tilde{N}_2^N\!\!\!=S_0\,\mbox{diag}\{0,0,1,0,0,0\}S_0^{-1}
=\tilde{\g}_{1|1}\!\!\ssmat{\tilde{\g}_{1|1}&\tilde{\g}_{1|2}&\cdots&\tilde{\g}_{1|6}\\ 
\tilde{\g}_{2|1}&\tilde{\g}_{2|2}&\cdots&\tilde{\g}_{2|6}\\
\vdots&\vdots&\ddots&\vdots\\
\tilde{\g}_{6|1}&\tilde{\g}_{6|2}&\cdots&\tilde{\g}_{6|6}}\!
=\tilde{\g}^2_{1|1}\!\!\sssmat{1\\ \tilde{d}_{2}\\ \vdots\\ \tilde{d}_{6}}\!\!\sssmat{1&\tilde{d}_{2}&\cdots&\tilde{d}_{6}}\label{Dmat}
\eea
The boundary and defect $\g$-factors are simply obtained by solving these equations. Since the fusion matrices $\tilde{N}_\mu$ form a basis for the matrix powers of $\tilde{N}_2$ and $\tilde{N}_{\mu 1}{}^{\mu'}=\delta_{\mu,\mu'}$, reading off the first row of (\ref{Dmat}) gives
\bea
\vec D=\tilde{\g}^2_{1|1}\!\!\sssmat{1\\ \tilde{d}_{2}\\ \vdots\\ \tilde{d}_{6}}\!\!\sssmat{1&\tilde{d}_{2}&\cdots&\tilde{d}_{6}}=\tilde{\g}^2_{1|1}\sum_{\mu=1}^6 \tilde{d}_\mu \tilde{N}_\mu
\eea
consistent with the fact that the quantum dimensions $\tilde{d}_\mu$ satisfy the coset graph fusion algebra. 

\subsubsection{Critical $(A_4,E_6)$ model}

\begin{figure}[htb]
\scriptsize
\begin{center}
\qquad\quad
\begin{pspicture}[shift=-.4](0,0)(4.75,3.6)
\psset{unit=.9cm}
\psline[linewidth=.5pt](0,0)(1,1)(2,2)(1,3)(0,2)(1,1)(2,0)
\psline[linewidth=.5pt](4,0)(3,1)(2,2)(3,3)(4,2)(3,1)(2,0)
\psline[linewidth=.5pt](2,3)(2,0)
\pscircle*(0,2){.06}
\pscircle*(0,0){.06}
\pscircle*(1,3){.06}
\pscircle*(1,1){.06}
\pscircle*(2,3){.06}
\pscircle*(2,2){.06}
\pscircle*(2,1){.06}
\pscircle*(2,0){.06}
\pscircle*(3,3){.06}
\pscircle*(3,1){.06}
\pscircle*(4,2){.06}
\pscircle*(4,0){.06}
\rput(-1.2,1.6){\normalsize $\tilde{G}$:}
\rput(.6,3.2){\pp{t}{$(1,2)$}}
\rput(2,-.15){\pp{t}{$(1,3)$}}
\rput(3.4,3.2){\pp{t}{$(1,4)$}}
\rput(-.4,0){\pp{t}{$(1,1)$}}
\rput(-.4,2){\pp{t}{$(2,1)$}}
\rput(4.4,0){\pp{t}{$(1,5)$}}
\rput(4.4,2){\pp{t}{$(2,5)$}}
\rput(2.4,1.15){\pp{t}{$(2,6)$}}
\rput(3.45,1.15){\pp{t}{$(2,4)$}}
\rput(2.45,2.15){\pp{t}{$(2,3)$}}
\rput(.55,1.15){\pp{t}{$(2,2)$}}
\rput(2,3.4){\pp{t}{$(1,6)$}}
\end{pspicture}\qquad\qquad
\psset{unit=.6cm}
\begin{pspicture}[shift=-2pt](0,0)(5,4)
\psframe[linewidth=0pt,fillstyle=solid,fillcolor=lightlightblue](0,0)(5,4)
\multirput(0,0)(0,2){2}{\multirput[bl](0,0)(2,0){3}{\psframe[linewidth=0pt,fillstyle=solid,fillcolor=lightpurple](0,0)(1,1)}}
\multirput(0,0)(1,0){3}{\multirput[bl](1,1)(0,2){1}{\psframe[linewidth=0pt,fillstyle=solid,fillcolor=darkpurple](0,0)(1,1)}}
\multirput(0,2)(1,0){3}{\multirput[bl](1,1)(0,2){1}{\psframe[linewidth=0pt,fillstyle=solid,fillcolor=darkpurple](0,0)(1,1)}}
\psgrid[gridlabels=0pt,subgriddiv=1](0,0)(5,4)
\rput(.5,.5){\small $\widehat 0$}
\rput(1.5,3.5){$\small -\!\widehat{\tfrac{3}{16}}$}
\rput(2.5,.5){$\small -\!\widehat{\tfrac{1}{6}}$}
\rput(3.5,3.5){$\small \widehat{\tfrac{1}{16}}$}
\rput(4.5,.5){$\small \widehat{\tfrac{1}{2}}$}
\rput(2.5,3.5){$\small \widehat{\tfrac{\,1'}{16}}$}
\rput(1.5,1.5){$\small -\!\widehat{\tfrac{13}{240}}$}
\rput(.5,2.5){$\small \widehat{\tfrac{3}{10}}$}
\rput(2.5,2.5){$\small -\!\widehat{\tfrac{1}{5}}$}
\rput(4.5,2.5){$\small -\!\widehat{\tfrac{\,1'}{5}}$}
\rput(3.5,1.5){$\small -\!\widehat{\tfrac{13'}{240}}$}
\rput(2.5,1.5){$\small -\!\widehat{\tfrac{11}{80}}$}
\rput(.5,-.3){$\small 1$}
\rput(1.5,-.3){$\small 2$}
\rput(2.5,-.35){$\small 3,\!6$}
\rput(3.5,-.3){$\small 4$}
\rput(4.5,-.3){$\small 5$}
\rput(5.3,-.3){$\small a$}
\rput(-.3,.5){$\small 1$}
\rput(-.3,1.5){$\small 2$}
\rput(-.3,2.5){$\small 2$}
\rput(-.3,3.5){$\small 1$}
\rput(-.3,4.3){$\small r$}
\end{pspicture}
\\[30pt]
\raisebox{30pt}{$\sqrt{\tfrac{5-\sqrt{5}}{40}}$
\!\!\sssmat{
&\sqrt{2\!+\!\sqrt{3}}&\sqrt{2}&\sqrt{2\!+\!\sqrt{3}}&\\[4pt]
\tfrac12(1\!+\!\sqrt{5})&&\tfrac12(1\!+\!\sqrt{5})(1\!+\!\sqrt{3})&&\tfrac12(1\!+\!\sqrt{5})\\[4pt]
&\tfrac12(1\!+\!\sqrt{5})\sqrt{2\!+\!\sqrt{3}}&\tfrac1{\sqrt{2}}(1\!+\!\sqrt{5})&\tfrac12(1\!+\!\sqrt{5})\sqrt{2\!+\!\sqrt{3}}&\\[4pt]
1&&1\!+\!\sqrt{3}&&1}}
\end{center}
\caption{The bipartite coset graph $\tilde{G}=A_4\otimes E_6/\mathbb{Z}_2=T_2\otimes E_6$ and Kac tables of conformal weights and 2-boundary $\g$-factors $\hat{\g}_{(1,1)|(r,a)}$ of the critical $(A_4,E_6)$ model with $c=-\tfrac{39}{10}$ and $c_\text{eff}=\tfrac9{10}$ where $T_2$ is the tadpole on 2 nodes. 
The nodes are $(r,a)$ with $r=1,2$ and $a=1,2,\ldots,6$. Alternatively, the ordered nodes $(r,a)=(1,1),(1,2),(1,3),(1,4),(1,5),(1,6),(2,1),(2,2),(2,3),(2,4),(2,5),(2,6)$ are labelled by $\mu=1,2,\ldots,12$. 
The $E_6$ graph intertwines with $A_{11}$~\cite{OcneanuInter,PZ93}. 
The extended conformal weights are 
$\widehat{0}=0+2$, $-\widehat{\tfrac{3}{16}}=-\tfrac{3}{16}+\tfrac{55}{48}+\tfrac{49}{16}$, $-\widehat{\tfrac16}=-\tfrac16+\tfrac12+2+\tfrac{13}3$, $\widehat{\tfrac1{16}}=\tfrac1{16}+\tfrac{55}{48}+\tfrac{93}{16}$, $\widehat{\tfrac12}=\tfrac12+\tfrac{15}2$, $\widehat{\tfrac{1'}{16}}=\tfrac1{16}+\tfrac{49}{16}$; 
$\widehat{\tfrac{3}{10}}=\tfrac3{10}+\tfrac{13}{10}$, $-\widehat{\tfrac{13}{240}}=-\tfrac{13}{240}+\tfrac{49}{80}+\tfrac{69}{80}$,$-\widehat{\tfrac{1}{5}}=-\tfrac{1}{5}+\tfrac{2}{15}+\tfrac{3}{10}+\tfrac{49}{30}$,
$-\widehat{\tfrac{13'}{240}}=-\tfrac{13}{240}-\tfrac{11}{80}+\tfrac{209}{80}$,$-\widehat{\tfrac{1}{5}}=-\tfrac{1'}{5}+\tfrac{19}{5}$,$-\widehat{\tfrac{11}{80}}=-\tfrac{11}{80}+\tfrac{69}{80}$.
\label{A4E6CosetGraph}}
\end{figure}

The coset graph of the critical $(A_4,E_6)$ model is $\tilde{G}=A_4\otimes E_6/\mathbb{Z}=T_2\otimes E_6$ as shown in Figure~\ref{A4E6CosetGraph}.
Explicitly, the nimrep fusion matrices are
\bea
\tilde{N}_{r,a}=N_r^{(T_2)}\!\otimes\! \hat{N}_a^{(E_6)},\  \tilde{N}_{1,a}\!=\!\!\!\ssmat{\hat{N}^{(E_6)}_a&0\\ 0&\hat{N}^{(E_6)}_a},\ 
\tilde{N}_{2,a}\!=\!\!\!\ssmat{0&\hat{N}^{(E_6)}_a\\ \hat{N}^{(E_6)}_a&\hat{N}^{(E_6)}_a},\quad r=1,2;\ a=1,2,\ldots,6
\eea
The nimreps of $T_2$ are given in (\ref{T2nimreps}) and the nimreps of $E_6$ are
\begin{align}
\hat{N}^{(E_6)}_a=\!\!
\sssmat{1&0&0&0&0&0\\ 0&1&0&0&0&0\\ 0&0&1&0&0&0\\ 0&0&0&1&0&0\\ 0&0&0&0&1&0\\ 0&0&0&0&0&1}, \!\!
\sssmat{0&1&0&0&0&0\\ 1&0&1&0&0&0\\ 0&1&0&1&0&1\\ 0&0&1&0&1&0\\ 0&0&0&1&0&0\\ 0&0&1&0&0&0},  \!\!
\sssmat{0&0&1&0&0&0\\ 0&1&0&1&0&1\\ 1&0&2&0&1&0\\ 0&1&0&1&0&1\\ 0&0&1&0&0&0\\ 0&1&0&1&0&0},  \!\!
\sssmat{0&0&0&1&0&0\\ 0&0&1&0&1&0\\ 0&1&0&1&0&1\\ 1&0&1&0&0&0\\ 0&1&0&0&0&0\\ 0&0&1&0&0&0},  \!\!
\sssmat{0&0&0&0&1&0\\ 0&0&0&1&0&0\\ 0&0&1&0&0&0\\ 0&1&0&0&0&0\\ 1&0&0&0&0&0\\ 0&0&0&0&0&1},  \!\!
\sssmat{0&0&0&0&0&1\\ 0&0&1&0&0&0\\ 0&1&0&1&0&0\\ 0&0&1&0&0&0\\ 0&0&0&0&0&1\\ 1&0&0&0&1&0}
\end{align}

The unitary matrix $\boldsymbol\Psi$ that diagonalizes $\tilde{G}$ and the coset graph fusion matrices is
\begin{align}
&\mbox{}\hspace{-16pt}\boldsymbol\Psi\!=\!{\cal S}\otimes \Psi, \ \  {\cal S}\!=\!\tfrac{2}{\sqrt{5}}\!\!\sssmat{\sin\tfrac{\pi}5&\sin\tfrac{2\pi}5\\[4pt] \sin\tfrac{2\pi}5&-\!\sin\tfrac{\pi}5},\ \  
\Psi^T\!\!=\!\Psi^{-1}\!\!=\!\!\!
\sssmat{
\sqrt{\frac{3-\sqrt{3}}{24}}&\sqrt{\frac{3+\sqrt{3}}{24}}&\sqrt{\frac{3+\sqrt{3}}{12}}&\sqrt{\frac{3+\sqrt{3}}{24}}&\sqrt{\frac{3-\sqrt{3}}{24}}&\sqrt{\frac{3-\sqrt{3}}{24}}\\[4pt]
\tfrac12&\tfrac12&0&-\tfrac12&-\tfrac12&0\\[4pt]
\sqrt{\frac{3+\sqrt{3}}{24}}&\sqrt{\frac{3+\sqrt{3}}{24}}&-\sqrt{\frac{3-\sqrt{3}}{12}}&\sqrt{\frac{3-\sqrt{3}}{24}}&\sqrt{\frac{3+\sqrt{3}}{24}}&-\sqrt{\frac{3+\sqrt{3}}{24}}\\[4pt]
\sqrt{\frac{3+\sqrt{3}}{24}}&-\sqrt{\frac{3-\sqrt{3}}{24}}&-\sqrt{\frac{3-\sqrt{3}}{12}}&-\sqrt{\frac{3-\sqrt{3}}{24}}&\sqrt{\frac{3+\sqrt{3}}{24}}&\sqrt{\frac{3+\sqrt{3}}{24}}\\[4pt]
\tfrac12&-\tfrac12&0&\tfrac12&-\tfrac12&0\\[4pt]
\sqrt{\frac{3-\sqrt{3}}{24}}&-\sqrt{\frac{3-\sqrt{3}}{24}}&\sqrt{\frac{3+\sqrt{3}}{12}}&-\sqrt{\frac{3+\sqrt{3}}{24}}&\sqrt{\frac{3-\sqrt{3}}{24}}&-\sqrt{\frac{3-\sqrt{3}}{24}}
}\!\!\!\!\!\!\!\!\\[6pt]
&\qquad\boldsymbol\Psi^{-1}\tilde{G}\,\boldsymbol\Psi=
\mbox{diag}\{\lambda_{r,a}\},\ \  \lambda_{1,a}=4\cos\tfrac{\pi}5 \cos\tfrac{a\pi}{12},\ \  \lambda_{2,a}=4\cos\tfrac{3\pi}5 \cos\tfrac{a\pi}{12},\ \  a\in \Exp(E_6)
\end{align}
where $\Exp(E_6)=\{1,\!4,\!5,\!7,\!8,\!11\}$. The 2-boundary $\g$-factors $\hat{\g}_{(1,1)|(r,a)}$ and defect $\g$-factors $\hat{d}_{(r,a)}$ are
\begin{align}
\hat{\g}_{(1,1)|(r,a)}\!=\!\sqrt{\tfrac{5-\sqrt{5}}{40}}\,\tilde{d}_{(r,a)},\quad
\hat{d}_{(r,a)}&\!=(\tfrac{1+\sqrt{5}}{2})^{r-1}\Big\{[1]_{x}, [2]_{x}, [3]_{x}, [2]_{x}, [1]_{x},\tfrac{[3]_{x}}{[2]_{x}}\Big\},\ \quad x=e^{\pi i/12}
\end{align}
where $\hat{\g}_{(1,1)|(1,1)}=\sqrt{\tfrac{5-\sqrt{5}}{40}}$.

In terms of the asymptotics of counting fusion paths, we find
\bea
\lim_{\substack{N\to\infty\\[1pt] \text{$N\!=\!\kappa$\,mod\,2}}} (\tfrac{1}{\lambda_+})^{N}\tilde{G}^{N}\!\!=\boldsymbol\Psi\,\text{diag}\{1,0,0,0,0,(-1)^\kappa,0,0,0,0,0,0\}\boldsymbol\Psi^{-1}=\!
\begin{cases}\vec D_+,&\mbox{$N$ even}\\ \vec D_-,&\mbox{$N$ odd}\end{cases}
\eea
with $\lambda_+=\tfrac{(1+\sqrt{3})(1+\sqrt{5})}{2\sqrt{2}}$. Combining the even and odd matrices, gives the rank-1 matrix $\vec D\!=\!\vec D_+\!+\!\vec D_-$\\
\begin{align}
\vec D\!&=\!\!\!\sssmat{\tfrac{1}{10}(5\!-\!\sqrt{5})&\tfrac{1}{\sqrt{5}}\\ \tfrac{1}{\sqrt{5}}&\tfrac{1}{10}(5\!+\!\sqrt{5})}\otimes 
\!\!\sssmat{
\tfrac{3-\sqrt{3}}{12} & \tfrac1{2\sqrt{6}}&\tfrac1{2\sqrt{3}} & \tfrac1{2\sqrt{6}}&\tfrac{3-\sqrt{3}}{12}&\sqrt{\tfrac{2-\sqrt{3}}{12}}\\[4pt]
\tfrac1{2\sqrt{6}} & \tfrac{3+\sqrt{3}}{12}&\sqrt{\tfrac{2+\sqrt{3}}{12}}& \tfrac{3+\sqrt{3}}{12}& \tfrac1{2\sqrt{6}}  &\tfrac1{2\sqrt{3}}\\[4pt]
\tfrac1{2\sqrt{3}}  & \sqrt{\tfrac{2+\sqrt{3}}{12}} & \tfrac{3+\sqrt{3}}{6}&\sqrt{\tfrac{2+\sqrt{3}}{12}}& \tfrac1{2\sqrt{3}} & \tfrac1{\sqrt{6}}\\[4pt]
\tfrac1{2\sqrt{6}} & \tfrac{3+\sqrt{3}}{12}&\sqrt{\tfrac{2+\sqrt{3}}{12}}& \tfrac{3+\sqrt{3}}{12}& \tfrac1{2\sqrt{6}}  &\tfrac1{2\sqrt{3}}\\[4pt]
\tfrac{3-\sqrt{3}}{12} & \tfrac1{2\sqrt{6}}&\tfrac1{2\sqrt{3}} & \tfrac1{2\sqrt{6}}&\tfrac{3-\sqrt{3}}{12}&\sqrt{\tfrac{2-\sqrt{3}}{12}}\\[4pt]
\sqrt{\tfrac{2-\sqrt{3}}{12}} &\tfrac1{2\sqrt{3}} &\tfrac1{\sqrt{6}} & \tfrac1{2\sqrt{3}} &\sqrt{\tfrac{2-\sqrt{3}}{12}}& \tfrac{3-\sqrt{3}}{6}
 }
\end{align}
Using the fact that $8 \|\boldsymbol\psi\|^{-2}\,\hat{\g}_{1|1}^2\!=\!\tfrac1{120}(3\!-\!\sqrt{3})(5\!-\!\sqrt{5})$ where $\boldsymbol\psi$ is the Perron-Frobenius eigenvector given by (\ref{PF}), the 2-boundary $\g$-factors $\hat{\g}_{\mu|\mu'}$ and defect $\g$-factors 
$\hat{d}_\mu$ are given by 
\bea
\vec D=8 \|\boldsymbol\psi\|^{-2}\,\hat{\g}_{1|1}\!\!\!
\sssmat{\hat{\g}_{1|1}&\hat{\g}_{1|2}&\ldots&\hat{\g}_{1|12}\\
\hat{\g}_{2|1}&\hat{\g}_{2|2}&\ldots&\hat{\g}_{2|12}\\
\vdots&\vdots&\ddots&\vdots\\
\hat{\g}_{12|1}&\hat{\g}_{12|2}&\ldots&\hat{\g}_{12|12}}
=8 \|\boldsymbol\psi\|^{-2}\,\hat{\g}^2_{1|1}\!\sssmat{1\\ \hat{d}_{2}\\[-4pt] \vdots\\ \hat{d}_{12}}\!\sssmat{1&\hat{d}_{2}&\ldots&\hat{d}_{12}}
=8 \|\boldsymbol\psi\|^{-2}\,\hat{\g}^2_{1|1} \sum_{\mu=1}^{12} \hat{d}_\mu\tilde{N}_\mu
\eea

\section{\ade RSOS Lattice Models\label{secLattice}}

In previous sections, we have focused on properties of the \ade minimal CFTs that are neatly encoded in coset graphs and given many examples. We now wish to turn our attention to (i) the conformal weights $\Delta_{r,s}^{m,m'}$, (ii) defects ${\cal L}_{(r,s)}$, labelled by the Kac labels $(r,s)$ as the continuum scaling limit of integrable seams ${\vec T}^{(r,s)}$ constructed on the lattice and (iii) boundary conditions and conformal cylinder partition functions as implemented by the action of the integrable seams on the vacuum boundary condition. For discussion of these properties, crucial insight comes from studies of exactly solvable \ade RSOS lattice models. To obtain expressions for the conformal weights, we will need the asymptotic limits of the $T$- and $Y$-systems and the Thermodynamic Bethe Ansatz equations (TBA). To obtain  integrable seams, we will use ``braid limits'' to construct column-to-column transfer matrices ${\vec T}^{(r,s)}$ labelled by the Kac labels $(r,s)$. Lastly, in this section, we will consider the action of integrable seams on the vacuum boundary and the role of the fusion rules on the lattice in determining conformal cylinder partition functions.

\subsection{\ade lattice models and their $T$- and $Y$-systems}

Consider the critical \ade\/ RSOS lattice models~\cite{ABF84,FB,Pasquier87a,Pasquier87b,Pasquier87c} with Coxeter numbers $(m,m')$ and  crossing parameter $\lambda=\tfrac{(m'\!-m)\pi}{m'}$. 
Starting in 1984, the first such models were built on the \mbox{$A$-type} Dynkin diagrams and solved off-criticality by Andrews, Baxter and Forrester~\cite{ABF84,FB}. 
The full family of critical \ade lattice models was subsequently introduced and studied by Pasquier~\cite{Pasquier87a,Pasquier87b,Pasquier87c}. 
These 2-dimensional lattice models on the square lattice are Yang-Baxter integrable~\cite{BaxBook82} so they are exactly solvable. 
The unitary and nonunitary $(A,G)$ minimal CFTs are obtained from these lattice models in the continuum scaling limit. 
In particular, the CFT defect lines ${\cal L}_\mu$ emerge as the continuum scaling limits of {\em integrable seams}~\cite{PRasmussen24,PRasmussen25,STRSaleur2025} implemented on the lattice as special column transfer matrices. 

The face weights of the critical \ade RSOS lattice models are
\bea
\face abcdu\;=\,\Wt abcdu=\,\frac{\sin(\lambda-u)}{\sin\lambda}\,\delta_{a,c}+\frac{g_c}{g_a}\frac{\sin u}{\sin\lambda}\, \frac{\sqrt{\psi_a}\sqrt{\psi_c}}{\psi_b}\,\delta_{b,d}
\label{faceWeights}
\eea
for $|a\!-\!b|=|b\!-\!c|=|c\!-\!d|=|d\!-\!a|=1$ but vanish otherwise. 
Usually authors consider the unitary \ade models, as in \cite{Pasquier87a,Pasquier87b,Pasquier87c}, or unitary and nonunitary $A$-type models, as in \cite{ArdonneEtAl2010}.
But this formulation (\ref{faceWeights}) applies to all unitary and nonunitary \ade models.
The spectral parameter is $u$, the crossing parameter is $\lambda=\tfrac{\pi(m'\!-m)}{m'}$ with $m'\!-m\in\Exp(G)$, $g_a$ are arbitrary gauge factors
and here $\psi_a$ are the components of the eigenvector of $G$ corresponding to the eigenvalue $2\cos\lambda$. We will work in the symmetric gauge with $g_a=1$. For $m, m'$ coprime, the components $\psi_b\neq 0$ are all non-zero. 
Some care needs to be taken with the principal branch square roots since, for nonunitary cases, some $\psi_a$ are negative and it is not generally true that $\sqrt{\psi_a\psi_c}=\sqrt{\psi_a}\sqrt{\psi_c}$. 
These face weights can be written in terms of the generators of the Temperley-Lieb algebra and satisfy the Yang-Baxter equation as shown in Appendix~\ref{ADEYBE}. 
In this formulation, the Temperley-Lieb generators $e_j$, given by setting $u=\lambda$, are symmetric but complex matrices. 
With $g_a=1$, the braid limits of the allowed \ade face weights are
\begin{subequations}
\begin{align}
\psset{unit=0.8cm}
\face abcd{-i\infty}\,=\Wtt Babcd&
\,= \!\lim_{u\to-i\infty}\frac{x^{-\frac 12}}{i\rho(u)} \,\Wt abcdu
=-i\bigg(\!x^{-\frac 12}\,\delta_{ac}-x^{\frac 12}\frac{\sqrt{\psi_a}\sqrt{\psi_c}}{\psi_b}\;\delta_{bd}\bigg)\\[6pt]
\label{invbraid}
\face abcd{i\infty}\,=\Wtt {\overline{B}}abcd&
\,=\lim_{u\to i\infty}\;\,\frac{i\,x^{\frac 12}}{\rho(u)}\;\, \Wt abcdu
=\,i\bigg(x^{\frac 12}\,\delta_{ac}\;-\,x^{-\frac 12}\frac{\sqrt{\psi_a}\sqrt{\psi_c}}{\psi_b}\;\delta_{bd}\bigg)
\end{align}
where $x=e^{i\lambda}$, $\rho(u)=\sin(\lambda\!-\!u)/\sin\lambda$ and bars denote complex conjugation. In the continuum scaling limit, these braid limits relate to the left and right chiral halves of the theory related to each other  by complex conjugation (see Figure~\ref{scaleLims}).
\label{braid}
\end{subequations}

To define transfer matrices, we will work on a square lattice on a cylinder with boundary conditions $(r,b)$ on the left and $(r'\!,c)$ on the right. The double row transfer matrix~\cite{BPO96} with $N$ faces is
\bea
\psset{unit=0.8cm}
\D^{(N)}_{(r,b)|(r',c)}(u)\,=\, \ \ 
\begin{pspicture}[shift=-1.1](-.7,.75)(6.7,3.3)
\facegrid{(0,1)}{(6,3)}
\pspolygon[fillstyle=solid,fillcolor=lightlightblue](0,2)(-1,1)(-1,3)
\pspolygon[fillstyle=solid,fillcolor=lightlightblue](6,2)(7,1)(7,3)
\rput(0.5,1.5){\small $u$}
\rput(1.5,1.5){\small $u$}
\rput(5.5,1.5){\small $u$}
\rput(0.5,2.5){\small $\lambda\!-\!u$}
\rput(1.5,2.5){\small $\lambda\!-\!u$}
\rput(5.5,2.5){\small $\lambda\!-\!u$}
\rput(3.5,1.5){\small $\dots$}
\rput(3.5,2.5){\small $\dots$}
\multirput(0,0)(1,0){6}{\psarc[linecolor=red,linewidth=.5pt](0,1){.15}{0}{90}}
\multirput(0,0)(1,0){6}{\psarc[linecolor=red,linewidth=.5pt](0,2){.15}{0}{90}}
\rput(-.57,2){\spos{c}{(r,b)}}
\rput(6.57,2){\spos{c}{(r'\!,c)}}
\rput(-1,.7){\spos{b}{b}}
\rput(0,.7){\spos{b}{b}}
\rput(1,.68){\spos{b}{a_2}}
\rput(2,.68){\spos{b}{a_3}}
\rput(3.5,.8){\small $\dots$}
\rput(5.1,.65){\spos{b}{a_{N\!-\!1}}}
\rput(-1,3.1){\spos{b}{b}}
\rput(0,3.1){\spos{b}{b}}
\rput(1,3.1){\spos{b}{b_2}}
\rput(2,3.1){\spos{b}{b_3}}
\rput(3.5,3.2){\small $\dots$}
\rput(5.1,3.06){\spos{b}{b_{N\!-\!1}}}
\rput(6,.7){\spos{b}{c}}
\rput(7,.7){\spos{b}{c}}
\rput(6,3.1){\spos{b}{c}}
\rput(7,3.1){\spos{b}{c}}
\end{pspicture}\qquad\quad r,r'\!\in A;\ \ b,c\in G
\label{D}
\eea
where the internal heights are summed over. The triangle boundary weights, dependent on a parameter $\xi$, are specified in \cite{BP2001}. 
As shown in \cite{BPO96}, these matrices form a 1-parameter family of commuting double row transfer matrices $[\D^{(N)}_{(r,b)|(r',c)}(u),\D^{(N)}_{(r,b)|(r',c)}(v)]=0$.\\[-6pt]

Next, let us consider the construction of integrable seams as periodic column-to-column transfer matrices labelled by the Kac labels $(r,s)$. In contrast to the double row transfer matrix, no boundary conditions appear in these periodic column transfer matrices. The construction starts with the fundamental integrable seams ${\vec T}^{(2,1)}$, ${\vec T}^{(1,2)}$ with Kac labels $(r,s)=(2,1)$ and $(r,s)=(1,2)$ respectively. 
The fundamental spectral-parameter-dependent $r$-type integrable seam is given by the column-to-column transfer matrix
\bea
\psset{unit=0.9cm}
\T^{(2,1)}(\xi)=\,\begin{pspicture}[shift=-.7](-.5,-.3)(6.5,1.3)
\psline[linewidth=.75pt,linestyle=dashed,dash=2pt 2pt](-.4,.5)(6.4,.5)
\facegrid{(0,0)}{(6,1)}
\multirput(0,0)(2,0){3}{\rput(.5,.5){\spos{c}{u+\xi}}}
\multirput(0,0)(2,0){3}{\rput(1.5,.5){\spos{c}{\lambda\!-\!u\!+\!\xi}}}
\multirput(0,0)(1,0){6}{\psarc[linecolor=red,linewidth=.5pt](1,0){.15}{90}{180}}
\rput(0,-.2){\spos{c}{a_1}}
\rput(1,-.2){\spos{c}{a_2}}
\rput(3.05,-.2){\small $\dots$}
\rput(5,-.2){\spos{c}{a_{M}}}
\rput(6,-.2){\spos{c}{a_1}}
\rput(0,1.2){\spos{c}{b_1}}
\rput(1,1.2){\spos{c}{b_2}}
\rput(3.05,1.2){\small $\dots$}
\rput(5,1.2){\spos{c}{b_{M}}}
\rput(6,1.2){\spos{c}{b_1}}
\end{pspicture}
\eea
with periodicity $a_{M\!+\!1}=a_1$, $b_{M\!+\!1}=b_1$. For display convenience, we have rotated the transfer matrix by 90 degrees anticlockwise. For this integrable seam, it is the parameter $\xi$ that plays the role of the spectral parameter and $u$ and $\lambda\!-\!u$ are alternating inhomogeneities. 
Using the local Yang-Baxter equations, it follows that these column transfer matrices form a commuting family \mbox{$[\T(\xi),\T(\xi')]=0$}. 
The braid limits $\xi\to\pm i\infty$ of these face weights are given by (\ref{braid}) independent of the inhomogeneities $u$ and $\lambda\!-\!u$. It follows that the fundamental $s$-type integrable seams 
\psset{unit=0.85cm}
\begin{align}
{\mathbf n}_2&=\mathbf B=\vec T^{(1,2)}\!=\lim_{\xi\to -i\infty} \Big(\frac{i\,x^{\frac 12}}{\rho(\xi)}\Big)^M \T^{(2,1)}(\xi)=\,
\begin{pspicture}[shift=-.7](-.5,-.3)(6.5,1.3)
\psline[linewidth=.75pt,linestyle=dashed,dash=2pt 2pt](-.4,.5)(6.4,.5)
\facegrid{(0,0)}{(6,1)}
\multirput(0,0)(2,0){3}{\rput(.5,.5){\spos{c}{-i\infty}}}
\multirput(0,0)(2,0){3}{\rput(1.5,.5){\spos{c}{-i\infty}}}
\multirput(0,0)(1,0){6}{\psarc[linecolor=red,linewidth=.5pt](1,0){.15}{90}{180}}
\rput(0,-.2){\spos{c}{a_1}}
\rput(1,-.2){\spos{c}{a_2}}
\rput(3.05,-.2){\small $\dots$}
\rput(5,-.2){\spos{c}{a_{N}}}
\rput(6,-.2){\spos{c}{a_1}}
\rput(0,1.2){\spos{c}{b_1}}
\rput(1,1.2){\spos{c}{b_2}}
\rput(3.05,1.2){\small $\dots$}
\rput(5,1.2){\spos{c}{b_{N}}}
\rput(6,1.2){\spos{c}{b_1}}
\end{pspicture}\\[8pt]
\bar{\mathbf n}_2&=\overline{\mathbf B}=\overline{\vec T}^{(1,2)}\!=\;\lim_{\xi\to i\infty}\Big( \frac{x^{-\frac 12}}{i\rho(\xi)}\Big)^M \T^{(2,1)}(\xi)\;=\begin{pspicture}[shift=-.7](-.5,-.3)(6.5,1.3)
\psline[linewidth=.75pt,linestyle=dashed,dash=2pt 2pt](-.4,.5)(6.4,.5)
\facegrid{(0,0)}{(6,1)}
\multirput(0,0)(2,0){3}{\rput(.5,.5){\spos{c}{i\infty}}}
\multirput(0,0)(2,0){3}{\rput(1.5,.5){\spos{c}{i\infty}}}
\multirput(0,0)(1,0){6}{\psarc[linecolor=red,linewidth=.5pt](1,0){.15}{90}{180}}
\rput(0,-.2){\spos{c}{a_1}}
\rput(1,-.2){\spos{c}{a_2}}
\rput(3.05,-.2){\small $\dots$}
\rput(5,-.2){\spos{c}{a_{N}}}
\rput(6,-.2){\spos{c}{a_1}}
\rput(0,1.2){\spos{c}{b_1}}
\rput(1,1.2){\spos{c}{b_2}}
\rput(3.05,1.2){\small $\dots$}
\rput(5,1.2){\spos{c}{b_{N}}}
\rput(6,1.2){\spos{c}{b_1}}
\end{pspicture}
\end{align}
precisely coincide with the torus Ocneanu integrable seams ${\mathbf n}_2$ and $\bar{\mathbf n}_2$ of \cite{PRasmussen24,PRasmussen25}. It therefore also follows that  ${\mathbf n}_2$ and $\bar{\mathbf n}_2$ separately  satisfy the graph fusion algebra (\ref{FusionAlgebras}) with the Verlinde structure constants
\bea
{\mathbf n}_1=\bar{\mathbf n}_1={\mathbf I},\quad {\mathbf n}_2={\mathbf B},\quad \bar{\mathbf n}_2=\overline{\mathbf B},\qquad {\mathbf n}_i{\mathbf n}_j=\sum _{k=1}^{m'-1} N_{ij}{}^k{\mathbf n}_k,\qquad \bar{\mathbf n}_i\bar{\mathbf n}_j=\sum _{k=1}^{m'-1} N_{ij}{}^k\bar{\mathbf n}_k
\eea
Note that the $s$-type fundamental integrable seams are the braid limits of the $r$-type fundamental integrable seams. These are independent of $\xi$ so there is no spectral parameter indicated in the notation ${\vec T}^{(1,2)}$, $\overline{\vec T}^{(1,2)}$ for these two $s$-type column seams related by complex conjugation.

Working with the fundamental column transfer matrices, compound $(r,s)$ integrable seams
\bea
\vec T^{(r,s)}(u)=\vec T^{(r,1)}(u)\,\vec T^{(1,s)},\qquad \overline{\vec T}^{(r,s)}(u)=\vec T^{(r,1)}(u)\,\overline{\vec T}^{(1,s)},\quad r,s=1,2,\ldots,m'\!-1
\eea
are constructed recursively using known fusion functional equations~\cite{BazhResh1989,KP92} and their braid limits~\cite{PRasmussen24,PRasmussen25}
\begin{align}
&\hspace{-6pt}\vec T_0^{(r,1)}\vec T_{r-1}^{(2,1)}\!=\!f_{r-1}\,\vec T_0^{(r-1,1)}\!+\!f_{r-2}\,\vec T_0^{(r+1,1)},\quad \vec T_0^{(0,1)}\!=\vec T_0^{(m',1)}\!=\!\vec 0,\ \  \vec T_0^{(1,1)}\!=\!f_{-1}\vec I,\ \ \vec T_0^{(m'\!-1,1)}\!=\!f_{m'\!-2}\boldsymbol\sigma\label{FusHierTr}\\[4pt]
&\quad \vec T^{(1,s)}\vec T^{(1,2)}=\vec T^{(1,s-1)}+\vec T^{(1,s+1)},\quad \vec T^{(1,0)}=\vec T^{(1,m')}=\vec 0,\quad \vec T^{(1,1)}=\vec I,\quad \vec T^{(1,m'\!-1)}=\boldsymbol\sigma\label{FusHierTs}
\end{align}
where $(r,s)\!=\!(2,1)$ and $(1,2)$ denote the fundamentals, the dependence on the inhomogeneities $u,\lambda\!-\!u$ is suppressed, $\boldsymbol\sigma$ is the $\mathbb{Z}_2$ height reversal operator and
\bea
\vec T^{(r,1)}_q\!=\!\vec T^{(r,1)}(\xi+q\lambda),\quad \vec T^{(r,1)}_0\!=\!\boldsymbol\sigma\, \vec T^{(m'\!-r,1)}_{r},
\quad f_r\!=\!\big(\tfrac{\sin(\xi+u+r\lambda)\sin(\xi-u+(r+1)\lambda)}{\sin^2\lambda}\big)^{M/2}
\eea 
It is seen that the Kac labels $r$ and $s$  label the fused integrable seams that appear at level $r$ and level $s$ in these two separate recursions. Note that in this fusion construction, at the level of the column transfer matrices, there is no need to obtain explicit fused face weights using the standard Wenzl-Jones projectors.

The $r$-type $T$- and $Y$-systems~\cite{KP92} are
\begin{align}
\frac{\vec T_0^{(r,1)}\vec T_1^{(r,1)}}{f_{-1}f_{r-1}}&=\vec I+\frac{\vec T_1^{(r-1,1)}\vec T_0^{(r+1,1)}}{f_{-1}f_{r-1}}\equiv\vec I+\vec Y_0^r,\qquad r=1,2,\ldots,m\!-\!1\label{Tsys}\\[4pt]
&\vec Y_0^r\vec Y_1^r=(\vec I+\vec Y_1^{r-1})(\vec I+\vec Y_0^{r+1}),\qquad r=1,2,\ldots,m-2\label{Ysys}
\end{align}
with $\vec Y^r_q\!=\!\vec Y^r(\xi+q\lambda)$. These functional equations were originally established~\cite{BazhResh1989,KP92} for unitary cases \mbox{($m'\!-m= 1$)} without inhomogeneities but can be established for unitary and nonunitary $A(m,m')$ models with inhomogeneities by the same methods. 
The $f$ functions remove the non-universal bulk free energies. 
In obtaining conformal defects in the continuum scaling limit, it suffices to consider the isotropic case with homogeneities $u=\lambda\!-\!u=\tfrac \lambda{2}$. 

The continuum scaling limit of the RSOS lattice models yields the associated CFT. More precisely, the scaling limit is $a\to 0$, $M,N\to\infty$, $aM\to L$, $aN\to R$ where $a$ is the lattice spacing, $L$ and $R$ are continuous coordinates and $M/N\to L/R$ is the aspect ratio. Within the CFT, the transfer matrices $\vec T^{(r,1)}(u)$ are replaced~\cite{BLZ} by a set $\mathbf{T}^r(u)$ of operator valued functions of the complex spectral parameter $u\in\mathbb{C}$. The operators $\mathbf{T}^r(u)$ satisfy~\cite{BLZ} precisely the same $T$- and $Y$-systems of functional equations as the lattice transfer matrices $\vec T^{(r,1)}(u)$. So the conformal topological defects ${\cal L}_r$ and ${\cal L}_s$ are operators corresponding to specializations of $\mathbf{T}^r(u)$ in the bulk and braid limits respectively as we explain in the next subsection. 

\subsection{Braid and bulk scaling limits of integrable seams}
\label{sec:rsSeams}

\begin{figure}[htb]
\psset{unit=1.2cm}
\begin{center}
{\begin{pspicture}[shift=-.7](0,0)(8,6.2)
\psellipse[fillstyle=solid,fillcolor=yellow,linecolor=yellow](3.5,3)(1.5,.3)
\psellipse[fillstyle=solid,fillcolor=lightblue,linecolor=lightblue](3.5,5)(1.5,.3)
\psellipse[fillstyle=solid,fillcolor=lightblue,linecolor=lightblue](3.5,1)(1.5,.3)
\psline[arrowsize=5pt,linewidth=.75pt]{->}(0,3)(8,3)
\psline[arrowsize=5pt,linewidth=.75pt]{->}(3,0)(3,6.1)
\psline[linewidth=.75pt,linestyle=dashed,dash=2pt 2pt](2,0)(2,6)
\psline[linewidth=.75pt,linestyle=dashed,dash=2pt 2pt](5,0)(5,6)
\psline[linewidth=.5pt](3.5,0)(3.5,6)
%
\rput(1.7,3.25){\scriptsize $-\tfrac{\lambda}{2}$}
\rput(3.7,3.25){\scriptsize $\tfrac{\lambda}{2}$}
\rput(5.25,3.25){\scriptsize $\tfrac{3\lambda}{2}$}
\rput(7.5,5.9){\large $\xi$}
\rput(3.7,5.9){\large $x$}
\rput(1.6,5){\scriptsize $\log M$}
\rput(1.5,1){\scriptsize $-\log M$}
\rput(4.3,4.6){\scriptsize $\bar s$}
\rput(4.3,2.6){\scriptsize $r$}
\rput(4.3,.6){\scriptsize $s$}
\end{pspicture}}
\end{center}
\caption{The complex plane of the spectral parameter $\xi$ showing the fundamental ($j=2$) analyticity strip $-\tfrac{\lambda}{2}<\re \xi <\tfrac{3\lambda}{2}$.
The scaling regimes for large $M$ are shown for $r$ (shaded yellow), $s$ and $\bar s$ (both shaded blue). 
The $s$, $\bar s$ scaling regimes are related by complex conjugation and are at a distance $\mp\log M$ from the real axis. For $A$-type models with diagonal modular invariants $\bar s=s$. Similar analyticity strips are related to each fusion level $j=2,3,\ldots,m'\!-\!2$. 
For the ground state eigenenergy and $j=2$, the zeros are dense on the dashed lines between the $s$ and $\bar s$ scaling regimes. 
The spectral parameter is written as $\xi=\tfrac{\lambda}{2}+\tfrac{\lambda}{\pi}\,i\, x$ and the braid limits are $x\to\mp\infty$. The lower/upper half planes relate to the left/right chiral halves of the associated CFT in the scaling limit. 
The $r$ scaling regime is reached by first taking the continuum scaling limit $x\sim \mp\log M$ with $M$ large thus moving to the $s$ scaling regime in the upper/lower half-plane, followed by taking the limit $x\to \pm\infty$ thus moving to the real axis. In \cite{KP92} this is called the ``bulk limit''. 
The braid and bulk limits in the upper half planes are directly related to the plateaux asymptotics at $x=\pm\infty$ in the Thermodynamic Bethe Ansatz (TBA). 
These bulk and braid limits are related to $(r,a)\in (A,G)$. That is, the braid limit is related to $a\in G$ (or $s\in G$ if $G$ is of $A$-type) and the bulk limit is related to $r\in A$. 
The $r$ scaling regime, centered on the real axis, is common to both left and right chiral halves of the CFT. In the continuum scaling limit, the integers $r,s,\bar s$ are good quantum numbers for the CFT.
\label{scaleLims}}
\end{figure}
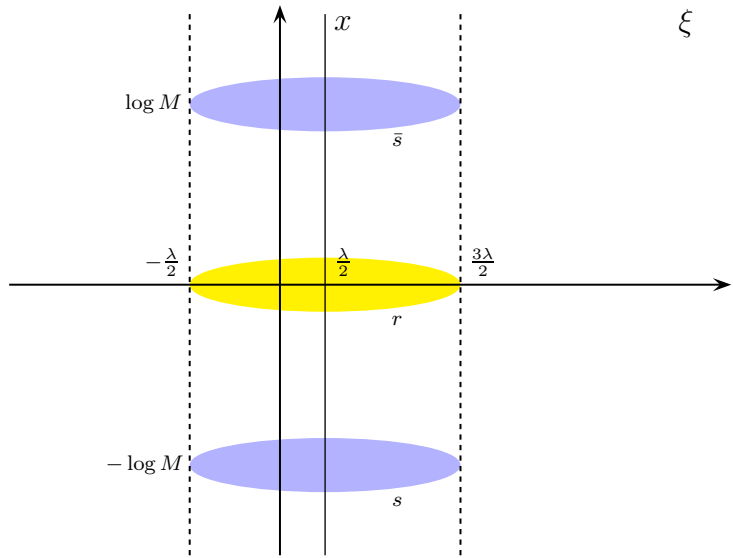

For finite-size commuting periodic transfer matrices ${\vec T}(\xi)={\vec T}^{(2,1)}(\xi)$, $s$ is a good quantum number so the eigenvalues separate into sectors labelled by $s$ or equivalently the braid limit eigenvalues $\tilde{d}_{(1,s)}=2\cos \tfrac{s\pi}{m'}$ where $m'$ is the Coxeter number.
In contrast, for finite-size transfer matrices $\vec T(\xi)$, $r$ is not a quantum number and the spectra does not separate into sectors labelled by $r$. In this case, $r$ only becomes a good quantum number in the continuum scaling limit, that is, first the scaling limit must be taken with $\xi=\tfrac\lambda{2}+\tfrac{\lambda}{\pi}\,ix$ and $x\sim \mp \log M$ with $M\to\infty$ followed by the limit $x\to\pm\infty$. 
Since ultimately $\xi\to \pm i\infty$, the bulk limits are constants independent of the inhomogeneities. 
The $r$ and $s$ scaling regimes are shown schematically in Figure~\ref{scaleLims}.
Considering the $A$ cases, replacing $r$ with $j$ and taking the bulk ($r$-type scaling) limit of the eigenvalues of the $T$- and $Y$-systems (\ref{Tsys})  and (\ref{Ysys}) using (3.2) of \cite{KP92} gives the recurrence relations
\begin{align}
&d_jd_2=d_{j-1}+d_{j+1},\qquad d_j^2=1+d_{j-1}d_{j+1}=1+\varepsilon_j, \qquad  j=1,2,\ldots,m\!-\!1\label{drecursion}\\
&\qquad\quad\varepsilon_j^2=d_{j-1}^2d_{j+1}^2=(1+\varepsilon_{j-1})(1+\varepsilon_{j+1}),\qquad   j=1,2,\ldots,m\!-\!2\label{epsrecursion}
\end{align}
where the generalized $r$-type quantum dimension $d_j$ is the scaling limit of $\vec T_0^{(j,1)}(u)$  in (\ref{FusHierTr}) and $\varepsilon_j$ is the bulk limit of $\vec Y_0^j$ in (\ref{Ysys}). The index $j$ indicates the fusion level so $j=1$ is the identity, $j=2$ is the spin-$\tfrac 12$ fundamental, $j=3$ is spin-1 and so on. 
The solutions of these equations are given by
\bea
d_j=d^m_{j,r}=\frac{\sin j\tau}{\sin\tau},\quad\  \varepsilon_j=\epsilon^m_{j,r}=\frac{\sin(j\!-\!1)\tau\sin(j\!+\!1)\tau}{\sin^2\tau},\quad\  \tau=\frac{r\pi}{m},\quad\  r=1,2,\ldots, m\!-\!1
\eea
where $r$ is a Coxeter exponent of $A_{m-1}$. These solutions imply $d^m_{j,r}=\pm 1$ for $j=m\!-\!1$.
The fundamental quantum dimensions are $d_{2,r}^m=\tilde{d}_{(1,r)}=2\cos \tfrac{r\pi}{m}$. 
Restricting to the upper half plane, the latter recurrence is also in accord with the bulk asymptotics of the TBA pseudoenergies $\varepsilon_j(x)$ given, in the unitary cases, by (3.92) of \cite{FPW2009}
\bea
\varepsilon_j(-\infty)^2=\big(1+\varepsilon_{j-1}(-\infty)\big)\big(1+\varepsilon_{j+1}(-\infty)\big),\qquad  j=1,2,\ldots,m\!-\!2
\eea
with $\varepsilon_j=\varepsilon_j(-\infty)$. This equation for the bulk asymptotics of the TBA pseudoenergies is universal independent of the boundary conditions and topology. 
The quantum dimensions $d^m_{j,r}$ and the analogs $d^{m'}_{j,s}$ for the $s$-type integrable seams, where $s$ is a Coxeter exponent of $A_{m'\!-1}$, are good quantum numbers of the CFT.  

\subsection{Dilogarithm identities}

In this section we show that the basic conformal data (central charges and conformal weights) can be expressed in terms of the generalized quantum dimensions $d^m_{j,r}$ and $d^{m'}_{j,s}$.

The unitary and nonunitary \ade lattice models are exactly solvable by the methods of Kl\"umper and Pearce~\cite{PK91,KP91,KP92,Zhou97}. 
For nonunitary models the details will be given in a separate paper. However, the details of these calculations for the (nonunitary) Lee-Yang model ${\cal M}(2,5)$ are given in \cite{BDP2015}. 
Yang-Baxter integrability means that it is possible to calculate the effective central charges and conformal weights in terms of dilogarithms~\cite{NahmRT93,Kirillov94,Kirillov95}. For the central charges, this was first carried out in \cite{BazhResh1989}. For the conformal dimensions, this involves analytic continuations of the dilogarithms and was first carried out in \cite{PK91,KP91,KP92}. 
In the current context, these considerations lead to the following formulas valid for all $(m,m')$ and for all unitary and nonunitary \ade models:
\begin{subequations}
\begin{align}
&\hspace{4.5cm} c^{m,m'}_\text{eff}\!\!=\!\tfrac{1}{m'\!-m}\,\Theta_{1,1}^{m,m'}\label{dilogceff}\\[4pt]
&\qquad\quad\Delta_{r,s}^{m,m'}=\tfrac 14(r\!-\!s)(r\!-\!s\!+\!m'\!\!-\!m)
-\tfrac{m'\!-m}{24}\Big[\Theta_{r,s}^{m,m'}\!-\Theta_{1,1}^{m,m'}\Big]\label{dilogdelta}\\
&\Theta_{r,s}^{m,m'}=\frac{6}{\pi^2}\Big[\sum_{j=1}^{m'\!-2} L_+(\varepsilon_{j,s}^{m'})-\sum_{j=1}^{m-2} L_+(\varepsilon_{j,r}^{m})\Big],\qquad 
\varepsilon_{j,t}^m\!=\!\frac{\sin (j\!-\!1)\tfrac{t\pi}m\sin (j\!+\!1)\tfrac{t\pi}m}{\sin^2\!\tfrac{t\pi}m}\quad
\end{align}
where the sums over dilogarithms relate to the braid and bulk limits in the strips labelled by the fusion level $j$. 
\label{Theta}
\end{subequations}
The standard Rogers dilogarithms are given by
\bea
L(x)\!=\!-\tfrac 12 \int_0^x \Big[\frac{\log(1\!-\!t)}{t}+\frac{\log t}{1\!-\!t}\Big] dt,\quad\ L_+(x)\!=\!\tfrac 12 \int_0^x \Big[\frac{\log(1\!+\!t)}{t}-\frac{\log t}{1\!+\!t}\Big] dt,
\quad\  0\le x\le 1 
\eea
For our purposes, we analytic continue the dilogarithm functions to the real line using
\bea
L_+(x)=\begin{cases}
L\big(\tfrac x{1+x}\big),&x\ge 0\\[2pt]
-L(-x),&x<0
\end{cases}\qquad\qquad L(x)=\begin{cases}
\tfrac{\pi^2}{3}-L(\tfrac 1x),& x>1\\[2pt]
L(\tfrac{1}{1-x})-\tfrac{\pi^2}{6},& x<0
\end{cases}\label{analytic}
\eea 
Notice that, as required, the formula for $c^{m,m'}_\text{eff}\!=1-\tfrac{6}{mm'}$ is in fact symmetric in $m$ and $m'$. 

For the analytic continuation~\cite{Kirillov94} of $L(x)$, we have
\bea
L_+(\varepsilon_j)=\begin{cases}
\;L(\tfrac{\varepsilon_j}{1+\varepsilon_j})\;
=L(1\!-\!\tfrac1{d_j^2})=\tfrac{\pi^2}6-L(\tfrac1{d_j^2})
=L(d_j^2)\!-\!\tfrac{\pi^2}6,\ \ & \varepsilon_j\ge 0\\[4pt]
-L(-\varepsilon_j)=-L(-d_{j-1}d_{j+1})=-L(1\!-\!d_j^2)=L(d_j^2)\!-\!\tfrac{\pi^2}6,& \varepsilon_j< 0\quad
\end{cases}
\eea
where we use (\ref{analytic}) and the identity $L(x)+L(1\!-\!x)=\tfrac{\pi^2}{6}$. This leads to formulas, equivalent to (\ref{Theta}), where $L_+(x)$ is replaced with $L(x)$ and squares of the quantum dimensions appear
\begin{subequations}
\begin{align}
&\qquad\qquad\qquad\ \qquad  c^{m,m'}_\text{eff}\!\!=\!\tfrac{1}{m'\!-m}\,\theta_{1,1}^{m,m'}-1\\[4pt]
&\quad\ \  \Delta_{r,s}^{m,m'}=\tfrac14(r\!-\!s)(r\!-\!s\!+\!m'\!\!-\!m)-\tfrac{m'\!-m}{24}\big[\theta_{r,s}^{m,m'}-\theta_{1,1}^{m,m'}\big]\\
&\theta_{r,s}^{m,m'}\!=\frac{6}{\pi^2}\Big[\sum_{j=1}^{m'\!-2}L((d_{j,s}^{m'})^2)-\sum_{j=1}^{m-2}L((d_{j,r}^{m})^2)\Big],\qquad 
d_{j,t}^{m}=\frac{\sin\tfrac{j t\pi}{m}}{\sin\tfrac{t\pi}{m}}
\end{align}
These identities precisely coincide with Corollary 3.8 of \cite{Kirillov94} after systematically making the replacement 
$L((d_{j,t}^{m})^2)=\tfrac{\pi^2}3-L({(d_{j,t}^{m})^{-2}})$.
\label{theta}
\end{subequations}

The results (\ref{Theta}) imply the master dilogarithm identities
\bea
1-\tfrac{6}{mm'}=\!\tfrac{1}{m'\!-m}\,\Theta_{1,1}^{m,m'},\qquad \Theta_{r,s}^{m,m'}\!-\Theta_{1,1}^{m,m'}=\tfrac{6(m'\!-m+mm'(r-s)+ms^2-m'r^2)}{mm'}
\eea
We call these master dilogarithm identities because they hold on the dense set $\tfrac{m'-m}{m'\!}\in (0,1)$. In fact, using the logarithmic limit of Rasmussen~\cite{RasLogLimit}, we find the further identities
\bea
c^{p,p'}_\text{log,eff}\!=\!\lim_{k\to\infty} c^{kp,kp'}_\text{eff}\!=\!1,\quad \Delta_{\text{log},r,s}^{p,p'}\!=\!\lim_{k\to\infty} \Delta^{kp,kp'}_{r,s}\!=\Delta_{r,s}^{p,p'}
\!=\!\frac{(rp'\!-\!sp)^2\!-\!(p'\!-\!p)^2}{4pp'},\quad r,s, k\in \mathbb{N}_{>0}\ \ 
\eea
for the effective central charges and infinitely extended set of conformal weights of the logarithmic minimal models ${\cal LM}(p,p')$~\cite{PRZ2006,PR2011,LogTY2014}. 
In the latter formula, $\Delta^{kp,kp'}_{r,s}$ 
stands for the dilogarithm terms on the RHS of (\ref{dilogdelta}) with $m=kp$, $m'=kp'$ and $p,p'$ coprime. The limit $k\to\infty$ is only needed to extend the range of $r$ and $s$.
Note that the RHS of (\ref{dilogdelta}) makes sense even when $m$ and $m'$ have a common factor $k$.

\subsection{Construction of integrable seams on the lattice}

Integrable seams for RSOS models on the lattice were first implemented in \cite{ChuiEtAl2001,ChuiEtAlOdyssey2001,ChuiEtAl2003} and studied more recently in \cite{BelleteteEtAl2023,TavaresEtAl2024,SYSRS24}. 
On the cylinder with symmetry algebra $\mbox{Vir}$, there are three relevant types of integrable seam, namely, (i)~Verlinde $\mathbf n_s$, (ii)~Pasquier $\widehat{\mathbf N}_a$ and (iii)~${\Bbb Z}_2$ automorphism seams $\boldsymbol\sigma$. In contrast, on the torus with symmetry algebra $\mbox{Vir}\otimes\mbox{Vir}$, there are additionally 
(iv) compound Ocneanu integrable seams $\widehat{\mathbf P}_{ab}=\widehat{\mathbf N}_a \overline{\widehat{\mathbf N}}_b$ where bars denote complex conjugation (left-right chiral conjugation). The Ocneanu integrable seams also exist on the cylinder but they can be reduced to Pasquier $\widehat{\mathbf N}_a$ type seams using the gluing condition (\ref{gluing}).

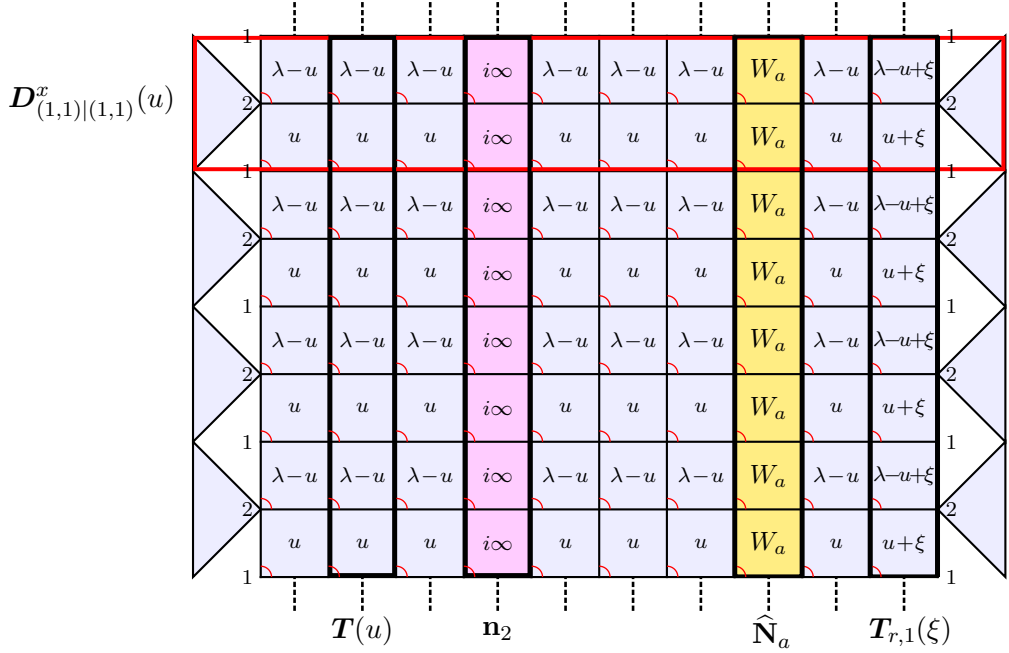
\begin{figure}[htb]
\begin{equation*}
\psset{unit=.9cm}
\begin{pspicture}[shift=-3.9](-2,-1)(8,7.75)
\multirput(0,0)(1,0){10}{\psline[linewidth=1.pt,linestyle=dashed,dash =2pt 1pt](-1.5,-1.5)(-1.5,7.5)}
\facegrid{(-2,-1)}{(8,7)}
\facegridp{(1,-1)}{(2,7)}
\multirput(0,0)(0,2){4}{\pspolygon[fillstyle=solid,fillcolor=lightlightblue](9,-1)(8,0)(9,1)}
\multirput(0,0)(0,2){4}{\pspolygon[fillstyle=solid,fillcolor=lightlightblue](-3,-1)(-2,0)(-3,1)}
\multirput(0,0)(0,2){4}{\multirput(0,0)(1,0){3}{\rput(-1.5,-.5){\scriptsize $u$}}}
\multirput(0,1)(0,2){4}{\multirput(0,0)(1,0){3}{\rput(-1.5,-.5){\scriptsize $\lambda\!-\!u$}}}
\multirput(4,0)(0,2){4}{\multirput(0,0)(1,0){5}{\rput(-1.5,-.5){\scriptsize $u$}}}
\multirput(4,1)(0,2){4}{\multirput(0,0)(1,0){5}{\rput(-1.5,-.5){\scriptsize $\lambda\!-\!u$}}}
\multirput(3,0)(0,1){8}{\multirput(0,0)(1,0){1}{\rput(-1.5,-.5){\scriptsize $i\infty$}}}
\multirput(4,0)(0,2){4}{\multirput(5,0)(1,0){1}{\rput(-1.5,-.5){\scriptsize $u\!+\!\xi$}}}
\multirput(4,0)(0,2){4}{\multirput(5,1)(1,0){1}{\rput(-1.5,-.5){\scriptsize $\lambda\!\!-\!\!u\!\!+\!\!\xi$}}}
\facegridy{(6,-1)}{(5,7)}
\psframe[fillstyle=none,linecolor=red,linewidth=1.5pt](-3,5)(9,7)
\psframe[fillstyle=none,linewidth=1.pt](5,-1)(6,7)
\psframe[fillstyle=none,linewidth=1.pt](7,-1)(8,7)
\psframe[fillstyle=none,linewidth=1.5pt](-1,-1)(0,7)
\psframe[fillstyle=none,linewidth=1.5pt](1,-1)(2,7)
\multirput(0,0)(0,1){8}{\multirput(0,0)(1,0){10}{\psarc[linewidth=0.5pt,linecolor=red]{-}(-2,-1){0.16}{0}{90}}}
\rput(-4.5,6){$\D_{(1,1)|(1,1)}^x(u)$}
\rput(-.5,-1.8){$\T(u)$}
\rput(7.6,-1.8){$\T_{\!r,1\,}\!(\xi)$}
\rput(1.5,-1.8){$\nn 2$}
\rput(5.55,-1.8){$\widehat{\mathbf N}_a$}
\multirput(1,.0)(0,1){8}{\rput(4.5,-.5){\small $W_a$}}
\psline[linewidth=1.75pt](5,-1)(5,7)
\psline[linewidth=1.75pt](6,-1)(6,7)
\psline[linewidth=1.75pt](7,-1)(7,7)
\psline[linewidth=1.75pt](8,-1)(8,7)
\multirput(-2.2,-1)(0,2){5}{\scriptsize $1$}
\multirput(-2.2,0)(0,2){4}{\scriptsize $2$}
\multirput(8.2,-1)(0,2){5}{\scriptsize $1$}
\multirput(8.2,0)(0,2){4}{\scriptsize $2$}
\end{pspicture}
\end{equation*}\\[0pt]
\caption{An $N\times M$ lattice on the cylinder with $(N,M)=(10,8)$ showing 
(i) the column/seam transfer matrices $\T(u)$, $\nn 2={\mathbf B}$, $\widehat{\mathbf N}_a$ and $\T_{r,1}(\xi)$ as explained in \cite{PRasmussen24,PRasmussen25} and 
(ii) the resulting double row transfer matrix $\D_{(1,1)|(1,1)}^x(u)$ (outlined in red) with $(1,2)$, $(1,a)$ and $(2,1)$ seam segments. 
The labels $W_a$ indicate that special face weights are assigned to the faces of the $(1,a)$ segment. 
The integrable seams commute with each other and $\T(u)$ so they can both be pushed to the left or right boundary.}
\label{TMs}
\end{figure}

The construction of various vertical integrable seams on the cylinder, as shown in Figure~\ref{TMs}, precisely coincides with the construction of Ocneanu integrable seams on the torus as explained in detail in \cite{PRasmussen24,PRasmussen25}.
We therefore forego giving further details of the construction here.
We observe, however, that the various integrable seams satisfy (i)~the Verlinde fusion algebra~\cite{Verlinde88}, (ii)~the Pasquier graph fusion algebra~\cite{PasquierThesis}, (iii)~the  internal symmetry (Dynkin graph) automorphism group and (iv) the Ocneanu graph fusion algebra~\cite{Ocneanu}. Among the \ade minimal models, only the $D_{2l}$ models require the ${\Bbb Z}_2$ automorphism to be added separately to the fusion algebra. 
Explicitly, we observe that the various integrable seams satisfy the Verlinde fusion algebra, the Pasquier graph fusion algebra and the Ocneanu graph fusion algebra for arbitrary systems sizes $M$
\begin{subequations}
\begin{align}
&\nn i\,\nn j=\sum_{k\in A_{m'\!-1}} N_{ij}{}^k\, \nn k,\quad 1\le i,j\le m'\!\!-\!1;\qquad \widehat{\mathbf N}_{a}\,\widehat{\mathbf N}_{b}=\sum_{c\in G} \widehat{N}_{ab}{}^c\, \widehat{\mathbf N}_{c},\quad a,b\in G\\[-8pt]
&\hspace{3cm}\widehat{{\mathbf P}}_\eta\, \widehat{{\mathbf P}}_{\mu}=\sum_{\nu=1}^{|\widetilde{G}|} \widetilde{N}_{\eta\mu}{}^\nu\, \widehat{{\mathbf P}}_{\nu},\quad 1\le \eta,\mu\le |\widetilde{G}|
\end{align}
\end{subequations}
where $\widetilde{G}$ is the Ocneanu graph and $\widehat{{\mathbf P}}_\eta=\widehat{{\mathbf P}}_{ab}$ denotes the Ocneanu seams. The integrable seams $\nn j$ are given as Chebyshev polynomials $\nn j=U_{j-1}(\tfrac12 \nn 2)$ of the second kind in the fundamental $\nn 2$. The integrable seams $\widehat{\mathbf N}_{a}$ are given as linear combinations of the $\nn j$. So, by construction, all these integrable seams commute with each other and with the transfer matrix $\vec T(u)$.

\subsection{Conformal defects as limits of integrable lattice seams}
\def\rrangle{\rangle\!\rangle}

In this section we first focus on $A$ type theories. The conformal defects ${\cal L}_{r,s}$ are then obtained as the continuum (braid and bulk) scaling limits of the finite lattice integrable seams $\vec T^{(r,s)}=\vec T^{(r,1)}\vec T^{(1,s)}$ as in Section~\ref{sec:rsSeams}. 

Since the integrable seams of Section~\ref{sec:rsSeams} are simultaneously  
diagonalizable, we conclude that the conformal $r$-type defects satisfy
\bea
{\cal L}_{(r,1)}^2={\cal I}+{\cal L}_{(r-1,1)}{\cal L}_{(r+1,1)},\quad\  {\cal L}_{(r,1)}{\cal L}_{(2,1)}={\cal L}_{(r-1,1)}+{\cal L}_{(r+1,1)},\quad\  {\cal L}_{(m-1,1)}{\cal L}_{(r,1)}={\cal L}_{(m-r,1)}
\eea
where ${\cal I}$ is the identity defect and the involution ${\cal L}_{(m-1,1)}$ is a height reversal operator.
Similarly, the continuum scaling limit of the integrable braid seams satisfy
\bea
 {\cal L}_{(1,s)}{\cal L}_{(1,2)}={\cal L}_{(1,s-1)}+{\cal L}_{(1,s+1)},\qquad {\cal L}_{(1,m'\!-1)}{\cal L}_{(1,s)}={\cal L}_{(1,m'-s)}
\eea
where the involution ${\cal L}_{(1,m'\!-1)}$ emerges from the height reversal operator. Taking the bulk and braid limits respectively of $\boldsymbol\sigma\,\vec T^{(1,1)}$ gives 
${\cal L}_{(m-1,1)}$ and ${\cal L}_{(1,m'\!-1)}$ with the product ${\cal L}_{(m-1,m'\!-1)}={\cal L}_{(m-1,1)}{\cal L}_{(1,m'\!-1)}={\cal I}$. So the Kac symmetry follows
\bea
{\cal L}_{(m-r,m'\!-s)}={\cal L}_{(m-r,1)}{\cal L}_{(1,m'\!-s)}={\cal L}_{(r,1)}{\cal L}_{(m-1,1)}{\cal L}_{(1,m'\!-1)}{\cal L}_{(1,s)}={\cal L}_{(r,1)}{\cal L}_{(1,s)}={\cal L}_{(r,s)}
\eea
These arguments follow the more detailed arguments found in \cite{STRSaleur2025}. 
Consequently, we deduce that the line defects ${\cal L}_\mu={\cal L}_{(r,s)}={\cal L}_{(r,1)}{\cal L}_{(1,s)}$ with $(r,s)\in {\mathbb K}$ satisfy the coset graph fusion algebra
\bea
{\cal L}_\mu{\cal L}_{\mu'}=\sum_{\mu''=1}^{|\mathbb K|}\widetilde{N}_{\mu\mu'}{}^{\mu''}{\cal L}_{\mu''}
\eea

The algebraic properties of the defect lines $\widehat{\cal L}_a$ and $\widehat{\cal L}_{(r,a)}$ follow from the existence and properties of rectangular intertwiners $C$~\cite{OcneanuInter,PZ93} satisfying
\bea
AC=CG
\eea
In particular, the $A_{11}$-$E_6$, $D_{10}$-$E_7$ and $A_{29}$-$E_8$ intertwiners are
\bea
E_{6,7,8}:\qquad  C=\ssmat{
 1 & 0 & 0 & 0 & 0 & 0 \\
 0 & 1 & 0 & 0 & 0 & 0 \\
 0 & 0 & 1 & 0 & 0 & 0 \\
 0 & 0 & 0 & 1 & 0 & 1 \\
 0 & 0 & 1 & 0 & 1 & 0 \\
 0 & 1 & 0 & 1 & 0 & 0 \\
 \hline\\[-8pt]
 1 & 0 & 1 & 0 & 0 & 0 \\
 0 & 1 & 0 & 0 & 0 & 1 \\
 0 & 0 & 1 & 0 & 0 & 0 \\
 0 & 0 & 0 & 1 & 0 & 0 \\
 0 & 0 & 0 & 0 & 1 & 0},\quad\ 
\ssmat{
 1 & 0 & 0 & 0 & 0 & 0 & 0 \\
 0 & 1 & 0 & 0 & 0 & 0 & 0 \\
 0 & 0 & 1 & 0 & 0 & 0 & 0 \\
 0 & 0 & 0 & 1 & 0 & 0 & 0 \\
 0 & 0 & 0 & 0 & 1 & 0 & 1 \\
 0 & 0 & 0 & 1 & 0 & 1 & 0 \\
 0 & 0 & 1 & 0 & 1 & 0 & 0 \\
  \hline\\[-8pt]
 0 & 1 & 0 & 1 & 0 & 0 & 0 \\
 1 & 0 & 0 & 0 & 0 & 0 & 1 \\
 0 & 0 & 1 & 0 & 0 & 0 & 0},\quad\ 
\ssmat{
 1 & 0 & 0 & 0 & 0 & 0 & 0 & 0 \\
 0 & 1 & 0 & 0 & 0 & 0 & 0 & 0 \\
 0 & 0 & 1 & 0 & 0 & 0 & 0 & 0 \\
 0 & 0 & 0 & 1 & 0 & 0 & 0 & 0 \\
 0 & 0 & 0 & 0 & 1 & 0 & 0 & 0 \\
 0 & 0 & 0 & 0 & 0 & 1 & 0 & 1 \\
 0 & 0 & 0 & 0 & 1 & 0 & 1 & 0 \\
 0 & 0 & 0 & 1 & 0 & 1 & 0 & 0 \\
  \hline\\[-8pt]
 0 & 0 & 1 & 0 & 1 & 0 & 0 & 0 \\
 0 & 1 & 0 & 1 & 0 & 0 & 0 & 1 \\
 1 & 0 & 1 & 0 & 1 & 0 & 0 & 0 \\
 0 & 1 & 0 & 1 & 0 & 1 & 0 & 0 \\
 0 & 0 & 1 & 0 & 1 & 0 & 1 & 0 \\
 0 & 0 & 0 & 1 & 0 & 1 & 0 & 1 \\
 0 & 0 & 0 & 0 & 2 & 0 & 0 & 0 \\
 0 & 0 & 0 & 1 & 0 & 1 & 0 & 1 \\
 0 & 0 & 1 & 0 & 1 & 0 & 1 & 0 \\
 0 & 1 & 0 & 1 & 0 & 1 & 0 & 0 \\
 1 & 0 & 1 & 0 & 1 & 0 & 0 & 0 \\
 0 & 1 & 0 & 1 & 0 & 0 & 0 & 1 \\
 0 & 0 & 1 & 0 & 1 & 0 & 0 & 0 \\
 0 & 0 & 0 & 1 & 0 & 1 & 0 & 0 \\
 0 & 0 & 0 & 0 & 1 & 0 & 1 & 0 \\
 0 & 0 & 0 & 0 & 0 & 1 & 0 & 1 \\
 0 & 0 & 0 & 0 & 1 & 0 & 0 & 0 \\
 0 & 0 & 0 & 1 & 0 & 0 & 0 & 0 \\
 0 & 0 & 1 & 0 & 0 & 0 & 0 & 0 \\
 0 & 1 & 0 & 0 & 0 & 0 & 0 & 0 \\
 1 & 0 & 0 & 0 & 0 & 0 & 0 & 0}
 \label{ECs}
\eea
For each exceptional $E_{6,7,8}$ lattice model, the rectangular intertwiner $C$ admits a (square) generalized left inverse $C^{-1}$ coinciding with the inverse of the top square block of $C$ shown in (\ref{ECs})
\bea
E_{6,7,8}:\quad\  C^{-1}=\ssmat{
 1 & 0 & 0 & 0 & 0 & 0 \\
 0 & 1 & 0 & 0 & 0 & 0 \\
 0 & 0 & 1 & 0 & 0 & 0 \\
 0 & -1 & 0 & 0 & 0 & 1 \\
 0 & 0 & -1 & 0 & 1 & 0 \\
 0 & 1 & 0 & 1 & 0 & -1},\quad\ 
\ssmat{
 1 & 0 & 0 & 0 & 0 & 0 & 0 \\
 0 & 1 & 0 & 0 & 0 & 0 & 0 \\
 0 & 0 & 1 & 0 & 0 & 0 & 0 \\
 0 & 0 & 0 & 1 & 0 & 0 & 0 \\
 0 & 0 & -1 & 0 & 0 & 0 & 1 \\
 0 & 0 & 0 & -1 & 0 & 1 & 0 \\
 0 & 0 & 1 & 0 & 1 & 0 & -1},
\quad\ 
\ssmat{
 1 & 0 & 0 & 0 & 0 & 0 & 0 & 0 \\
 0 & 1 & 0 & 0 & 0 & 0 & 0 & 0 \\
 0 & 0 & 1 & 0 & 0 & 0 & 0 & 0 \\
 0 & 0 & 0 & 1 & 0 & 0 & 0 & 0 \\
 0 & 0 & 0 & 0 & 1 & 0 & 0 & 0 \\
 0 & 0 & 0 & -1 & 0 & 0 & 0 & 1 \\
 0 & 0 & 0 & 0 & -1 & 0 & 1 & 0 \\
 0 & 0 & 0 & 1 & 0 & 1 & 0 & -1}
\eea
So, from (\ref{FusionAlgebras}), it follows that
\bea
n_s=\sum_{b\in G} C_{sb} \widehat{N}_b,\quad \widehat{N}_a=\sum_{s=1}^{L} C_{as}^{-1} n_s,\quad
\nn s=\sum_{b\in G} C_{sb} \widehat{\mathbf N}_b,\quad \widehat{\mathbf N}_{a}=\sum_{s=1}^{L} C_{as}^{-1} \nn s,
\qquad L=6,7,8
\eea
and, in the continuum scaling limit, the $a$-type $E$ defect lines are given by
\bea
\widehat{\cal L}_{a}=\sum_{s=1}^{L} C_{as}^{-1} {\cal L}_{(1,s)},\qquad \widehat{\cal L}_{(r,a)}=\sum_{s=1}^{L} C_{as}^{-1} {\cal L}_{(r,s)}
\eea
Since ${\cal L}_{(r,s)}$ satisfies the Verlinde algebra, it follows that the defect lines $\widehat{\cal L}_{a}$ and $\widehat{\cal L}_{(r,a)}$ satisfy the Pasquier graph fusion algebra and the coset graph fusion algebra respectively.

\subsection{Boundary conditions on the lattice}
The nonzero  $a$-type right-boundary weights $|a\rangle$ are constructed~\cite{BP2001} by fusing braid seams to the vacuum boundary
\begin{subequations}
\label{bdy}
\bea
\psset{unit=0.8cm}
\begin{pspicture}[shift=-1.2](0,-.3)(1.5,2.3)
\facegrid{(0,0)}{(1,2)}
\pspolygon[fillstyle=solid,fillcolor=lightlightblue](1.5,0)(1,1)(1.5,2)(1.5,0)
\multirput(0,0)(0,1){2}{\psline[linewidth=1.5pt,linecolor=blue](0,.5)(.4,.5)}
\multirput(0,0)(0,1){2}{\psline[linewidth=1.5pt,linecolor=blue](.6,.5)(1,.5)}
\psline[linewidth=1.5pt,linecolor=blue](.5,0)(.5,2)
\psbezier[linewidth=1.5pt,linecolor=blue](1,.5)(1.4,.6)(1.4,1.4)(1,1.5)
\end{pspicture}\;=\;
\begin{pspicture}[shift=-1.2](0,-.3)(1.5,2.3)
\facegrid{(0,0)}{(1,2)}
\pspolygon[fillstyle=solid,fillcolor=lightlightblue](1.5,0)(1,1)(1.5,2)(1.5,0)
\psarc[linewidth=1.5pt,linecolor=blue](0,1){.5}{-90}{90}
\psarc[linewidth=1.5pt,linecolor=blue](1,0){.5}{90}{180}
\psarc[linewidth=1.5pt,linecolor=blue](1,2){.5}{180}{270}
\psbezier[linewidth=1.5pt,linecolor=blue](1,.5)(1.4,.6)(1.4,1.4)(1,1.5)
\end{pspicture}\qquad\qquad
\begin{pspicture}[shift=-1.2](0,-.3)(1.5,2.3)
\facegrid{(0,0)}{(1,2)}
\pspolygon[fillstyle=solid,fillcolor=lightlightblue](1.5,0)(1,1)(1.5,2)(1.5,0)
\psarc[linewidth=0.025]{-}(0,0){0.16}{0}{90}
\multirput(.5,.5)(1,0){1}{\spos{c}{i\infty}}
\multirput(.5,1.5)(1,0){1}{\spos{c}{i\infty}}
\rput(-.2,1){\spos{c}{c}}
\rput(0,-.1){\spos{t}{a}}
\rput(1,-.1){\spos{t}{b}}
\rput(1.5,-.1){\spos{t}{b}}
\rput(0,2.1){\spos{b}{a'}}
\rput(1,2.1){\spos{b}{b}}
\rput(1.5,2.1){\spos{b}{b}}
\end{pspicture}\;=\delta_{a,a'}\ \ \;
\begin{pspicture}[shift=-1.2](0,-.3)(.5,2.3)
\pspolygon[fillstyle=solid,fillcolor=lightlightblue](.5,0)(0,1)(0.5,2)(0.5,0)
\rput(-.2,1){\spos{c}{c}}
\rput(.5,-.1){\spos{t}{a}}
\rput(.5,2.1){\spos{b}{a'}}
\end{pspicture}\qquad\qquad
\begin{pspicture}[shift=-1.2](0,-.3)(1.5,2.6)
\facegrid{(0,0)}{(2,2)}
\pspolygon[fillstyle=solid,fillcolor=lightlightblue](2.5,0)(2,1)(2.5,2)(2.5,0)
\psarc[linewidth=0.025]{-}(0,0){0.16}{0}{90}
\multirput(.5,.5)(1,0){2}{\spos{c}{i\infty}}
\multirput(.5,1.5)(1,0){2}{\spos{c}{i\infty}}
\rput(-.2,1){\spos{c}{c}}
\rput(0,-.13){\spos{t}{a}}
\rput(1,-.1){\spos{t}{b}}
\rput(2,-.13){\spos{t}{a}}
\rput(2.5,-.13){\spos{t}{a}}
\rput(0,2.1){\spos{b}{a}}
\rput(1,2.1){\spos{b}{b}}
\rput(2,2.1){\spos{b}{a}}
\rput(2.5,2.1){\spos{b}{a}}
\end{pspicture}\qquad\ =\ 
\begin{pspicture}[shift=-1.2](0,-.3)(.5,2.3)
\pspolygon[fillstyle=solid,fillcolor=lightlightblue](.5,0)(0,1)(0.5,2)(0.5,0)
\rput(-.2,1){\spos{c}{c}}
\rput(.5,-.1){\spos{t}{a}}
\rput(.5,2.1){\spos{b}{a}}
\end{pspicture}\label{bdya}\qquad\quad\\[8pt]
\psset{unit=0.8cm}
|1\rangle=\ \ 
\begin{pspicture}[shift=-1.2](0,-.3)(.5,2.3)
\pspolygon[fillstyle=solid,fillcolor=lightlightblue](.5,0)(0,1)(0.5,2)(0.5,0)
\rput(-.2,1){\spos{c}{2}}
\rput(.5,-.1){\spos{t}{1}}
\rput(.5,2.1){\spos{b}{1}}
\end{pspicture}\; = \sqrt{\frac{\psi_2}{\psi_1}},\quad\ \ 
|a\rangle=\ \;
\begin{pspicture}[shift=-1.2](0,-.3)(.5,2.3)
\pspolygon[fillstyle=solid,fillcolor=lightlightblue](.5,0)(0,1)(0.5,2)(0.5,0)
\rput(-.2,1){\spos{c}{b}}
\rput(.5,-.1){\spos{t}{a}}
\rput(.5,2.1){\spos{b}{a}}
\end{pspicture}
\ =\ \;
\begin{pspicture}[shift=-1.2](0,-.3)(5.5,2.3)
\facegrid{(0,0)}{(5,2)}
\pspolygon[fillstyle=solid,fillcolor=lightlightblue](5.5,0)(5,1)(5.5,2)(5.5,0)
\psarc[linewidth=0.025]{-}(0,0){0.16}{0}{90}
\multirput(.5,.5)(1,0){5}{\spos{c}{i\infty}}
\multirput(.5,1.5)(1,0){5}{\spos{c}{i\infty}}
\rput(-.2,1){\spos{c}{b}}
\rput(0,-.25){\spos{b}{a}}
\rput(1,-.33){\spos{b}{a_{n-1}}}
\rput(2,-.3){\spos{b}{\cdots}}
\rput(3,-.3){\spos{b}{a_3}}
\rput(4,-.3){\spos{b}{2}}
\rput(5,-.3){\spos{b}{1}}
\rput(0,2.175){\spos{b}{a}}
\rput(1,2.1){\spos{b}{a_{n-1}}}
\rput(2,2.1){\spos{b}{\cdots}}
\rput(3,2.1){\spos{b}{a_3}}
\rput(4,2.1){\spos{b}{2}}
\rput(5,2.1){\spos{b}{1}}
\rput(5.5,-.1){\spos{t}{1}}
\rput(5.5,2.1){\spos{b}{1}}
\end{pspicture}\;=
\sqrt{\frac{\psi_{b}}{\psi_a}},\qquad G_{ab}=1\label{bdyb}
\eea
\end{subequations}
Internal heights are summed out. From (\ref{bdya}), it follows that the boundary weights $|a\rangle$ are independent of the choice of path $(a,a_{n-1},\ldots,a_3,2,1)$ from $a$ to $1$ on $G$ and that the upper and lower paths are identical and can be chosen to be 
the shortest path on $G$ from $a$ to 1. 
The 2-column braid matrix $\widehat B$, as in (\ref{Bhat}), is the $N=2$ periodic braid seam acting on the Hilbert space spanned by the cyclic paths of length 2
\bea
{\cal H}_B=\mbox{span}(\{(1,2,1),(2,1,2),(2,3,2),\ldots\}),\qquad \mbox{dim}{\cal H_B}=2\,(\#\,\mbox{edges of $G$})
\eea 
Consequently, for type I theories, it satisfies~\cite{PRasmussen24,PRasmussen25} the Pasquier graph fusion algebra
\bea
\psset{unit=0.8cm}
\widehat B_1=I,\ \ \ \widehat B_2=\widehat B,\ \ \ \widehat B\widehat B_b=\sum_{c\in G} G_{bc} \widehat B_c,
\ \ \ \widehat B_a\widehat B_b=\sum_{c\in G} \widehat{N}_{ab}{}^c \widehat B_c,\quad 
\widehat B_{\,(a_1,a_2,a_1)}^{\,(b_1,b_2,b_1)}=\quad\begin{pspicture}[shift=-1.2](0,-.3)(1.5,2.3)
\facegrid{(0,0)}{(1,2)}
\psarc[linewidth=0.025]{-}(0,0){0.16}{0}{90}
\multirput(.5,.5)(1,0){1}{\spos{c}{i\infty}}
\multirput(.5,1.5)(1,0){1}{\spos{c}{i\infty}}
\rput(0,2.1){\spos{b}{a_1}}
\rput(-.2,1){\spos{c}{a_2}}
\rput(0,-.1){\spos{t}{a_1}}
\rput(1,-.1){\spos{t}{b_1}}
\rput(1.2,1){\spos{c}{b_2}}
\rput(1,2.1){\spos{b}{b_1}}
\end{pspicture}\label{Bhat}
\eea
Setting $|a\rangle=\widehat B_a|1\rangle$ and acting on the vacuum $|1\rangle$, it follows that 
\bea
\widehat B |b\rangle\;=\;\sum_{c\in G} G_{bc} |c\rangle,\qquad \widehat B_a|b\rangle\;=\;\sum_{c\in G} \widehat{N}_{ab}{}^c |c\rangle 
\eea

More generally, for all type I and II \ade lattice models, the nonzero right $s$-type boundary weights $|s\rrangle=B_s|1\rrangle$ are defined recursively by
\begin{subequations}
\bea
B_1=I,\ \ \ B_2=B=\widehat{B},\ \ \ B B_s=B_{s-1}+B_{s+1},\quad B_s B_{s'}=\sum_{s''\in A_{m'\!-1}} \!\! N_{ss'}{}^{s''} B_{s''}\\
\psset{unit=0.8cm}
|1\rrangle=|1\rangle,\ \ |2\rrangle=|2\rangle,\ \ B|s\rrangle\;
=\ |s\!-\!1\rrangle+|s\!+\!1\rrangle,\quad B_s|s'\rrangle=\sum_{s''\in A_{m'\!-1}} \!\! N_{ss'}{}^{s''} |s''\rrangle
\eea
\end{subequations}
These are the relations of the Verlinde fusion algebra with $B_s=U_{s-1}(\tfrac 12 B)$ where $U_n(z)$ are the Chebyshev polynomials of the second kind. 
The solution to this recursion is
\bea
|s\rrangle= \sum_{a\in G} C_{sa}  |a\rangle,\qquad C_{sa}=n_{s1}{}^a,\qquad s=1,2,\ldots,|G|\label{sbdy}
\eea
where $C_{sa}$ are the entries of the rectangular fundamental intertwiner $C$~\cite{OcneanuInter,PZ93}.
For the $A_L$ lattice models, $B_L$ is the $\mathbb{Z}_2$ automorphism $B_L |s\rrangle=|L\!+\!1\!-\!s\rrangle$ and the $s$-type boundary conditions are
\bea
\{|s\rrangle: s\in A_L\}
\eea
For the $D_L$ lattice models, the $s$-type boundary conditions are
\bea
\{|1\rangle, |2\rangle, \ldots,|L\!-\!2\rangle, |L\!-\!1\rangle\!+\!|L\rangle,|L\!-\!2\rangle,\ldots ,|2\rangle,|1\rangle \}
\eea
If $L$ is even, then $B_s=B_{2L-2-s}$ and the $\mathbb{Z}_2$ automorphism needs to be added separately. If $L$ is odd, then $\sigma=B_{2L-3}$ is the $\mathbb{Z}_2$ automorphism with the action to interchange the nodes $L\!-\!1$ and $L$ and $B_s=\sigma B_{2L-2-s}$. 
For the $E_6$, $E_7$ and $E_8$ lattice models, the $s$-type boundary conditions $|s\rrangle$ are all given by (\ref{sbdy}) but, for $E_7$, $|G|$ is replaced with $10$.
Actually, (\ref{sbdy}) can be inverted giving
\bea
E_{6,7,8}:\qquad|a\rangle=\sum_{s=1}^L C^{-1}_{as} |s\rrangle,\qquad L=6,7,8
\eea
where $C^{-1}$ is the inverse of the top $L\times L$ square block of $C$ in (\ref{ECs}).
Alternatively, the $s$-type boundary conditions can be constructed directly using the Wenzl-Jones projectors~\cite{Wenzl87,Jones97}.

Actually, the vertical integrable seams can be taken to be general Ocneanu seams~\cite{PRasmussen24,PRasmussen25} with independent commuting left- and right-chiral components $B_s$ and $\overline{B}_{s'}$ (or $\widehat{B}_a$ and $\overline{\widehat{B}}_b$ in type I cases) related by complex conjugation. However, acting on the right boundary with $B_s$ or $\overline{B}_s$ produces the same real boundary $|s\rrangle$. This is the ``gluing condition'' 
\bea
(B_s-\overline{B}_s)|1\rrangle=0\label{gluing}
\eea
So any $\overline{B}_s$ seam can, without loss of generality, be replaced with a $B_s$ seam. 
Since this can be lifted to the action of the integrable defect seam ${\mathbf B}_s$ acting on the vacuum state, it guarantees that there is only one copy of $\mbox{Vir}$ on the cylinder in the continuum scaling limit. 
The consequent reduction of the torus Ocneanu graph fusion algebras are consistent with the graph fusion algebras on the cylinder.

\subsection{Boundary conditions and fusion rules for double row transfer matrices}

In this section we fix $r=r'=1$. The 1-parameter family of commuting double row transfer matrices (\ref{D}) are normal and so they are simultaneously unitarily similar to diagonal matrices which we denote by 
$\diagD_{(1,b)|(1,c)}^{(N)}(u)$. 
The fusion rules for the finite-size $(1,s')\times (1,s'')$ and $(1,b)\times(1,c)$ double row transfer matrices are then respectively
\begin{subequations}
\label{DuFusionRules}
\bea
\D_{(1,s')|(1,s'')}^{(N)}(u)\ \sim
\, \disp\bigoplus_{s}\, N_{ss'}{}^{s''}\, \D_{(1,1)|(1,s)}^{(N)}(u)\sim
\, \bigoplus_{s}\, N_{ss'}{}^{s''}\, \diagD_{(1,1)|(1,s)}^{(N)}(u),\quad \mbox{$G$ is type I or II}\quad\\
\D_{(1,b)|(1,c)}^{(N)}(u)\ \sim\begin{cases}
\, \disp\bigoplus_{s}\, n_{sb}{}^{c}\, \D_{(1,1)|(1,s)}^{(N)}(u)\sim
\, \bigoplus_{s}\, n_{sb}{}^{c}\, \diagD_{(1,1)|(1,s)}^{(N)}(u),\qquad\ \  \mbox{$G$ is type I or II}\\[14pt]
\, \disp\bigoplus_{a\in G}\, \widehat{N}_{ab}{}^{c}\, \D_{(1,1)|(1,a)}^{(N)}(u)\sim
\, \bigoplus_{a\in G}\, \widehat{N}_{ab}{}^{c}\, \diagD_{(1,1)|(1,a)}^{(N)}(u),\qquad \mbox{$G$ is type I}
\end{cases}
\eea
\end{subequations}
where $\sim$ denotes equivalence up to similarity transformations.

The direct sum decompositions are a mathematical consequence of the following facts:
\begin{enumerate}
\item[(i)] the boundary states are created by acting with the integrable defect seams on the vacuum boundary states on the left and right
\bea
\langle (1,a)|=\langle(1,1)| \widehat{\mathbf N}_a,\qquad |(1,b)\rangle=\widehat{\mathbf N}_{b} |(1,1)\rangle
\eea
For the vacuum state on the lattice, the heights along the left and right edges alternate between heights 1 and  2. 
\item[(ii)] The seams $\widehat{\mathbf N}_a$ are topological, propagate freely and commute with each other.
\item[(iii)] The integrable seams $\widehat{\mathbf N}_a=\widehat{\mathbf N}_{(1,a)}$ satisfy the graph algebra fusion rules. 
\end{enumerate}
This means that, as in Figure~\ref{PartitionFunctDecomp}, the left seam can propagate from the left boundary to the right boundary where it is decomposed according to the fusion rules for seams. Essentially, this implements the local operator product expansion.

For $A$ type theories, the spectra of the double row transfer matrices in the $(s,s')$ sector is given by finitized Virasoro characters
\bea
Z^{(N)}_{s,s'}(q)=\mathop{\mbox{Tr}}\D_{(1,s)|(1,s')}^{(N)}(u)=\hspace{-.5cm}
\sum_{{s'\!'=|s-s'|+1}\atop s+s'+s'\!'=1\text{\ mod 2}}^{s_\text{max}}\hspace{-.5cm} Z^{(N)}_s(q),\qquad 
Z^{(N)}_s(q)= \mathop{\mbox{Tr}} \D_{(1,1)|(1,s)}^{(N)}(u)=\chi_{1,s}^{(N)}(q)
\eea
where the modular parameter is $q=\exp(-\tfrac{2\pi}{N} \sin \tfrac{\pi u}{\lambda})$. 
Taking the braid limit in (\ref{DuFusionRules}), it also follows that
\bea
\D_{(1,s)|(1,s')}^{(N)}(\pm i\infty)\ \sim\!\!\!\! \bigoplus_{{s'\!'=|s-s'|+1}\atop s+s'+s'\!'=1\text{\ mod 2}}^{s_\text{max}} \D_{(1,1)|(1,s'\!')}^{(N)}(\pm i\infty)\sim
\!\!\!\! \bigoplus_{{s'\!'=|s-s'|+1}\atop s+s'+s'\!'=1\text{\ mod 2}}^{s_\text{max}} \diagD_{(1,1)|(1,s'\!')}^{(N)}(\pm i\infty)
\eea
where the diagonal matrices $\diagD_{(1,1)|(1,s'\!')}^{(N)}(\pm i\infty)$ are multiples of the identity with the single eigenvalue $2\cos s''\lambda$.

The proof of these direct sum decompositions (\ref{DuFusionRules}) follows from properties of the boundary weights presented in the previous subsection. Since {\color{red} $r$ is not a good quantum number on a finite lattice}, there is no simple correspondence between $r$-type integrable seams on the lattice and the topological $r$-type conformal defects that emerge in the continuum scaling limit so these simple arguments break down for $r$-type boundary conditions.

\section{Conclusion}

In this paper we considered the unitary and nonunitary $(A,G)$ coset minimal CFTs with $G=A,D,E$. For these theories, we defined coset graph fusion algebras based on universal coset graphs $A\otimes G/\mathbb{Z}_2$. Representations of this algebra include (i) the fusion matrices (nimreps), (ii) the coset quantum dimensions and (iii) the topological defects of the associated CFT. 
Indeed, we argued that much of the CFT data is encoded by these coset graphs and the associated generalized quantum dimensions $d^m_{j,r}$ and $d^{m'}_{j,s}$. More specifically, the coset fusion graphs were shown to encode (i) the fusion matrices (nimreps), (ii) the Affleck-Ludwig boundary and defect $\g$-factors and entropies, (iii) the relative Symmetry Resolved Affleck-Ludwig Entropies (SRALEs) and (iv) the central charges and conformal weights through analytically continued dilogarithm functions of the factorized quantum dimensions. 
We presented numerous prototypical examples applying these considerations to unitary and nonunitary cases for both diagonal and non-diagonal theories.

Separately, working on the cylinder, we constructed Yang-Baxter integrable seams on the lattice as mutually commuting column transfer matrices and argued that, in the continuum scaling limit, these produce the various kinds of topological line defects $\widehat{\cal L}_{(r,a)}={\cal L}_{(r,1)}\widehat{\cal L}_{(1,a)}$ of the associated minimal CFTs satisfying either the Verlinde, graph fusion or Ocneanu graph fusion algebras. Due to a lattice ``gluing condition'', the action of the integrable seam $\overline{\mathbf{B}}_s$ on the vacuum state reproduces $\mathbf{B}_s$ thus ensuring there is just a single copy of Vir on the cylinder in the continuum scaling limit. Lastly, we observed throughout that, remarkably due to integrability, many of the known CFT discrete structures already exist at the level of the lattice model.

\subsection*{Acknowledgments}

This paper is dedicated in memory of the life and beautiful works of Rodney Baxter. 
We thank J{\o}rgen Rasmussen and Andreas Kl\"umper for helpful discussions. The research of JH was supported by an Australian Government Research Training Program (RTP) Scholarship. 

\appendix

\section{Coset graphs and coset nimreps}
\label{App:Coset}

Coset graphs are constructed mathematically by combining the standard constructs of tensor product graphs and quotient graphs from graph theory (see for example \cite{AlgGraphTheory}) as explained in this appendix. 
Recall that a graph $G$ is defined as a set of nodes $V$ with a set of edges $E$ consisting of a set of pairs of nodes of $G$ specifying node adjacency. We assume doubly-directed edges (also called undirected edges) and allow for loops such that a node forms an edge or is connected to itself. A graph is uniquely represented by its adjacency matrix.

Given two graphs $G$ and $G'$, the nodes of the tensor product (or Kronecker product) graph $G\otimes G'$ are the pairs $(a,a')$ where $a\in G$ and $a'\in G'$. 
There is an edge between the nodes $(a,a')$ and $(b,b')$ on $G\otimes G'$ if $(a,b)$ is an edge of $G$ and $(a'\!,b')$ is an edge of $G'$. 
The adjacency matrix of $G\otimes G'$ is just the tensor product of the adjacency matrices of $G$ and $G'$. 
The tensor product of matrices is associative. It is also commutative up to relabelling, that is, a simultaneous permutation of the rows and columns.

Given an equivalence relation $a\equiv b$ between the nodes of a graph $G$, the graph can be partitioned into equivalence classes. Recall that nodes $a$ and $b$ belong to the same equivalence class $C_j\subset G$ if and only if $a\equiv b$ with $\mathop{\cup}_j C_j=G$. The nodes of the quotient graph $Q:=G/R$, where $R$ is the equivalence operation, are the equivalences classes $C_j$. Two equivalence classes $C_j$ and $C_k$ are adjacent in $Q$ if some node in $C_j$ is adjacent to some node in $C_k$ on the original graph $G$. This process effectively ``glues together'' sets of nodes and edges from the original graph thereby simplifying its structure by representing groups of the original nodes with single nodes in the quotient graph. A simple example is the quotient of $A_{m-1}$ with $m$ odd under the $\mathbb{Z}_2$ equivalence relation $r\equiv m\!-\!r$. This quotient folds the $A_{m-1}$ diagram to yield $A_{m-1}/\mathbb{Z}_2=T_{(m\!-\!1)/2}$. Note that the folding of $A$ diagrams to form $T$ diagrams leads to the loops of the tadpole diagrams.

In the context of our $(A,G)$ coset graphs, we first take the tensor product $A\otimes G$ where $A$ and $G$ are restricted to the Dynkin diagrams with the standard labelling of the nodes as in Figure~\ref{fig:Graphs}. 
The coset graph $A\otimes G/\mathbb{Z}_2$ is subsequently constructed by taking the quotient of $A\otimes G$ with respect to the $\mathbb{Z}_2$ equivalence under the Kac table symmetry
\begin{align}
&(r,s)\equiv (m\!-\!r,m'\!\!-\!s),\quad G=A\\
&(r,a)\equiv (m\!-\!r,a),\qquad\ \ \, G\neq A
\end{align}
Importantly, we find that the coset graphs always factorize in terms of tadpoles
\bea
A\otimes G/\mathbb{Z}_2=\begin{cases} 
A_{m-1}\otimes A_{m'-1}/\mathbb{Z}_2=T_{(m-1)/2}\otimes A_{m'\!-\!1},\quad&\mbox{$G=A$, $m$ odd}\\[4pt]
A_{m-1}\otimes A_{m'-1}/\mathbb{Z}_2=A_{m-1}\otimes T_{(m'\!-\!1)/2},&\mbox{$G=A$, $m'$ odd}\\[4pt]
A_{m-1}\otimes G/\mathbb{Z}_2=T_{(m-1)/2}\otimes G,&\mbox{$G=D$ or $E$, $m$ odd}
\end{cases}
\eea
where this list exhausts all distinct cases.

The \ade diagrams in Figure~\ref{fig:Graphs} have distinguished nodes corresponding to the identity and fundamental labelled by $a=1,2$ respectively. The coset identity node of the $(A,G)$ coset graphs is $(r,a)=(1,1)$. The coset fundamental is the unique coset node adjacent to $(1,1)$. This is $(r,a)=(2,2)$ if $m>3$ and $(r,a)=(1,2)$ otherwise. In the latter case with $m\leq 3$, the coset graph is a linear graph, namely, $A_{m'-1}$ ($m=3$) or $T_{(m'-1)/2}$ ($m=2$). It follows that the coset graph fusion matrices (nimreps) relevant to the cylinder are
\begin{align}
\tilde{N}_{r,s}&=\begin{cases}
N^{(T_{\frac12\!(m\!-\!1)})}_r\!\otimes N^{(A_{m'\!-\!1})}_s,&\quad\mbox{$(A_{m\!-\!1},A_{m'\!-\!1)}$, $m$ odd}\\[2pt]
N^{(A_{m\!-\!1})}_r\!\otimes N^{(T_{\frac12\!(m'\!-\!1)})}_s,&\quad\mbox{$(A_{m\!-\!1},A_{m'\!-\!1)}$, $m'$ odd}\\[2pt]
N^{(T_{\frac12\!(m\!-\!1)})}_r\!\otimes n^{(G)}_s,&\quad\mbox{$(A_{m\!-\!1},G)$, $G$ is type I or II, $m$ odd}
\end{cases}\\
\tilde{N}_{r,a}&\,=\,
N^{(T_{\frac12\!(m\!-\!1)})}_r\!\otimes \widehat{N}^{(G)}_a,\qquad\qquad\ \mbox{$(A_{m\!-\!1},G)$, $G$ is type I, $m$ odd}
\end{align}

\begin{figure}[htb]
\begin{center}
\psset{unit=0.8cm}
\qquad\qquad\begin{pspicture}[shift=-.40](0,0)(4,7.25)
\psbezier[linewidth=1pt,linestyle=dashed,dash= 2pt 2pt](1.2,6.4)(3.27,7)(5.34,7)(7.4,6.4)
\rput(0,-2){\psbezier[linewidth=1pt,linestyle=dashed,dash= 2pt 2pt](1.2,6.4)(3.27,7)(5.34,7)(7.4,6.4)}
\rput(0,-4){\psbezier[linewidth=1pt,linestyle=dashed,dash= 2pt 2pt](1.2,6.4)(3.27,7)(5.34,7)(7.4,6.4)}
\rput(-1.6,-1.4){\psbezier[linewidth=1pt,linestyle=dashed,dash= 2pt 2pt](1.2,6.4)(3.27,7)(5.34,7)(7.4,6.4)}
\rput(-1.6,-3.4){\psbezier[linewidth=1pt,linestyle=dashed,dash= 2pt 2pt](1.2,6.4)(3.27,7)(5.34,7)(7.4,6.4)}
\rput(-1.6,-5.4){\psbezier[linewidth=1pt,linestyle=dashed,dash= 2pt 2pt](1.2,6.4)(3.27,7)(5.34,7)(7.4,6.4)}
\rput(2.3,-1.4){\psbezier[linewidth=1pt,linestyle=dashed,dash= 2pt 2pt](1.2,6.4)(3.27,7)(5.34,7)(7.4,6.4)}
\rput(2.3,-3.4){\psbezier[linewidth=1pt,linestyle=dashed,dash= 2pt 2pt](1.2,6.4)(3.27,7)(5.34,7)(7.4,6.4)}
\rput(2.3,-5.4){\psbezier[linewidth=1pt,linestyle=dashed,dash= 2pt 2pt](1.2,6.4)(3.27,7)(5.34,7)(7.4,6.4)}
\rput(.6,-2.8){\psbezier[linewidth=1pt,linestyle=dashed,dash= 2pt 2pt](1.2,6.4)(3.27,7)(5.34,7)(7.4,6.4)}
\rput(.6,-4.8){\psbezier[linewidth=1pt,linestyle=dashed,dash= 2pt 2pt](1.2,6.4)(3.27,7)(5.34,7)(7.4,6.4)}
\rput(.6,-6.8){\psbezier[linewidth=1pt,linestyle=dashed,dash= 2pt 2pt](1.2,6.4)(3.27,7)(5.34,7)(7.4,6.4)}
\psline[linewidth=1.pt,linecolor=red](-.5,1)(-.5,5)
\multirput(0,0)(0,-2){3}{\psline[linewidth=1.pt,linecolor=red](-.5,5)(1.2,6.4)}
\psline[linewidth=1.pt,linecolor=blue](3.5,1)(3.5,5)
\multirput(0,0)(0,-2){3}{\psline[linewidth=1.pt,linecolor=blue](3.5,5)(1.2,6.4)}
\psline[linewidth=1.pt,linecolor=blue](1.77,-.5)(1.77,3.5)
\psline[linewidth=1.pt,linecolor=red](1.83,-.5)(1.83,3.5)
\multirput(0,0)(0,-2){3}{\psline[linewidth=1.pt,linecolor=red](3.5,5)(1.8,3.6)}
\multirput(0,0)(0,-2){3}{\psline[linewidth=1.pt,linecolor=blue](-.5,5)(1.8,3.6)}
\multirput(-.5,1)(0,2){3}{\pscircle[linewidth=.5pt,fillstyle=solid,fillcolor=black](0,0){.13}}
\multirput(1.2,2.4)(0,2){3}{\pscircle[linewidth=.5pt,fillstyle=solid,fillcolor=black](0,0){.13}}
\multirput(3.5,1)(0,2){3}{\pscircle[linewidth=.5pt,fillstyle=solid,fillcolor=lightyellow](0,0){.13}}
\multirput(1.8,-.4)(0,2){3}{\pscircle[linewidth=.5pt,fillstyle=solid,fillcolor=lightyellow](0,0){.13}}
\rput(1.3,6.8){$1$}
\rput(-.9,5){$2$}
\rput(-.9,3){$3$}
\rput(-.9,1){$4$}
\rput(1.3,2.8){$5$}
\rput(1.3,4.8){$6$}
\rput(4.,5){$7$}
\rput(2.2,3.3){$8$}
\rput(2.2,1.3){$9$}
\rput(2.2,-.7){$10$}
\rput(4.,1){$11$}
\rput(4.,3){$12$}
\end{pspicture}\hspace{1.6cm}
\raisebox{0cm}{\begin{pspicture}[shift=-.40](0,0)(4,7.25)
\psline[linewidth=1.pt,linecolor=red](-.5,1)(-.5,5)
\multirput(0,0)(0,-2){3}{\psline[linewidth=1.pt,linecolor=red](-.5,5)(1.2,6.4)}
\psline[linewidth=1.pt,linecolor=blue](3.5,1)(3.5,5)
\multirput(0,0)(0,-2){3}{\psline[linewidth=1.pt,linecolor=blue](3.5,5)(1.2,6.4)}
\psline[linewidth=1.pt,linecolor=blue](1.77,-.5)(1.77,3.5)
\psline[linewidth=1.pt,linecolor=red](1.83,-.5)(1.83,3.5)
\multirput(0,0)(0,-2){3}{\psline[linewidth=1.pt,linecolor=red](3.5,5)(1.8,3.6)}
\multirput(0,0)(0,-2){3}{\psline[linewidth=1.pt,linecolor=blue](-.5,5)(1.8,3.6)}
\multirput(-.5,1)(0,2){3}{\pscircle[linewidth=.5pt,fillstyle=solid,fillcolor=black](0,0){.13}}
\multirput(1.2,2.4)(0,2){3}{\pscircle[linewidth=.5pt,fillstyle=solid,fillcolor=black](0,0){.13}}
\multirput(-.5,1)(0,2){3}{\pscircle[linewidth=.75pt,linecolor=black](0,.25){.25}}
\multirput(1.2,2.4)(0,2){3}{\pscircle[linewidth=.75pt,linecolor=black](0,.25){.25}}
\multirput(3.5,1)(0,2){3}{\pscircle[linewidth=.75pt,linecolor=black](0,.25){.25}}
\multirput(1.8,-.4)(0,2){3}{\pscircle[linewidth=.75pt,linecolor=black](0,.25){.25}}
\multirput(3.5,1)(0,2){3}{\pscircle[linewidth=.5pt,fillstyle=solid,fillcolor=lightyellow](0,0){.13}}
\multirput(1.8,-.4)(0,2){3}{\pscircle[linewidth=.5pt,fillstyle=solid,fillcolor=lightyellow](0,0){.13}}
\rput(1.3,6.8){\small $13$}
\rput(1.3,4.8){\small $18$}
\rput(1.3,2.8){\small $17$}
\rput(-1.,5){\small $14$}
\rput(-1.,3){\small $15$}
\rput(-1.,1){\small $16$}
\rput(2.2,3.4){\small $20$}
\rput(2.2,1.4){\small $21$}
\rput(2.2,-.6){\small $22$}
\rput(4.,5){\small $19$}
\rput(4.,3){\small $24$}
\rput(4.,1){\small $23$}
\end{pspicture}}
\vspace{.3cm}
\end{center}
\caption{The coset Ocneanu fusion graph is given by the Cartesian product $A_4\times \mbox{Oc}(E_6)/\mathbb{Z}_2=T_2\times \mbox{Oc}(E_6)$ with 24 vertices $\{\widehat{\mathbf{P}}_\mu\}_{\mu=1}^{24}=\{\widehat{\mathbf{P}}_{a,b},\boldsymbol{\tau}\widehat{\mathbf{P}}_{a,b}\}_{a\in E_6;b=1,2}$ where $\boldsymbol{\tau}\widehat{\mathbf{P}}_\mu=\widehat{\mathbf{P}}_{\mu+12}$ for $\mu=1,2,\ldots,12$. 
The solid red/blue lines show the action of the Ocneanu fundamentals $\widehat{\mathbf{P}}_{2,1}=\widehat{\mathbf{P}}_2$ and $\widehat{\mathbf{P}}_{1,2}=\widehat{\mathbf{P}}_7$. 
The dashed black lines and solid black loops show the action of $r$-type fundamental $\boldsymbol{\tau}=\mathbf{N}_2=\widehat{\mathbf{P}}_{13}$ satisfying 
$\boldsymbol{\tau}^2=I+\boldsymbol{\tau}$ with $\boldsymbol{\tau}=T_2\otimes I$.
 \label{E6CosetOcneanuGraph}}
\end{figure}
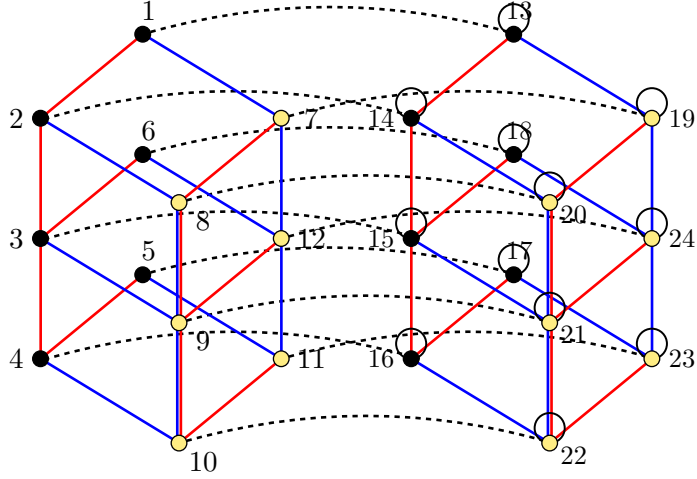

On the torus, there are left- and right-chiral copies of the Virasoro algebra and the coset graph fusion matrices (nimreps) are replaced by coset Ocneanu graph fusion matrices. For $(A,A)$ cases, the coset Ocneanu graph fusion matrices coincide with the cylinder coset graph fusion matrices. For the $(A,G)$ cases, the coset Ocneanu nimreps are
\bea
\tilde{P}_{r,\eta}=N^{(T_{\frac12\!(m\!-\!1)})}_r\!\otimes \widehat{P}^{(G)}_\eta,\qquad\qquad\ \mbox{$(A_{m\!-\!1},G)$, $G$ is type I, $m$ odd}
\eea
where $\widehat{P}^{(G)}_\eta$ are the known nimreps associated for the \ade Ocneanu graphs. As an example, the coset Ocneanu fusion graph of $(A_4,E_6)$ is shown in Figure~\ref{E6CosetOcneanuGraph}.

\section{Yang-Baxter equation for critical \ade models}
\label{ADEYBE}

\def\facer#1#2#3#4#5{\begin{pspicture}[shift=-.97](-.2,-.6)(1.6,1.6)
\psline[linecolor=blue,linewidth=1.5pt](.35,-.6)(.35,1.6)
\psline[linecolor=blue,linewidth=1.5pt](1.05,-.6)(1.05,1.6)
\rput{45}(0.707,-.2){\facegrid{(0,0)}{(1,1)}
\psarc[linecolor=red](0,0){.15}{0}{90}
}
\rput(.707,.507){\scriptsize $#5$}
\rput(.707,-.3){\scriptsize $#1$}
\rput(1.507,0.507){\scriptsize $#2$}
\rput(0.707,1.314){\scriptsize $#3$}
\rput(-0.1,.507){\scriptsize $#4$}
\rput(0.35,-0.75){\scriptsize$j$}
\rput(1.05,-0.75){\scriptsize$j\!+\!1$}
\end{pspicture}}

\def\Idr#1#2#3#4{
\begin{pspicture}[shift=-1.17](-.2,-.8)(1.7,1.6)
\rput(0.707,-.35){\scriptsize $#1$}
\rput(1.514,.507){\scriptsize $#2$}
\rput(.707,1.314){\scriptsize $#3$}
\rput(-.1,.507){\scriptsize $#4$}
\psline[linecolor=blue,linewidth=1.5pt](.35,-.6)(.35,1.6)
\psline[linecolor=blue,linewidth=1.5pt](1.05,-.6)(1.05,1.6)
\rput{45}(0.707,-0.25){
\facegrid{(0,0)}{(1,1)}
\psline[linestyle=dashed](0,0)(1,1)
\psarc[linewidth=1.5pt,linecolor=blue](1,0){0.5}{90}{180}
\psarc[linewidth=1.5pt,linecolor=blue](0,1){0.5}{270}{360}
\psarc[linecolor=red](0,0){0.15}{0}{90}
}
\rput(0.35,-0.75){\scriptsize$j$}
\rput(1.05,-0.75){\scriptsize$j\!+\!1$}
\end{pspicture}
}

\def\Ejr#1#2#3#4{\begin{pspicture}[shift=-1.17](-.2,-.8)(1.2,1.6)
\rput(0.707,-.3){\scriptsize $#1$}
\rput(1.514,.507){\scriptsize $#2$}
\rput(0.707,1.364){\scriptsize $#3$}
\rput(-.1,.507){\scriptsize $#4$}
\psline[linecolor=blue,linewidth=1.5pt](.35,-.6)(.35,1.6)
\psline[linecolor=blue,linewidth=1.5pt](1.05,-.6)(1.05,1.6)
\rput{45}(0.707,-0.2){
\facegrid{(0,0)}{(1,1)}
\pspolygon[fillstyle=solid,fillcolor=lightblue,linewidth=0pt](1,1)(0,1)(1,0)(1,1)
\psline[linestyle=dashed](0,1)(1,0)
\psarc[linewidth=1.5pt,linecolor=blue](0,0){0.5}{0}{90}
\psarc[linewidth=1.5pt,linecolor=blue](1,1){0.5}{180}{270}
\psarc[linecolor=red](0,0){0.15}{0}{90}
}
\rput(0.35,-0.75){\scriptsize$j$}
\rput(1.05,-0.75){\scriptsize$j\!+\!1$}
\end{pspicture}
}
\def\Esquare{
\begin{pspicture}[shift=-1.17](-.2,-.8)(1.2,1.6)
\rput{45}(0.707,-0.2){
\facegrid{(0,0)}{(1,1)}
\pspolygon[fillstyle=solid,fillcolor=lightblue,linewidth=0pt](1,1)(0,1)(1,0)(1,1)
\psline[linestyle=dashed](0,1)(1,0)
\psarc[linewidth=1.5pt,linecolor=blue](0,0){0.5}{0}{90}
\psarc[linewidth=1.5pt,linecolor=blue](1,1){0.5}{180}{270}
\psarc[linecolor=red](0,0){0.15}{0}{90}
}
\end{pspicture}
}
\def\halffaced{\begin{pspicture}[shift=-.57](-.2,-.2)(1.7,0.9)
\rput[bl]{135}(1.414,0.2){\pspolygon[fillstyle=solid,fillcolor=lightblue,linewidth=0pt](0,0)(1,0)(1,1)(0,0)\psline(0,0)(1,0)(1,1)\psline[linestyle=dashed](0,0)(1,1)\psarc[linewidth=1.5pt,linecolor=blue](1,0){0.5}{90}{180}}
\rput(.707,1.07){\scriptsize $a$}
\rput(-0.1,0.2){\scriptsize $b$}
\rput(1.514,0.2){\scriptsize $b$}
\end{pspicture}
}
\def\halffacel{\begin{pspicture}[shift=-.57](-.2,-.2)(1.7,1.2)
\rput[br]{-45}(0,0.707){\pspolygon[fillstyle=solid,fillcolor=lightlightblue,linewidth=0pt](0,0)(1,0)(1,1)(0,0)\psline(0,0)(1,0)(1,1)\psline[linestyle=dashed](0,0)(1,1)\psarc[linewidth=1.5pt,linecolor=blue](1,0){0.5}{90}{180}}
\rput(.707,-.1){\scriptsize $a$}
\rput(-0.1,0.707){\scriptsize $b$}
\rput(1.514,0.707){\scriptsize $b$}
\end{pspicture}
}
\def\TLia{
\begin{pspicture}[shift=-10](0,2.2)(2,4)
\psline[linewidth=1.5pt,linecolor=blue](0.65,0.3)(0.65,1.5)
\psline[linewidth=1.5pt,linecolor=blue](1.35,0.3)(1.35,1.5)
\psline[linewidth=1.5pt,linecolor=blue](0.65,3.2)(0.65,4.1)
\psline[linewidth=1.5pt,linecolor=blue](1.35,3.2)(1.35,4.1)
\pscircle[linewidth=1.5pt,linecolor=blue](1,2.225){0.53}
\rput(0.8,1.4){\Esquare}
\rput(0.8,2.825){\Esquare}
\rput(0.65,0.1){\scriptsize$j$}
\rput(1.35,0.1){\scriptsize$j\!+\!1$}
\psline[linestyle=dashed](0.3,1.5)(0.3,2.95)
\psline[linestyle=dashed](1.7,1.5)(1.7,2.95)
\pscircle[fillstyle=solid,fillcolor=black](1,2.225){0.05}
\rput(1,0.7){\scriptsize$a$}
\rput(0.2,1.5){\scriptsize$b$}
\rput(1.8,1.5){\scriptsize$b$}
\rput(0.8,2.2){\scriptsize$d$}
\rput(0.2,2.9){\scriptsize$b$}
\rput(1.8,2.9){\scriptsize$b$}
\rput(1,3.75){\scriptsize$c$}
\end{pspicture}
}
\def\TLib{
\begin{pspicture}[shift=-2](0,0)(2,1.3)
\rput(.8,0){\Ejr abcb}
\end{pspicture}
}
\def\TLLoop{
\begin{pspicture}[shift=-.77](1.8,1.8)
\rput[br]{-45}(0.18,1.614){\pspolygon[fillstyle=solid,fillcolor=lightlightblue,linewidth=0pt](0,0)(1,0)(1,1)(0,0)\psline(0,0)(1,0)(1,1)\psline[linestyle=dashed](0,0)(1,1)\psarc[linewidth=1.5pt,linecolor=blue](1,0){0.5}{90}{180}}
\rput{135}(1.6,0.2){\pspolygon[fillstyle=solid,fillcolor=lightblue,linewidth=0pt](0,0)(1,0)(1,1)(0,0)\psline(0,0)(1,0)(1,1)\psline[linestyle=dashed](0,0)(1,1)\psarc[linewidth=1.5pt,linecolor=blue](1,0){0.5}{90}{180}}
\psline[linestyle=dashed](0.2,0.2)(0.2,1.6)
\psline[linestyle=dashed](1.6,0.2)(1.6,1.6)
\pscircle[linecolor=blue,linewidth=1.5pt](0.9,0.9){0.53}
\rput(0.1,0.1){\scriptsize$b$}
\rput(1.7,0.1){\scriptsize$b$}
\rput(0.1,1.7){\scriptsize$b$}
\rput(1.7,1.7){\scriptsize$b$}
\rput(0.7,0.9){\scriptsize$d$}
\pscircle[fillstyle=solid,fillcolor=black](0.9,0.9){0.05}
\end{pspicture}
}
\def\TLiiA{
\begin{pspicture}(0,2.2)(2.8,4.2)
\psline[linecolor=blue,linewidth=1.5pt](0.65,.4)(.65,1.2)
\psline[linecolor=blue,linewidth=1.5pt](1.35,.4)(1.35,1.2)
\psline[linecolor=blue,linewidth=1.5pt](0.65,3.2)(.65,4)
\psline[linecolor=blue,linewidth=1.5pt](1.35,3.2)(1.35,4)
\psline[linecolor=blue,linewidth=1.5pt](2.075,.4)(2.075,4)
\rput(0.8,1.4){\Esquare}
\rput(0.8,2.814){\Esquare}
\rput(1.51,2.11){\Esquare}
\rput(.65,.2){\scriptsize{$j$}}
\rput(1.35,.2){\scriptsize{$j\!+\!1$}}
\rput(2.05,.2){\scriptsize{$j\!+\!2$}}
\psarc[linewidth=1.5pt,linecolor=blue](1,2.225){0.505}{135}{225}
\rput(1,0.7){\scriptsize$a$}
\rput(1.8,1.4){\scriptsize$b$}
\rput(.2,1.5){\scriptsize$b$}
\rput(.2,2.95){\scriptsize$b$}
\rput(1.8,3){\scriptsize$b$}
\rput(.8,2.2){\scriptsize$c$}
\rput(2.55,2.2){\scriptsize$c$}
\rput(1,3.8){\scriptsize$d$}
\psline[linestyle=dashed](0.3,1.5)(0.3,2.95)
\end{pspicture}
}
\def\TLiiB{
\begin{pspicture}(3,2)
\rput[l](0.1,0){\Ejr abdb}
\psline[linecolor=blue,linewidth=1.5pt](2.05,-1)(2.05,1.2)
\rput(2.05,-1.15){\scriptsize$j\!+\!2$}
\end{pspicture}
}
\def\TLUII{
\begin{pspicture}(2.6,1)
\rput(0,-.8){
\pspolygon[fillstyle=solid,fillcolor=lightlightblue,linewidth=0pt](.2,1.614)(.9,.9)(1.614,1.614)(.2,1.614)
\rput(1.4,.8){\Esquare}
\pspolygon[fillstyle=solid,fillcolor=lightblue,linewidth=0pt](1.6,.2)(.2,.2)(.9,.9)(1.6,.2)
\psline[linestyle=dashed](1.614,0.2)(.2,.2)(.2,1.614)(1.614,1.614)
\psline(.2,.2)(.9,.9)(.2,1.614)
\psarc[linecolor=blue,linewidth=1.5pt](0.9,.9){.51}{45}{315}
\psarc[linecolor=red](.9,.9){.15}{45}{135}
\rput(.1,.1){\scriptsize$b$}
\rput(1.7,.1){\scriptsize$b$}
\rput(.7,.9){\scriptsize$c$}
\rput(2.4,.9){\scriptsize$c$}
\rput(.1,1.7){\scriptsize$b$}
\rput(1.7,1.7){\scriptsize$b$}
}
\end{pspicture}
}

The Yang-Baxter equation for critical \ade lattice models can be proved diagrammatically by extending the arguments of \cite{BaxOwcz87,GillThesis}. 
The local face transfer operator associated with the face weights (\ref{faceWeights}) can be written in terms of the Temperley-Lieb (TL) generators $e_j$ as
\bea
X_j(u)=\sin(\lambda\!-\!u)I+\sin u\; e_j\label{faceTransfer}
\eea
or diagrammatically by
\bea
X_j(u) = \facer abcdu = \sin(\lambda-u)\, \delta_{a,c}\; \Idr abcd + \sin u\, \delta_{b,d}\;\Ejr abcd
\eea
where the first face is the identity $I$ and the second face is the operator $e_{j}$ for $j = 1,2,\ldots,n-1$. 
The face transfer matrices (\ref{faceTransfer}) automatically satisfies the Yang-Baxter equations if the generators $e_j$ satisfy the Temperley-Lieb algebra
\bea
e_j^2=2\cos\lambda\;e_j,\qquad e_je_{j\pm 1}e_j=e_j,\qquad e_je_k=e_ke_j,\quad |j-k|\ge 2
\eea
The Temperley-Lieb generators factor into triangles
\begin{subequations}
\begin{align}
\halffaced &= h'_{ab}G_{ab},\qquad h'_{ab}=\tilde{g}_a\frac{\psi_a}{\psi_b}\\[-12pt]
\halffacel &= h_{ab}G_{ab},\qquad h_{ab}=\tilde{g}^{-1}_a
\end{align}
\end{subequations}
where $G$ is the adjacency matrix, $h_{ab}$ and $h_{ab}'$ are triangle weights, $\psi_a$ are the components of the eigenvector corresponding to the eigenvalue $2\cos\lambda$ of $G$ with $\lambda=\tfrac{(m'-m)\pi}{m'}$ and $\tilde{g}_a$ are arbitrary gauge factors. 
For symmetry, it is better to choose $\tilde{g}_a=g_a/\sqrt{\psi_a}$ as in (\ref{faceWeights}) with $g_a=1$. 
The first TL relation follows from 
\begin{subequations}
\begin{align}
\TLia &=2\cos\lambda\; \TLib\\[45pt]
\TLLoop = \sum_d G_{bd} h'_{db} h_{db} &=\sum_{d}G_{bd}\frac{\psi_{d}}{\psi_{b}} = 2\cos\lambda,\qquad \lambda=\tfrac{(m'-m)\pi}{m'},\quad m'-m\in\Exp(G)
\end{align}
\end{subequations}
where the weight of the central square is determined by the eigenvalue equation $\sum_{d}G_{bd}\,\psi_{d} = 2\cos\lambda\;{\psi_{b}}$. 
The second TL relations follow from
\begin{subequations}
\begin{align}
\centering
\TLiiA &= \TLiiB\\[45pt]
\TLUII = h_{bc}h'_{bc}h_{cb}h'_{cb}\,G_{bc} &= \frac{\psi_{c}}{\psi_{b}}\frac{\psi_b}{\psi_c}\,G_{bc} = G_{bc} = 1\nonumber\\[-10pt]
\end{align}
\end{subequations}
where the scalar weight is 1 since $b$ and $c$ are allowed neighbours. The third TL relation is trivial.



\begin{thebibliography}{99}
%

	\bib{BaxBook82}
	R.J.~Baxter, 
	{\it Exactly Solved Models in Statistical Mechanics},
	Academic Press (1982).
	
	\bibitem{FMS97}
	P.~Di Francesco, P.~Mathieu, D.~S\'en\'echal, 
	{\em Conformal Field Theory},
	Springer (1997).
		
	\bib{ABF84} 
	G.E.~Andrews, R.J.~Baxter, P.J.~Forrester, 
	{\em Eight-vertex SOS model and generalized Rogers-Ramanujan-type identities}, 
	J.~Stat.~Phys.~{\bf 35} (1984) 193--266.
	
	\bib{FB} 
 	P.J.~Forrester, R.J.~Baxter, 
 	{\em Further exact solutions of the eight-vertex SOS model and generalizations of the Rogers-Ramanujan identities}, 
 	J. Stat. Phys. {\bf 38} (1985) 435--472.
	
	\bib{Pasquier87a}
	V.~Pasquier,
	{\em Two-dimensional critical systems labelled by Dynkin diagrams},
	Nucl. Phys. {\bf B285} (1987) 162--172.
	
	\bib{Pasquier87b}
	V.~Pasquier,
	{\em Operator content of the ADE lattice models},
	J. Phys. A: Math. Gen. {\bf 20} (1987) 5707--5717.
	
	\bib{Pasquier87c}
	V.~Pasquier,
	{\em Lattice derivation of modular invariant partition functions on the torus},
	J. Phys. A: Math. Gen. {\bf 20} (1987) L1229--L1237.
	
	\bib{BPZ84} 
	A.A.~Belavin, A.M.~Polyakov, A.B.~Zamolodchikov,
	{\em Infinite conformal symmetry in two-dimensional quantum field theory},
	Nucl.~Phys.~{\bf B241} (1984) 333--380.
	
	\bibitem{MooreSeiberg}
 	G. Moore, N. Seiberg, 
 	{\em Classical and quantum conformal field theory}, 
 	Commun. Math. Phys. {\bf 123} (1989) 177--254.
	
	\bib{GKO1985}
	P.~Goddard, A.~Kent, and D.~Olive, 
	{\em Virasoro algebras and coset space models}, 
	Phys. Lett. {\bf B152} (1985) 88--92.

	\bibitem{CIZ87}
	A.~Cappelli, C.~Itzykson, J.~B. Zuber.
	{\em The ${A}$-${D}$-${E}$ classification of minimal and ${A}_1^{(1)}$ conformal invariant theories}, 
	{Commun. Math. Phys.} {\bf 113} (1987) 1--26.
	
	\bib{BPPZ} 
 	R.E. Behrend, P.A. Pearce, V.B. Petkova, J.-B. Zuber, 
 	{\em Boundary conditions in rational conformal field theories}, 
 	Nucl. Phys. {\bf B579} (2000) 707--773.
	
	\bib{PetkovaZuber2001}
	V.B.~Petkova, J.-B.~Zuber, 
	{\em Generalised twisted partition functions}, 
	Physics Letters {\bf B504} (2001) 157--164;
	{\em Conformal field theories, graphs and quantum algebras},
	MathPhys Odyssey 2001: Integrable Models and Beyond, 415--435.
	
	\bib{PetkovaZuberFaces2001}
	V.B.~Petkova, J.-B.~Zuber, 
	{\em The many faces of Ocneanu cells}, 
	Nucl. Phys. {\bf B603} (2001) 449--496.
	
	\bibitem{PRasmussen24}
	P.A.~Pearce, J.Rasmussen, 
	{\em Ocneanu algebra of seams: Critical unitary $E_6$ RSOS lattice model}, 
	Matrix Annals, \arxiv{hep-th}{2409.06236} (2025).
	
	\bibitem{PRasmussen25}
	P.A.~Pearce, J.Rasmussen, 
	{\em Ocneanu algebra of seams: Critical unitary \ade RSOS lattice models},
	in preparation (2025).
	
	\fontfamily{ptm}\selectfont
	\bib{Ocneanu}  A.~Ocneanu, 
	{\em Paths on Coxeter Diagrams: From Platonic Solids and Singularities to
	Minimal Models and Subfactors}, in Lectures on Operator Theory, Fields Institute,
	Waterloo, Ontario, April 26–30, 1995, (Notes taken by S. Goto) Fields Institute Monographs, 
	AMS 1999, Rajarama Bhat et al, eds;\\
	{\em Quantum symmetries for SU(3) CFT Models}, 
	Lectures at Bariloche Summer School, Argentina, Jan 2000,
	AMS Contemporary Mathematics, R.~Coquereaux, A.~Garcia and R.~Trinchero, eds;\\
	{\em The classification of subgroups of quantum SU(N)}, Bariloche, 2000:\\
	https://www.cpt.univ-mrs.fr/{\textasciitilde}coque/Bariloche2000/Bariloche2000/Bariloche2000.html
	
	\bibitem{Frohlich:2010maph.conf..608F}
	J.~{Fr{\"o}hlich}, J.~{Fuchs}, I.~{Runkel}, C.~{Schweigert},
	 \href{http://dx.doi.org/10.1142/9789814304634_0056}{{\em Defect Lines,
	 Dualities and Generalised Orbifolds},} in {\em XVIth International Congress on Mathematical Physics, 3-8 August 2009, Prague}, P.~{Exner}, ed.,
	 pp.~608--613,
	\newblock 2010.
	\newblock \href{http://arxiv.org/abs/0909.5013}{{\ttfamily arXiv:0909.5013
	 [math-ph]}}.

	\bibitem{Gaiotto:2015JHEP...02..172G}
	D.~{Gaiotto}, A.~{Kapustin}, N.~{Seiberg}, B.~{Willett}, {\em Generalized
	 global symmetries}, \href{http://dx.doi.org/10.1007/JHEP02(2015)172}{{
	 J. High Energy Phys.} (2015) 172},
	 \href{http://arxiv.org/abs/1412.5148}{{\ttfamily arXiv:1412.5148 [hep-th]}}.

	\bibitem{Bhardwaj:2018JHEP...03..189B}
	L.~{Bhardwaj} and Y.~{Tachikawa}, {\em On finite symmetries and their gauging in
	 two dimensions}, \href{http://dx.doi.org/10.1007/JHEP03(2018)189}{{
	 J. High Energy Phys.} (2018) 189},
	 \href{http://arxiv.org/abs/1704.02330}{{\ttfamily arXiv:1704.02330
	 [hep-th]}}.

	\bibitem{Fateev:1985mm}
	V.A. Fateev, A.B. Zamolodchikov, {\em Parafermionic Currents in the
	 Two-Dimensional Conformal Quantum Field Theory and Selfdual Critical Points
	 in Z(n) Invariant Statistical Systems}, {Sov. Phys. JETP} {\bfseries
	 62} (1985) 215--225.

	\bibitem{Affleck:1988NuPhB.305..582A}
	I.~{Affleck}, {\em Critical behaviour of $SU(n)$ quantum chains and topological
	 non-linear {$\sigma$}-models},
 	\href{http://dx.doi.org/10.1016/0550-3213(88)90117-4}{{Nucl. Phys.}
	 {\bfseries B305} (1988) 582--596}.

	\bibitem{Affleck:1989JPhA...22..511A}
	I.~{Affleck}, D.~{Gepner}, H.J. {Schulz}, T.~{Ziman}, {\em Critical
	 behaviour of spin-s Heisenberg antiferromagnetic chains: analytic and
	 numerical results},
	 \href{http://dx.doi.org/10.1088/0305-4470/22/5/015}{{J. Phys.} {\bfseries
	 A22} (1989) 511--529}.
	 
	\bib{BazhResh1989}
	V.V.~Bazhanov, N.Yu.~Reshetikhin, 
	{\em Critical RSOS models and conformal field theory}, 
	Int.~J.~Mod.~Phys.~{\bf A4} (1989) 115--142.	
	
	\bib{KP92}
	A.~Kl\"umper, P.A.~Pearce, 
	{\em Conformal weights of RSOS lattice models and their fusion hierarchies}, 
	Physica~{\bf A183} (1992) 304--350.	

	\bibitem{Fuchs:2002NuPhB.646..353F}
	J.~{Fuchs}, I.~{Runkel}, C.~{Schweigert}, {\em TFT construction of RCFT
 	correlators I: partition functions},
	 \href{http://dx.doi.org/10.1016/S0550-3213(02)00744-7}{{\em Nucl. Phys.} 
	 {\bfseries B646} no.~3, (2002) 353--497},
	 \href{http://arxiv.org/abs/hep-th/0204148}{{\ttfamily arXiv:hep-th/0204148
 	[hep-th]}}.
			
	\bibitem{STRSaleur2025}
	M.~Sinha, T.S. Tavares, A. Roy, H.~Saleur, 
	{\em Integrability and lattice discretizations of all topological defect lines in minimal CFTs}, 
	\arxiv{}{2509.04257} (2025).
	
	\bib{Verlinde88} 
 	E. Verlinde, 
 	{\em Fusion rules and modular transformations in 2D conformal field theory}, 
 	Nucl. Phys. {\bf B300} (1988) 360--376.
	
	\bib{PasquierThesis}
	V. Pasquier, {\em These d'Etat}, Orsay (1988).
	
	\bibitem{AL91}
	I.~Affleck, A.W.W.~Ludwig,
	{\em Universal noninteger ``ground-state degeneracy'' in critical quantum systems},
	Phys. Rev. Lett. {\bf 67} (1991) 161--164.
	
	\bibitem{DGMN23} G.~Di~Giulio, R.~Meyer, C.~Northe, H.~ Scheppach, S.~Zhao, 
	{\em On the boundary conformal field theory approach to symmetry-resolved entanglement}, 
	SciPost~Physics~Core, {\bf 6} (3) (2023) 049.
	
	\bibitem{Northe23} C.~Northe, 
	{\em Entanglement resolution with respect to conformal symmetry}, 
	Phys. Rev. Lett. {\bf 131} (2023) 151601.
	
	
	\bibitem{KMOP23} Y.~Kusuki, S.~Murciano, H.~Ooguri,  S.~Pal, 
	{\em Symmetry-resolved entanglement entropy, spectra and boundary conformal field theory}, 
	Journal of High Energy Physics, 2023 (11) (2023) 216.
	
	\bibitem{DMVSB24} A.~Das, J.~Molina-Vilaplana, P.~Saura-Bastida, 
	{\em Generalized symmetry resolution of entanglement in conformal field theory for twisted and anyonic sectors}, 
	Phys.~Rev. {\bf D110} (12) (2024) 125005.
	
	\bibitem{HQ24} J.~Heymann, T.~Quella, 
	{\em Revisiting the symmetry-resolved entanglement for non-invertible symmetries in 1+1d conformal field theories}, 
	to appear (2024).

	\bibitem{CRZ24} 
	Y.~Choi, B.C. Rayhaun, Y.~Zheng, 
	{\em Noninvertible symmetry-resolved Affleck-Ludwig-Cardy formula and entanglement entropy from the boundary tube algebra}, 
	Phys.~Rev.~Lett. {\bf 133} (25) (2024) 251602.
	
	\bibitem{SBDSMV24} P.~Saura-Bastida, A.~Das, G.~Sierra, J.~Molina-Vilaplana,
	{\em Categorical-symmetry resolved entanglement in conformal field theory}, 
	Phys.~Rev. {\bf D109} (10) (2024) 105026.
	
	\bibitem{BGPS25}
	A.~Bhattacharyya, S.~Ghosh, S.~Pal, J.~Santara,
	{\em On the resolution of categorical symmetries in (non-) unitary rational CFTs}, 
	\arxiv{}{2511.16363} (2025).
	
	\bibitem{BPO96} 
	R.E.~Behrend, P.A. Pearce, D. L. O'Brien, 
	{\em Interaction-round-a-face models with fixed boundary conditions: The ABF fusion hierarchy}, 
	J. Stat. Phys. {\bf 84} (1996) 1--48.
	
	\bibitem{BP2001}
	R.E. Behrend, P.A. Pearce, 
	{\em Integrable and conformal boundary conditions for $\widehat{s\ell}(2)$ \ade lattice models and unitary minimal conformal field theories}, 
	J.~Stat.~Phys. {\bf 102} (2001) 577--640.	
 
	\bibitem{AlgGraphTheory}
	C.~Godsil, G.~Royle,
	{\em Algebraic Graph Theory}, 
	Springer-Verlag, New York (2001).
	
	\bibitem{BPZ1998}
	R.E.~Behrend, P.A.~Pearce, J.-B.~ Zuber,
	{\em Integrable boundaries, conformal boundary conditions and \ade fusion rules},
	J. Phys. A: Math. Gen. {\bf 31} (1998) L763--L770.
	
	\bibitem{KP2006}
	A.~Kitaev, J. Preskill, 
	{\em Topological entanglement entropy}, 
	Phys. Rev. Lett. {\bf 96} (2006) 110404.
	
	\bibitem{FNO2017}
	L.~Fiedler, P. Naaijkens, T.J.~Osborne, 
	{\em Jones index, secret sharing and total quantum dimension}, 
	New J. Phys. {\bf 19} (2017) 023039 pp1-24.
	
	\bibitem{Tu2017}
	H.-H. Tu,
	{\em Universal entropy of conformal critical theories on a Klein bottle},
	Phys. Rev. Lett. {\bf 119} (2017) 261603.
	
	\bibitem{FZ93}
	P.~Di~Francesco, J.-B.~Zuber, 
	{\em Fusion potentials. I}, 
	J. Phys. A {\bf 26} (1993) 1441

	\bibitem{RasmussenStudent}
	J.~Fyfield,
	{\em Infinite dimensional Lie algebras: characters and fusion},
	Bachelor of Advanced Science Honours Thesis; Supervisor: J.~Rasmussen, 
	University of Queensland (2018).
	
	\bibitem{PRZ2006}
	P.A. Pearce, J. Rasmussen, J.-B. Zuber, 
	{\em Logarithmic minimal models}, 
	J. Stat. Mech. (2006) P11017.
	
	\bibitem{PR2011}
	P.A. Pearce, J. Rasmussen, 
	{Coset graphs in bulk and boundary logarithmic minimal models}, 
	Nucl. Phys. {\bf B846} (2011) 616--649.
	
	\bibitem{LogTY2014}
	A.~Morin-Duchesne, P.A. Pearce, J. Rasmussen, 
	{\em Fusion hierarchies, $T$-systems, and $Y$-systems of logarithmic minimal models}, 
	J. Stat. Mech. (2014) P05012.
	
	\bibitem{NahmRT93}
	W.~Nahm, A.~Recknagel, M.~Terhoeven,
	{\em Dilogarithm identities in conformal field theory},
	Mod. Phys. Lett. {\bf 08} (1993) 1835--1847.
	
	\bibitem{vBS} 
	J.-M. Vanden Broeck, L.W. Schwartz, 
 	{\em A one-parameter family of sequence transformations}, 
 	SIAM J. on Math. Anal. {\bf 10} (1979) 658--666.
	
	\bibitem{L13} A.M.~L\"auchli,
	{\em Operator content of real-space entanglement spectra at conformal critical points}, 
	\arxiv{}{1303.0741} (2013).
	
	\bibitem{OT15} K.~Ohmori, Y.~Tachikawa,
	{\em Physics at the entangling surface}, 
	J. Stat. Mech. {\bf 1504} (2015) P04010.
	
	\bibitem{CT16} J.~Cardy, E.~Tonni,
	{\em Entanglement Hamiltonians in two-dimensional conformal field theory}, 
	J.~Stat.~Mech. {\bf 1612} (12) (2016) 123103.
	
	\bib{PK91}
	P.A.~Pearce, A.~Kl\"umper, 
	{\em Finite-size corrections and scaling dimensions of solvable lattice models: An analytic method},
	Phys. Rev. Lett. {\bf 66} (1991) 974--977.

	\bibitem{KP91}
	A.~Kl\"umper, P.A.~Pearce,
	{\em Analytic calculation of scaling dimensions: Tricritical hard squares and critical hard hexagons}, 
	 J. Stat. Phys. {\bf 64} (1991) 13--76.
	 
	\bibitem{LMC91}
	M.~L\"assig, G.~Mussardo, L.L.~Cardy,
	{\em The scaling region of the tricritical Ising model in two dimensions}, 
	Nucl. Phys. {\bf B348} (1991) 591--618.
	
	\bibitem{DKMM94}
	S.~Dasmahapatra, R.~Kedem, M.M.~McCoy, E.~Melzer,
	{Virasoro characters from Bethe equations for the critical ferromagnetic three-state Potts model}, 
	J. Stat. Phys. {\bf 74} (1994) 239--274.

	\bibitem{SaleurEtAl2024}
	M.~Sinha, L.~Grans-Samuelsson, A.~Roy, H.~Saleur, 
	{\em Lattice realizations of topological defects in the critical (1+1)-d three-state Potts model}, 
	J. High Energy Phys. 225 (2024) 1--55.
	
	\bibitem{OBP95}
	D.~O'Brien, P.A.~Pearce, 
	{\em Lattice realizations of unitary minimal modular invariant partition functions}, 
	J. Phys. A {\bf 28} (1995) 4891--4905.

	\bibitem{Iino21}
	S.~Iino, 
	{\em Boundary CFT and tensor network approach to surface critical phenomena of the tricritical 3-state Potts model}, 
	J. Stat. Phys. {\bf 182} (2021) 56.	
	
	\bibitem{LeeYang} 
	T.D. Lee and C.N. Yang, 
	{\em Statistical theory of equations of state and phase transitions. I. Theory of condensation; II. Lattice gas and Ising model}, 
	Phys. Rev. {\textbf 87} (1952) 404; 410.
	
	\bibitem{BLZ97}
	V.V. Bazhanov, S.L. Lukanov, A.B. Zamolodchikov,
	{\em Quantum field theories in finite volume: Excited state energies},
	Nuccl. Phys. {\bf B489} (1997) 497--531.

	\bibitem{DPTW} 
	P. Dorey, A.J. Pocklington, R. Tateo, G. Watts, 
	{\em TBA and TCSA with boundaries and excited states},
	Nucl.Phys. \textbf{B525} (1998) 641--663.

	\bibitem{DRTW} 
	P. Dorey, I. Runkel, R. Tateo and G. Watts, 
	{\em g-function flow in perturbed boundary conformal field theories}, 
	Nucl. Phys. {\bf B578} (2000) 85--122.
	
	\bibitem{BDP2015}
	Z.~Bajnok, O. el Deeb, P.A.~Pearce,
	{\em Finite-volume spectra of the Lee-Yang model}, 
	JHEP {\bf 04} (2015) 073 pp~1--42.
	
	\bibitem{OcneanuInter}
	A.~Ocneanu,
	{\em Quantized groups, string algebras and Galois theory for algebras in operator algebras and applications}, 
	Lond. Math. Soc. Lecture Series {\bf 136} (1988) 119--170.
	
	\bibitem{PZ93}
	P.A.~Pearce,Y-K,~Zhou, 
	{\em Intertwiners and $A$-$D$-$E$ lattice models},
	Int. J. Mod. Phys. {\bf B7} (1993) 3649-3705.
	
	\bibitem{ArdonneEtAl2010}
	E.~Ardonne ,J.~Gukelberger, A.W.W.~Ludwig, S.~Trebst, M.~Troyer,
	{\em Microscopic models of interacting Yang-Lee anyons}, 
	New J. Phys. {\bf 13} (2011) 045006.
		
	\bib{BLZ}
	V.V. Bazhanov, S.L. Lukyanov, A.B. Zamoldchikov, 
	{\em Integrable structure of conformal field theory, quantum KdV theory and thermodynamic Bethe ansatz}, 
	Commun. Math. Phys. {\bf 177} (1996) 381--398; \\
	{\em Integrable structure of conformal field theory II. $Q$-operator and DDV equation}, 
	Commun. Math. Phys. {\bf 190} (1997) 247--278; \\
	{\em Integrable structure of conformal field theory III. The Yang-Baxter relation}, 
	Commun. Math. Phys. {\bf 200} (1999) 297--324; \\
	{\em Quantum field theories in finite volume: Excited state energies}, 
	Nucl. Phys. {\bf B489} (1997) 487--531.
	
	\bibitem{FPW2009}
	G.~Feverati, P.A.~Pearce, N.S~Witte,
	{\em Physical combinatorics quasiparticles}, 
	J. Stat. Mech. (2009) P10013.
	
	\bib{Zhou97}
	Y.~Zhou. W.~Sun,
	{\em Solving the fusion hierarchies of $D$ and $E$ lattice models}, 
	Nucl. Phys. {\bf B504} (1997) 719--737.
	
	\bibitem{Kirillov94}
	A.N.~Kirillov, 
	{\em Dilogarithm identities, partitions, and spectra in conformal field theory, Part I}, 
	Algebra i Analiz {\bf 6} (1994) 152--175; \arxiv{hep-th}{9212150} (1992).
	
	\bibitem{Kirillov95}
	A.N. Kirillov,
	{\em Dilogarithm identities},
	Prog. Theor. Phys. Supplement, {\bf 118} (1995) 61--142.
	
	\bibitem{RasLogLimit}
	 J.~Rasmussen,
	 {\em Logarithmic limits of minimal models},
	 Nucl. Phys. {\bf B701} (2004) 516--528; 
	 
	\bib{ChuiEtAl2001}
	C.H.~Otto~Chui, C.~Mercat, W.P.~Orrick, P.A.~Pearce,
	{\em Integrable lattice realizations of conformal twisted boundary conditions},
	Phys. Lett. {\bf B517} (2001) 429--435.
	
	\bib{ChuiEtAlOdyssey2001}
	C.H.~Otto~Chui, C.~Mercat, P.A.~Pearce,
	{\em Integrable boundaries and universal TBA functional equations},
	MathPhys Odyssey 2001: Integrable Models and Beyond, 391--413.
	
	\bib{ChuiEtAl2003}
	C.H.~Otto~Chui, C.~Mercat, P.A.~Pearce,
	{\em Integrable and conformal twisted boundary conditions for $sl(2)$ $A$-$D$-$E$ lattice models},
	J. Phys. A: Math. Gen. {\bf 36} (2003) 2623--2662.
	
\def\arxiv#1#2#3{\href{https://arxiv.org/abs/#2.#3}{\textsf{arXiv:#2.#3\,[#1]}}}
	\bib{BelleteteEtAl2023}
	J.~Bellet\^ete, A.M.~Gainutdinov, J.L~Jacobsen, H.~Saleur, T.S.~Tavares,
	{\em Topological defects in periodic RSOS models and anyonic chains}, 
	\arxiv{math-ph}{2003}{11293};\\
	{\em Topological defects in lattice models and affine Temperley-Lieb algebra},
	Commun. Math. Phys. {\bf 400} (2023) 1203--1254.
	
\def\arxiv#1#2#3{\href{https://arxiv.org/pdf/#2.#3}{\textsf{arXiv:#2.#3\,[#1]}}}
	\bib{TavaresEtAl2024}
	T.S.~Tavares, M.~Sinha, L.~Grans-Samuelsson, A.~Roy, H.~Saleur, 
	{\em Integrable RG flows on topological defect lines in 2D conformal field theories}, 
	\arxiv{hep-th}{2408}{08241} (2024).
	
	\bib{SYSRS24}
	M.~Sinha, F.~Yan, L.~Grans-Samuelsson, A.~Roy, H.~Saleur,
	{\em Lattice realizations of topological defects in the critical (1+1)-d three state Potts model}, 
	J.~High Energy Phys, {\bf 07} (2024) 225.
	
	\bib{Wenzl87}
	H.~Wenzl,
	{\em On sequences of projections}, 
	C. R. Math. Rep. Acad. Sci. Canada, 9(1), 5–9 (1987).
	
	\bib{Jones97}
	V.F.R.~Jones, 
	{\em A polynomial invariant for knots via von Neumann algebras}, 
	In Fields Medallists’ lectures, Volume 5, World Sci. Ser. 20th Century Math. 448–458, 
	World Sci. Publ., River Edge, NJ (1997).
	
	\bibitem{BaxOwcz87}
	R.J.~Baxter, A.L.~Owczarek,
	{\em A class of interaction-round-a-face models and its equivalence with an ice-type model}, 
	J. Stat. Phys. {\bf 49} (1987) 1093--1115.
	
	\bibitem{GillThesis}
	J.~Gill, 
	{\em Towards logarithmic $D$-type models},
	University of Melbourne Master of Science Thesis, supervised by Paul Pearce (2015).
		
\end{thebibliography}
\end{document}